\documentclass{aa}  

\usepackage{graphicx}
\usepackage{xcolor}
\usepackage{longtable}
\usepackage{txfonts}
\usepackage{amsmath}
\usepackage{geometry}
\usepackage{soul}
\usepackage{hyperref}
\usepackage{multirow}
\usepackage{algorithm2e}
\usepackage{multicol}
\usepackage{ragged2e}
\usepackage{caption}

\usepackage{neuralnetwork}
\usepackage{placeins}

\def\Teff{T_{\text{eff}}}
\def\kms{\mathrm{km~s}^{-1}}
\def\logg{log(g)}

\usepackage{tikz}
\usetikzlibrary{shapes,arrows}
\usetikzlibrary{matrix}
\usetikzlibrary{arrows.meta, positioning}

\tikzstyle{decision} = [diamond, draw, fill=blue!20, 
    text width=8em, text badly centered, node distance=1.5cm, inner sep=0pt]
\tikzstyle{block} = [rectangle, draw, fill=blue!20, 
    text width=30em, text centered, rounded corners, minimum height=2em]
\tikzstyle{block2} = [rectangle, draw, fill=gray!20, 
    text width=30em, text centered, rounded corners, minimum height=2em]
\tikzstyle{block3} = [rectangle, draw, fill=orange!20, 
    text width=30em, text centered, rounded corners, minimum height=2em]
\tikzstyle{block4} = [rectangle, draw, fill=orange!20, 
    text width=30em, text centered, rounded corners, minimum height=2em]
\tikzstyle{line} = [draw, -latex']
\tikzstyle{cloud} = [draw, ellipse,fill=red!20, node distance=1.5cm,
    minimum height=2em]

\tikzset{
  font=\sffamily,
  red arrow/.style={
    midway,red,sloped,fill, minimum height=3cm, single arrow, single arrow head extend=.5cm, single arrow head indent=.25cm,xscale=0.3,yscale=0.15,
    allow upside down
  },
  black arrow/.style 2 args={-stealth, shorten >=#1, shorten <=#2},
  black arrow/.default={1mm}{1mm},
  tree box/.style={draw, rounded corners, inner sep=1em},
  node box/.style={white, draw=black, text=black, rectangle, rounded corners},
}

\begin{document} 

   \title{Machine Learning for Exoplanet Detection in High-Contrast Spectroscopy}
   
    \subtitle{Revealing Exoplanets by Leveraging Hidden Molecular Signatures in Cross-Correlated Spectra with Convolutional Neural Networks}

   \author{Emily O. Garvin\inst{\ref{ethz},\ref{ethz_stat}}, Markus J. Bonse\inst{\ref{ethz}}, Jean Hayoz\inst{\ref{ethz}}, Gabriele Cugno  \inst{\ref{umich}}, Jonas Spiller \inst{\ref{ethz}}, Polychronis A. Patapis\inst{\ref{ethz}}, Dominique Petit Dit de la Roche\inst{\ref{unige}}, Rakesh Nath-Ranga\inst{\ref{STAR}}, Olivier Absil\inst{\ref{STAR}}, Nicolai F. Meinshausen \inst{\ref{ethz_stat}}, Sascha P. Quanz \inst{\ref{ethz}}}

   \institute{
   Institute for Particle Physics and Astrophysics, ETH Zürich, Wolfang-Pauli-Strasse 27, 8093 Zürich, Switzerland\label{ethz}
   \and
    Seminar für Statistik, ETH Zürich, Raemistrasse 101, 8092 Zürich, Switzerland\label{ethz_stat}
    \and Department of Astronomy, University of Michigan, Ann Arbor, MI 48109, USA\label{umich}
    \and STAR Institute, University of Liège, 19 Allée du Six Août, 4000 Liège, Belgium\label{STAR}
    \and Département d’Astronomie, Université de Genève, Sauverny, Switzerland\label{unige}}

\titlerunning{Improving exoplanet detections using machine learning on cross-correlated spectra (MLCCS)}
\authorrunning{Garvin et al.}

   \date{Received January 2, 2024; revised May 17, 2024} 

 
  \abstract
{The new generation of observatories and instruments (VLT/ERIS, JWST, ELT) motivate the development of robust methods to detect and characterise faint and close-in exoplanets. Molecular mapping and cross-correlation for spectroscopy use molecular templates to isolate a planet's spectrum from its host star. However, reliance on signal-to-noise ratio (S/N) metrics can lead to missed discoveries, due to strong assumptions of Gaussian independent and identically distributed noise.}
{We introduce machine learning for cross-correlation spectroscopy (MLCCS); the method aims to leverage weak assumptions on exoplanet characterisation, such as the presence of specific molecules in atmospheres, to improve detection sensitivity for exoplanets.}
{MLCCS methods, including a perceptron and unidimensional convolutional neural networks, operate in the cross-correlated spectral dimension, in which patterns from molecules can be identified. They flexibly detect a diversity of planets by taking an agnostic approach towards unknown atmospheric characteristics. The MLCCS approach is implemented to be adaptable and modulable for a variety of instruments and modes. We test on mock datasets of synthetic planets inserted into real noise from SINFONI at K-band.}
{The results from MLCCS show outstanding improvements. The outcome on a grid of faint synthetic gas giants shows that for a false discovery rate up to $5\%$, a perceptron can detect about $26$ times the amount of planets compared to an S/N metric. This factor increases up to $77$ times with convolutional neural networks, with a statistical sensitivity (completeness) shift from $0.7\%$ to $55.5\%$. In addition, MLCCS methods show a drastic improvement in detection confidence and conspicuity on imaging spectroscopy.}
{Once trained, MLCCS methods offer sensitive and rapid detection of exoplanets and their molecular species in the spectral dimension. They handle systematic noise and challenging seeing conditions, can adapt to many spectroscopic instruments and modes, and are versatile regarding planet characteristics, enabling the identification of various planets in archival and future data.}


\maketitle
\section{Introduction}
 
Spectroscopic observations of substellar companions are crucial for advanced characterisation of exoplanet and brown dwarf atmospheres from emission and transmission spectra. The primary objectives in characterising these atmospheres consist of constraining the molecular composition, abundances, clouds and thermal structure of exoplanet atmospheres \citep[e.g.,][]{line2016no, brogi2019retrieving}. 
These measurements offer valuable insights into the formation history of exoplanets \citep[e.g.,][]{nowak2020peering, Mollière_2022}, as well as the evolution and migration of planets with regards to snowlines \citep{madhusudhan2014toward, oberg2011effects}.

The characterisation of exoplanet atmospheres is usually conducted with dedicated methods such as grid fitting of self consistent atmospheric models \citep[e.g.,][]{charnay2018self, petrus2023jwst, morley2024sonora}, Bayesian free retrievals \citep[e.g.,][]{madhusudhan2014toward}, cross-correlation for spectroscopy (CCS) \citep[e.g.,][]{brogi2014carbon, ruffio2019radial} or even machine learning (ML) \citep[e.g.,][]{waldmann2016dreaming}. Such methods can also be merged, where \citet{vasist2023neural} implemented Bayesian retrievals with ML, \citet[][]{brogi2019retrieving, xuan2022clear, hayoz2023_crocodile} unified retrievals and CCS, \citet[and][]{marquez2018supervised, fisher2020interpreting} combined CCS and ML to characterise exoplanet atmospheres. While retrievals are usually favoured to retrieve molecular abundances, CCS methods have proven useful to detect individual molecules on exoplanets \citep[e.g.,][]{konopacky2013detection} when the planet's continuum can not be preserved during data reduction.

\subsection{Cross-Correlation Spectroscopy and Molecular Mapping}

\begin{figure*}[htb!]
\centering
\includegraphics[width=\hsize]{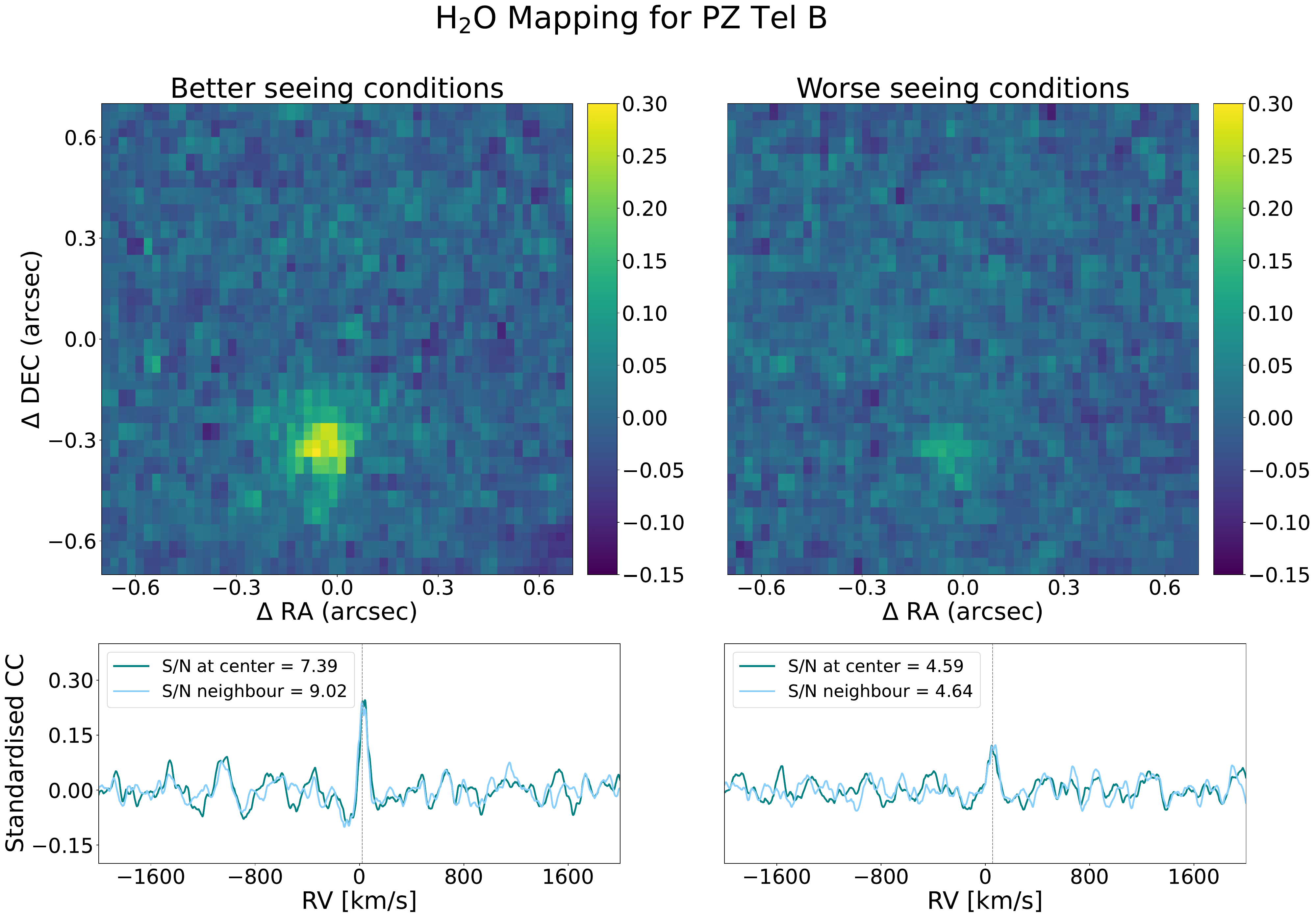}
\caption{Molecular maps of H$_2$O for real PZ\,Tel\,B data using cross-correlation for spectroscopy. This figure shows a real case example where the noise structures may reduce detection capabilities of cross-correlation methods. The brown dwarf was observed under good conditions (airmass: $1.11$, Seeing start to end: $0.77-0.72$) and lower conditions (airmass: $1.12$, Seeing: $1.73-1.54$), c.f. Appendix~\ref{supp:molmap} for full details on observing conditions. Upper plots show molecular maps of PZ\,Tel\,B, while the lower plots show the cross-correlation series along the radial velocity (RV) support for pixels at the centre of the object, and within the object's brightness area. While the brown dwarf should appear at the same spatial coordinates for respective RV locations in both cases (c.f. vertical lines), it is clearly visible when conditions are good, but hardly visible on equal scales under lower conditions.}
\label{fig:molmap_PZTEL}
\end{figure*}
The CCS method consists of applying cross-correlation of a spectral template with a planet's observed spectrum over a range of radial velocities. Depending on the similarity of the template in regards to the measured spectrum, the resulting cross-correlation series is expected to show a peak at the radial velocity (RV) of the planet (c.f. lower panels in Fig.~\ref{fig:molmap_PZTEL}). Since CCS is a good way to test whether two spectra are similar (or if synthetic spectra are accurately generated), this method can be adapted to detect individual molecules in the spectra from exoplanet atmospheres \citep[e.g.,][]{konopacky2013detection, de2013detection}

Molecular mapping \citep{hoeijmakers2018medium} is a special case of CCS, where the latter is usually applied to integral field spectroscopy (IFS) observations. It involves the cross-correlation of every spaxel (i.e. a spatial pixel with a wavelength dimension) of an IFS cube with a single molecular template. By taking a slice into the resulting cross-correlated cube at the RV of the planet, it should be possible to map molecular species (e.g. top right panel, Fig.~\ref{fig:molmap_PZTEL} and Fig.~\ref{fig:molmap_PZTEL_CO}). This approach aims to separate the planet's molecular signals from the stellar spectrum by relying on differences between molecular and atomic spectral lines. This method was applied and tested on real and simulated data from several instruments at different resolutions and spectral bands (\citealp[e.g. VLT/SINFONI,][]{hoeijmakers2018medium, petrus2021medium, cugno2021molecular}; \citealp[Keck/OSIRIS,][]{dit2018molecule}; \citealp[JWST/MIRI,][]{patapis2021direct, malin2023simulated}; \citealp[ELT/HARMONI,][]{houlle2021direct}).

Recent work involving the use of molecular mapping \citep[e.g.,][with VLT/SINFONI]{hoeijmakers2018medium}, and CCS (\citealp[e.g.,][with CRIRES]{snellen2010exoplanet, brogi2014carbon}, \citealp[and][with KECK/OSIRIS]{agrawal2023detecting}) on medium and high resolution spectra demonstrate that cross-correlation methods offer great potential for exoplanet detection. This approach serves a double purpose: detecting closer-in and fainter planets while gathering summary characterisation information which can enable a better planning for follow up observations.

\begin{figure*}[tbh!]
    \centering
    \includegraphics[width=18.2cm]{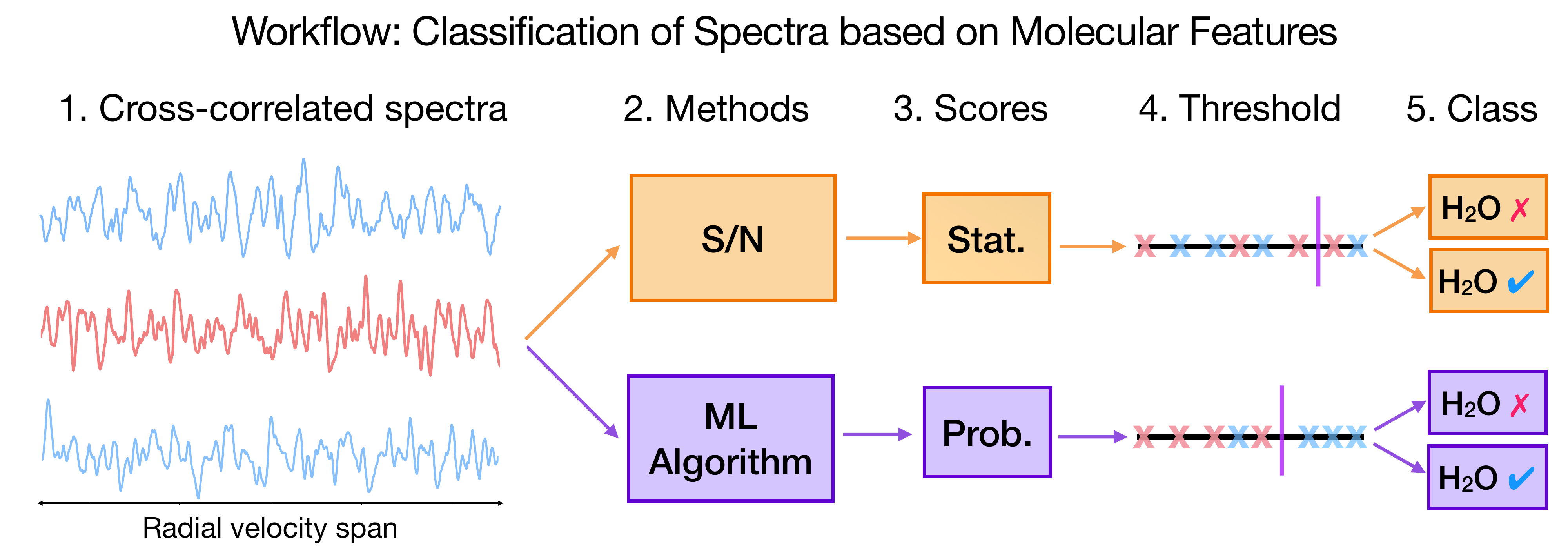}
    \caption{Flowchart representing the methods, scoring and classification workflows presented across sections. Each cross-correlated spatial pixel (spaxel) is treated as an independent instance and is passed through a classifier of static (statistic) or a dynamic (learning algorithm) type. The methods will evaluate the RV series and yield scoring metrics (e.g. a statistic or probability score). In order to perform classification, the scores need to first be separated using a meaningful threshold. The current standard classification scheme is yield by the signal-to-noise ratio (S/N) on the cross-correlation peak at the planet's RV. We propose to analyse the RV series in a holistic approach using machine learning (ML) to detect the planets and their molecules, and use the resulting probability scores.}
    \label{fig:flow}
\end{figure*}

\subsection{Current Signal Detection Standards}\label{subsect:standards}

Thus far, detections with molecular mapping have been conducted using S/N metrics to assess the cross-correlation peak strength in relation to the noise and indicate the extent similarity with a given template. For instance, \citet{dit2018molecule, cugno2021molecular} use the signal and noise from a same spaxel, while \citet{hoeijmakers2018medium, petrus2021medium, patapis2021direct} use the peak value of the signal, over the cross-correlated noise from another spaxel in an annulus situated a few pixels and RV steps away from the central peak. \citet{cugno2021molecular, patapis2021direct} investigate the use of a corrected S/N, to account for template auto-correlation effects. Under the strong assumption of Gaussian independent and identically distributed (i.i.d.) residual spectral noise, a S/N of CCS reaching the value of 3 is commonly accepted as a weak detection; a strong detection is confirmed over the threshold value of 5 (interpreted respectively as $3\sigma$ and $5\sigma$ detections). However, the lack of consensus in the literature affects the comparability and the interpretability of the S/N scores attributed to detections. 

In fact, many non-systematic or hidden systematic effects influence the noise of the data, such as instrumental noise, observing conditions or residual stellar contamination \citep[c.f.][]{malin2023simulated}. This is especially true for cases of close-in planets or for ground based observations with persisting telluric effects (e.g. left panels in Fig.~\ref{fig:molmap_PZTEL} and Fig.~\ref{fig:molmap_PZTEL_CO}). In addition, residual molecular systematic \citep[e.g. harmonics and overtones, c.f.][]{hoeijmakers2018medium, malin2023simulated} should also be considered, and may be particularly prominent in cases where single molecular templates are used. As a consequence, the signals are embedded in non-Gaussian and/or non i.i.d. noise, reducing the cross-correlation peak and signals. In addition, non Gaussian i.i.d. noise will lead to a misinterpretation of the uncertainties related to the classical $3\sigma$ and $5\sigma$ thresholds \citep{Bonse_2023}.

\subsection{Contributions}

This paper and its companion \citep{nath2024}, introduce the concept of combining ML with CCS to improve detection sensitivities to hidden, noisy and faint exoplanet signals in spectroscopic data. In this paper, we show that one dimensional (1D) convolutional neural networks (CNNs) are able to effectively leverage the full RV dimension to learn hidden deterministic patterns from molecular features and overcome detection challenges for various planet types (c.f. Fig.~\ref{fig:flow}, steps 1 and 2). Alternatively, \citet{nath2024} use multi-dimensional CNNs to investigate the spatial and temporal features in cross-correlation cubes from IFS datasets. Thus, both studies take complementary ML approaches to demonstrate that learning relevant features from cross-correlated data enable higher detection rates than traditional S/N based metrics.

Our one-dimensional approach provides useful qualities which are worth addressing. The MLCCS methods can incorporate uncertainties regarding properties of exoplanet atmospheres by relying simultaneously on multiple molecular templates. This minimises assumptions about chemical composition and incorporates variability in atmospheric parameters. Thus, the approach is versatile and agnostic towards diverse exoplanets and brown dwarfs, which makes it valuable for identifying new candidates in a variety of datasets. Another key aspect of our implementation involves focusing the search in the spectral dimension to preserve the spatial independence of detections. We train supervised classification algorithms exclusively on the full RV extent of individual cross-correlated spaxels, to ensure adaptability towards spectroscopic data from various instruments, including both integral field and slit spectroscopy. This approach proves valuable for identifying new candidates in a variety of datasets.

To demonstrate the capabilities of MLCCS methods, we apply them to real K-band SINFONI IFS noise at spectral resolution R$\sim$5000 with insertions of synthetic gas giant and brown dwarf atmospheres. We use this foundation to build two mock datasets, namely an unstructured stack of individual spectra (\textit{the extracted spectra of companions dataset}), and spatially structured spectra as a stack of flattened IFS cubes (\textit{the imaged companions dataset}). We evaluate the MLCCS methods in comparison to the S/N baseline by looking at two aspects. Firstly, we evaluate the scoring confidence and conspicuity (i.e. contrast effect), by examining the separation of score distributions between planets of interest and noise (Fig.\ref{fig:flow}, step 3). Secondly, we investigate gains in detection sensitivity (equivalently: statistical sensitivity or completeness) after classification, for instance by quantifying the true positive detections after setting a threshold which controls the proportion of false discoveries (Fig.\ref{fig:flow}, steps 4 and 5). Then, we test our framework on realistic IFS data to investigate the applicability of MLCCS in challenging noise environments and bad observing conditions. Finally, we put the results into perspective by addressing the interpretability and explainability of the framework and identify areas requiring additional research.

\section{Methodology}\label{sec:method} 

This section provides the methodology from the cross-correlation method to the ML algorithms. We start by providing a short preamble, in order to motivate our choice to work with cross-correlated spectra, and to describe the required dataset shape for a spatially independent 1D CNN. Then, we explain the cross-correlation step of a given set of spectra with a molecular template (c.f. column 1 from Fig~\ref{fig:flow}). Subsequently, we present the baselines used to benchmark the performance of our method and the architecture of the CNNs (c.f. column 2 from Fig~\ref{fig:flow}).

Our approach aims to classify spectra individually, based on signature molecules in exoplanet atmospheres. Yet, our previous attempts to classify raw spectra directly with ML algorithms have been inconclusive, similarly to \citet{nath2024}. However, it is possible to learn a transformation of those spectra, resulting into the CCS method. In fact, \citet{hoeijmakers2018medium} emphasise that the CCS has the advantage to co-add the planet's absorption lines while ignoring the stellar and telluric features. This provides a first level of disentangling of the information in the data, which can be used by a conventional statistic or a learning algorithm. However, to assess a detection, classical metrics like S/N rely on Gaussian i.i.d. noise \citep{ruffio2019radial}, as well as a strong cross-correlation peak at the planet's RV. Conversely, neural networks can considerably improve the framework in specific cases of faint signals with unclear cross-correlation peaks or non-Gaussian i.i.d. noise. They provide a holistic analysis of the transformed spectra by considering contributions at every RV step. Hence, we employ 1D CNNs to learn molecular signatures in the transformed spectral dimension, thus using the cross-correlation values along RV features. While the cross-correlation step could be integrated into CNNs, we keep this separate to ensure comparable inputs with baseline methods.

Treating spectra individually allows to train and learn independently from the initial dataset shape and nature. Hence, by doing so, we enable the MLCCS methods to adapt the training and operate equivalently to long or single slit spectroscopy for different spectral resolutions (e.g., VLT/SINFONI, VLT/SPHERE,  GPI, JWST/NIRSpec, CRIRES+, Keck/OSIRIS, Keck/NIRSpec), and generalise to emission or transmission spectra. To achieve such one-dimensional operation, datasets need reshaping to incorporate spectra as row elements and wavelength bins as columns. To analyse an IFS cube or long slit images, each spaxel (spatial pixel with a wavelength dimension) must be stacked vertically. Spectroscopic datasets tend to have multiple exposures, introducing a time dimension across wavelength cubes or frames. Our method is designed to perform detection on individual (or sub-combined) exposure units, eliminating the need to combine cubes. Actually, working with uncombined cubes is beneficial for ML tasks, as it increases the available data while providing a more complex and variable noise structure. It is only crucial to maintain sensible indexing for the spatial and time dimensions to preserve integrity of the training and testing sets, and ensure reliable spatial and temporal reconstruction of the results.

\subsection{Cross-Correlation of the Spectra with the Templates}\label{subsec:ccfdata}

To obtain a dataset of transformed spectra, every spectral element of the dataset is cross-correlated with a template. As variations of chemical composition and atmospheric parameters are large and intrinsic to each planet in the stack, it would require as many different templates as there are planet variations to obtain exact cross-correlation fits. Thus, the use of full atmospheric templates are useful when searching for one (or few) particular companion(s) with known properties. However, in practical applications, when searching for unknown and previously undetected candidates, one will not have prior knowledge about spectra which do contain a planet, nor about which template parameters provide the best fit to a candidate planet. Thus, attributing an exact template fit to each spectrum in a large stack without such prior knowledge is unfeasible. Nevertheless, optimal template fit is not necessary for detection. An imperfectly matching template, that is sensitive to weak signals, is sufficient for detecting the molecule and its associated sub-stellar companion.

\begin{figure}[]
{\begin{center}
\begin{tikzpicture}
    \matrix (M) [matrix of nodes,
        nodes={minimum height = 5mm, minimum width = 0.5cm, outer sep=0.5, anchor=center, draw},
        column 1/.style={nodes={draw=none}, minimum width = 4cm},
        column 4/.style={nodes={draw=none}, minimum width = 4cm},
        column 6/.style={nodes={draw=none}, minimum width = 4cm},
        column 9/.style={nodes={draw=none}, minimum width = 4cm},
        row 1/.style={nodes={draw=none}, minimum width = 4cm},
        row sep=1mm, column sep=-\pgflinewidth, nodes in empty cells,
        e/.style={fill=yellow!10}, f/.style={fill=blue!10}
      ]
      {
         & -2000 & -1999 & ... & 0 & ... & 1998 & 1999 &  Y \\
         CC spaxel 1 & & &  ... & & ... & & &  1 \\
         CC spaxel 2 & & &  ... & & ... & & &  0 \\
         CC spaxel 3 & & &  ... & & ... & & &  0 \\
         CC spaxel (...) & & &  ... & & ... & & &  ... \\
         CC spaxel 24\,312 & & &  ... & & ... & & &  1\\
      };
      \draw (M-1-2.north west) ++(0,2mm) coordinate (LT) edge[|<->|, >= latex] node[above]{Radial velocity steps} (LT-|M-1-8.north east);
\end{tikzpicture}
 \caption{\small{Illustration of the shape and size of one cross-correlated dataframe. Each row is a sample, and it represents a cross-correlated spaxel (CC spaxel). The RV steps are called features, and the elements of the last column Y is the categorical labelling indicating the presence of a planet or molecule of interest in the spectra. One whole cross-correlated dataframe as above is named a template channel; it results from the cross-correlation of the whole spectral dataset (i.e. all samples) with a unique template.}}
  \label{fig:dt}
 \end{center}}
\end{figure}
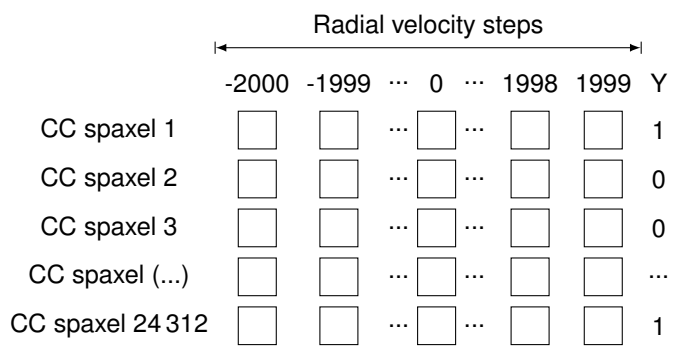

In our case, the primary goal is to detect exoplanets by broadly leveraging candidate characteristics. To achieve this, we need the MLCCS methods to remain as agnostic as possible regarding the chemical composition and atmospheric physics defined in the templates. The aim is to maximise the amount of candidate discoveries across a variety of planets and brown dwarfs, while using a minimal amount of templates and parameters. For instance, using a full chemical composition in an atmospheric template is too restrictive towards any planet which does not match the criterion. Instead, we will widen the search by relaxing assumptions on characterisation: we use a single molecule of interest which is generally able to indicate the presence of a substellar companion (e.g. H$_2$O, CO, etc). Following our agnostic approach, we only make very general approximations by selecting arbitrary atmospheric parameters such as effective temperature ($\Teff$) and surface gravity ($\logg$), in a way that roughly covers the parameter space of the class of planets we search for (e.g., gas giants with detectable amounts of water). Taking this into consideration, we use the template to repeatedly cross-correlate each spectral series (row element) of the dataset. This results in a cross-correlated dataset as represented in Fig.~\ref{fig:dt}.

We name the resulting cross-correlated data frame a \textit{template channel}, in lieu of the usual RGB colour channel used for CNNs on standard white light images. Here, each template channel of the same spectral dataset results from its cross-correlation with a different template. If multiple templates are required to incorporate flexibility regarding assumptions on atmospheric characteristics, we use several template channels for the same spectral dataset: 

\begin{itemize}
\item[$\bullet$] Channel 1 [N $\times$ 4001]: (Dataset$_k$ $\ast$ Template 1) 
\item[$\bullet$] Channel 2 [N $\times$ 4001]: (Dataset$_k$ $\ast$ Template 2) 
\item[$\bullet$] ... 
\item[$\bullet$] Channel M [N $\times$ 4001]: (Dataset$_k$ $\ast$ Template M)
\end{itemize}
where (Dataset$_k$) is a dataset composed of a stack of spectra from one or several data cubes. $N$ is the total number of spectra in the Dataset$_k$. Thus, if we use $M$ different templates on Dataset$_k$, it will result in $M$ different template channels, which are cross-correlated variations of the same Dataset$_k$. 

The cross-correlation function was applied between the templates and each spectrum of every datasets for 4000 RV steps. A series of cross-correlation values are obtained between $RV=[-2000;+2000]~\kms$, as the template is Doppler-shifted in steps of $1~\kms$. A cross-correlation peak is expected to appear at the planet's RV relative to Earth (e.g., $RV=x~\mathrm{km~s}^{-1}$), provided the template's composition matches the spectrum. We adapted the \texttt{crosscorrRV} function from the \texttt{PyAstronomy} library in \texttt{Python} by including a standardisation factor. This results in the following equation: 

\begin{equation}\label{eq:ccf}
 CCF_{w,i,m}=\frac{\sum_{j=1}^{J}(S_{j,i} \times T_{j,w,m})}{\sqrt{\sum_{j=1}^{J}(S_{j,i} \times S_{j,i}) \times \sum_{j=1}^{J}(T_{j,w,m} \times T_{j,w,m})}}
 , \end{equation}
 
with $CCF_{w, i, m}$ the standardised cross-correlation between a spectrum $S_{j,i}$ and a template $T_{j,w,m}$ at $RV=w~\kms$ ($\forall w \in \mathbb{Z}: [-2000,+2000]~\kms$). For $S_{i,j}$, $j$  ($\forall j \in \mathbb{N}: [1,J]$) is the j\textsuperscript{th} element of the spectrum vector in data row $i$ ($\forall i \in \mathbb{N}: [1,N]$). For $T_{j,w,m}$, $j$ is the j\textsuperscript{th} element of the Doppler-shifted and interpolated template spectrum $m$ ($\forall m \in \mathbb{N}: [1,M]$) at $RV=w~\kms$. The template spectrum $m$ is one template among a variety of $M$ templates (if multiple need to be used to construct the CNN channels). Every cross-correlation point $w$ is calculated for a template shifted at $RV=w~\kms$. Thus, $CCF_{i, m}$ is the i\textsuperscript{th} cross correlated vector (row) for template channel $m$. 

There are several advantages of standardising the cross-correlation, as in Eq.~\ref{eq:ccf}. First, the standardised cross-correlation peak can only reach a maximum of 1 in the case of an exact match, which happens when a series is cross-correlated with itself. This standardisation makes the peak values comparable between cross-correlated spaxels, allowing interpretation of the signal strength by the ML methods. In addition, normalisation ensures robustness of the classifications against contrast and brightness variations in the image noise, which can affect the absolute cross-correlation peak strength \citep{briechle2001template}. Note, in this case, that the cross-correlation noise is centered around a mean of 0, which makes normalising and standardising equivalent.

\subsection{Performance Benchmarks}

In order to make performance assessments of the MLCCS methods, we define the S/N as the primary baseline. Moreover, to ensure an informative benchmark for the CNNs, we evaluate the performance of a single layer neural network, called perceptron. 

\subsubsection{Signal-to-noise Ratio Statistic}

In order to compute the signal-to-noise ratio (S/N) of a cross-correlated series, we follow \citet{de2013detection} and \citet{dit2018molecule}. Hence, we evaluate the peak strength at the RV of the planet, namely $RV = x~\kms$, over a noise interval $z$ in the same series, situated at least $\pm 200~\kms$ away from the peak centre $x$:
\begin{equation}\label{eq:sn} S/N_{i,m}=\frac{CCF_{x,i,m}}{\sigma(CCF_{z,i,m})}, \end{equation}
where $CCF_{x, i, m}$ is the cross-correlation value at $x~\kms$, ($\forall x \in \mathbb{Z}$), for an empirical standardised cross-correlation vector ($\forall i \in \mathbb{N}: [1,N]$) between spectra of planets inserted in noise and a template channel $m$ ($\forall m \in \mathbb{N}: [1,M]$). We note that $x=0~\kms$ for a companion at rest frame. $CCF_{z,i,m}$ represents the series of noise values taken $\pm \delta~\kms$ away from the cross-correlation peak. The interval of $[x-\delta; x+\delta]~\kms$ corresponds to the point where the strong signals generally appear to fade out and the cross-correlation wings mimic randomness (c.f. Fig~\ref{fig:molmap_PZTEL}). Thus, for H$_2$O, we would have $z \in \mathbb{Z}: [-2000; 2000] \setminus [x-200; x+200]~\kms$ for the same cross-correlation vector series $i$, in the same template channel $m$. 

By construction, a fundamental assumption of the S/N metric is the Gaussian i.i.d. behaviour of residual cross-correlation noise (whether taken from cross-correlation wings or any other spaxel in an image), inherited by the assumed Gaussianity of the spectral noise. However, we highlight two reasons for the disputable nature of the commonly accepted detection thresholds and their resulting confidence intervals, namely for  $T=3$ for $3\sigma$  and $T=5$ for $5\sigma$ detections. 

Primarily, inconsistencies in S/N computation across the literature (c.f. Sect.~\ref{subsect:standards}) and in the choice of intervals impact the interpretation of detection thresholds and confidence intervals. Under the (asymptotic) Gaussian noise assumption, a Z-statistic (or respectively t-statistic) incorporating a proper variance-stabilising factor are more reliable. Therefore, we emphasise that we have also tested the methods against alternative S/N measures, that is, using the noise from a different spaxel as in \citet{hoeijmakers2018medium, patapis2021direct}, correcting the cross-correlation for template auto-correlation as in \citet{cugno2021molecular, patapis2021direct}, or even using a proper t-test statistic. As the scores of those tests did not affect the results and conclusions of this study, they are left out of the paper for the sake of readability. 

Secondly, spectral noise tends to be non-Gaussian or at least non-i.i.d. due to various noise sources and their effects on the cross-correlation (c.f. Sect.~\ref{subsect:standards}). While cross-correlation noise is optimal and equivalent to maximum likelihood estimator under well-behaved spectral noise \citep{ruffio2019radial, brogi2019retrieving}, non-Gaussian i.i.d. spectral noise causes cross-correlation to be sub-optimal. Fortunately, neural networks don't rely on Gaussian or i.i.d. assumptions to yield correct classifications, thereby offering an alternative approach to improving detection performance in the presence of complex spectral noise.

\subsubsection{The Perceptron as a MLCCS Baseline}\label{subsec:pct}

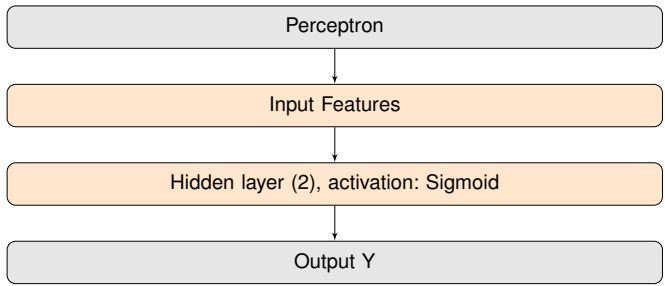
\begin{figure}[]
{\begin{center}
\scalebox{0.8}{
\begin{tikzpicture}[node distance = 1.3cm, auto, scale = 0.25]
    \node [block2] (input) {Perceptron};
    \node [block3, below of=input] (inputlay) {Input Features};
    \node [block4, below of=inputlay] (dense) {Hidden layer (2), activation: Sigmoid};
    \node [block2, below of=dense] (output) {Output Y};
    \path [line] (input) -- (inputlay);
    \path [line] (inputlay) -- (dense);
    \path [line] (dense) -- (output);
\end{tikzpicture}
}
\end{center}}
\caption{\small{Architecture of the perceptron. This simple one-layer neural network analyses the values of a whole cross-correlation series in a holistic approach, to detect the presence of a molecular signal.}}
\label{fig:NN1}
\end{figure}

The perceptron is a simple linear neural network with no hidden layer, and only one activation function. In our case, it contains only the sigmoid activation function, which makes it similar to a logistic regression. Hence, it is the simplest form of a binary neural network classifier. We use this simple architecture, shown in Fig.~\ref{fig:NN1}, as a baseline performance to track improvements of more complex and non-linear neural networks such as the CNNs, presented in ~\ref{subsec:cnn}.

The perceptron takes a stack of cross-correlated spectra as input, and outputs a vector of probabilities, relating to the presence of a molecule of interest in each spectrum (c.f. Fig.~\ref{fig:flow}). In order to learn this task, it is given a portion of the input dataset as training set, together with binary labels informing about the presence or absence of a molecule of interest in each cross-correlation vector series. The algorithm learns the patterns linking features of the input data to the label. The validation and test sets are separate portions of the dataset which are kept aside to evaluate how well the trained model generalises its classification skills to a new dataset. Hence, we can perform an iterative search process, varying the hyperparameter set, to evaluate which perceptron model trains best on a validation set. This hyperparameter search process is done using a meta-heuristic algorithm. Once the best model is found, it is evaluated on a test set, which reserved a last portion of data for model evaluation. All those steps can be applied in a cross-validated fashion, to investigate the stability of the test results across different portions of a data set. Further details on the splitting of our mock data sets for training, validation and testing sets are described in Sect.~\ref{subsec:respp} and ~\ref{subsec:resultsimgplanet}.

The perceptron was developed and trained using the \texttt{keras} library in python \citep{chollet2015keras, gulli2017deep}. We include early stopping, which is similar to low $L_2$ type regularisation on the RV features. The model's solution on the log-loss term is found with the RMSprop\footnote{\url{https://keras.io/api/optimizers/rmsprop/}\label{ft2}} optimiser. As for the hyperparameters, they are optimised by a heuristic evolutionary algorithm. Thus, for every new dataset, we set prior bounds on hyperparameters listed below and let the process converge: 
\begin{enumerate}
\item \textbf{Batch Size:} This hyperparameter regulates the trade-off between the training speed and the accuracy of the gradient estimates. We allow for five possible values from the set: $B_{size}=\{16, 32, 64, 128, 256\}$. 
\item \textbf{Epochs:} Optimising for the number of epochs corresponds to an "early stopping" regularisation of an $L_2$ type. We let the network learn over a continuous range of possible values: $E = [100, 200]$ with $E \in \mathbb{N}$. 
\end{enumerate}

\subsection{Convolutional Neural Networks}\label{subsec:cnn}

\begin{figure*}[]
    \centering
    \includegraphics[width=\hsize]{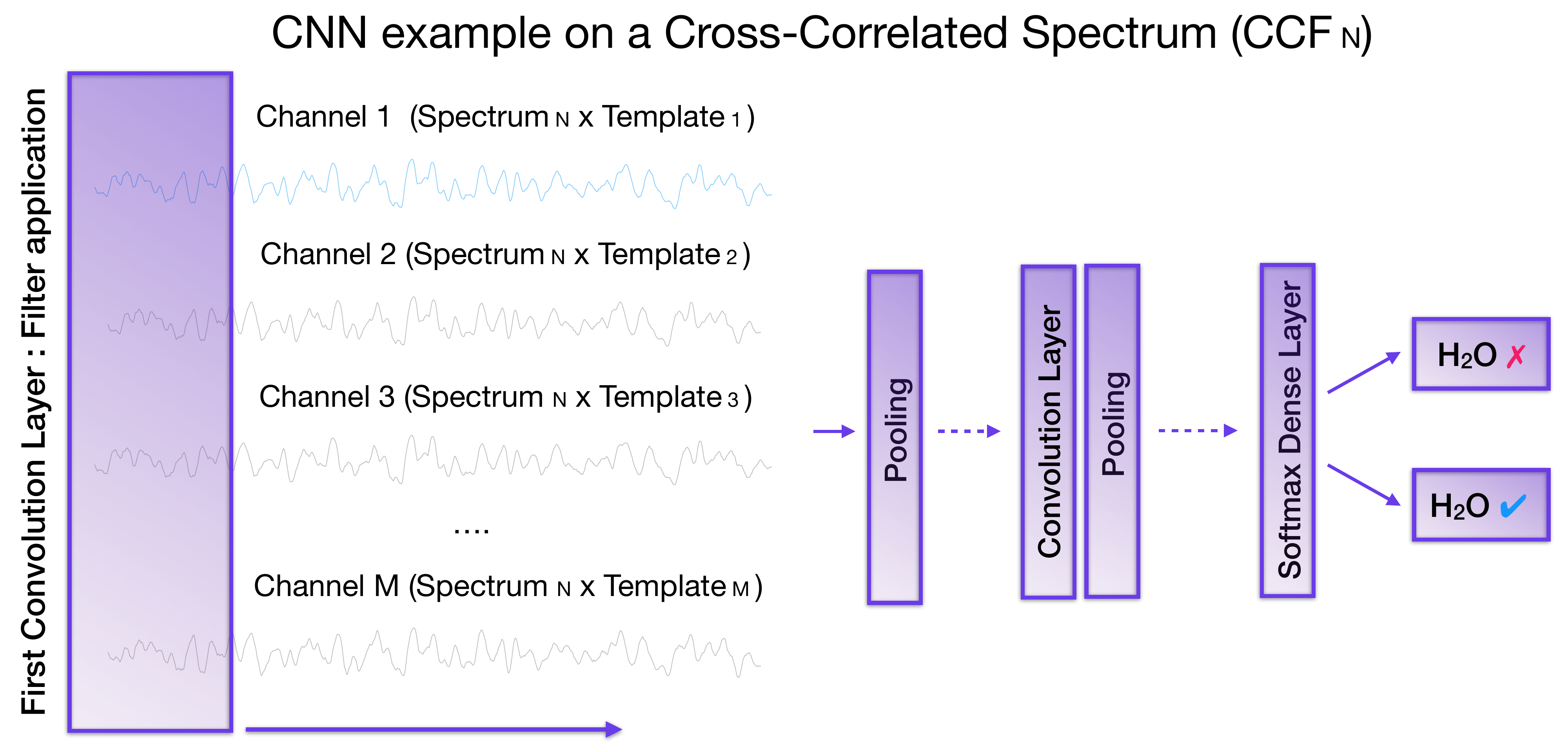}
    \caption{Example of the application of a Convolutional Neural Network (CNN) on one cross-correlated spectrum. For each sample of a cross-correlation N, and for all M template channels, the convolution filter runs across the channels and along the RV series. The filter depth is M, and its size is optimised according to the training. Hence, for a same series cross-correlated with M different templates, those convolutional layers allow to filter out important and recurrent patterns.}
    \label{fig:CNN}
\end{figure*} 

Convolutional neural networks \citep{krizhevsky2012imagenet, o2015introduction, gu2018recent} are a class of models which have proven to achieve formidable results in pattern recognition tasks. In their two-dimensional form, they are typically used for image recognition and classification, as they are robust to small pattern shifts and variations in space. In our framework, we apply 1D CNNs \citep{malek2018one} on the RV dimension of the samples. 

While a S/N statistic is only able to evaluate a spectrum based on a single given point (i.e. the cross-correlation peak), ML algorithms are able to take a holistic evaluation of patterns by considering all RV features in each cross-correlation series. However, what make the CNNs important for our case application are the benefits towards uncertainties of exoplanet atmospheric compositions. While the S/N and the perceptron can only be given one template channel at a time, the CNNs are able to use multiple template channels simultaneously, presenting different atmospheric properties. This means that we can consider several cross-correlated variations for one spectrum, as shown in Fig.~\ref{fig:CNN}. This enables to incorporate uncertainties regarding atmospheric characteristics when searching for previously uncharacterised planets. In fact, the number of template channels are used as filter depth for the CNN (c.f. Fig.~\ref{fig:CNN}). Then, the CNN uses convolution and pooling layers to downsample those channels. This allows to reduce the matrices and extract important patterns efficiently. The two CNNs we use are presented below.

\subsubsection{Convolutional Neural Network Architecture}

\begin{figure}[tbh!]
{\begin{center}
\scalebox{0.8}{
\begin{tikzpicture}[node distance = 1cm, auto]
    \node [block2] (input) {CNN1};
    \node [block3, below of=input] (inputlay) {Input Layer};
    \node [block, below of=inputlay,node distance=1.6cm] (conv1) {1D Convolutional layer, \\ Input shape: (4000 features, 9 channels), \\ Filters: (depth = number of channels), \\ Kernel: (size = optimised)};
    \node [block, below of=conv1,node distance=1.6cm] (pool1) {Max Pooling (pooling size = optimised)};
    \node [block, below of=pool1, node distance=1.4cm] (conv2) {1D Convolutional layer, \\ Filters: (depth = optimised 1/2), \\ Kernel: (size = optimised)};
    \node [block, below of=conv2,node distance=1.4cm] (pool2) {Max Pooling (pooling size = optimised)};
    \node [block, below of=pool2] (flat) {Flattening};
    \node [block4, below of=flat] (dense) {Dense output layer (2), activation: Sigmoid};
    \node [block2, below of=dense] (output) {Output Y};
    \path [line] (input) -- (inputlay);
    \path [line] (inputlay) -- (conv1);
    \path [line] (conv1) -- (pool1);
    \path [line] (pool1) -- (conv2);
    \path [line] (conv2) -- (pool2);
    \path [line] (pool2) -- (flat);
    \path [line] (flat) -- (dense);
    \path [line] (dense) -- (output);
\end{tikzpicture}
}
\end{center}}
\caption{\small{Architecture of CNN1. This figure illustrates the architecture of our CNN with several convolution layers and one dense layer. This model tests the effect of adding convolutional layers and template channels in addition to the sigmoid activation.}}
\label{fig:CNN1}
\end{figure}
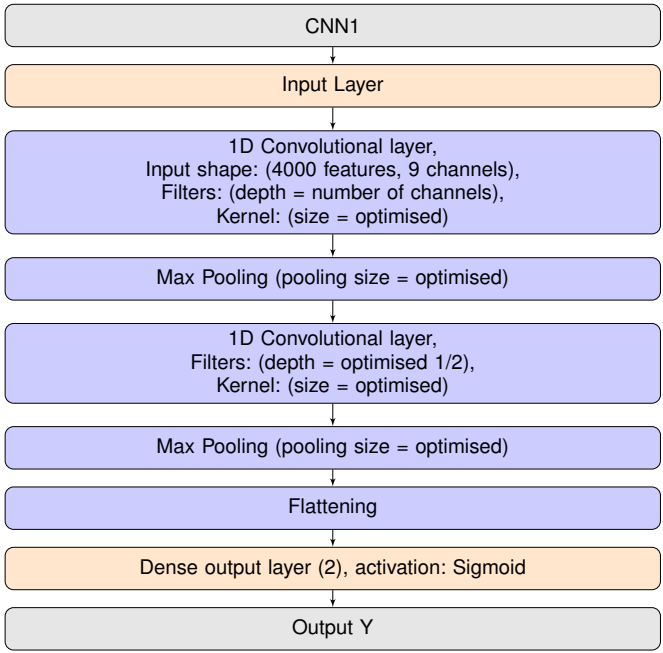

Our main model (CNN1), as shown in Fig.~\ref{fig:CNN1}, is made of convolutional and pooling layers, followed by one final dense layer which is similar to the perceptron. With this network, we can test and isolate the effect of adding convolutional layers in comparison to the perceptron.  Thus, CNN1 does not include any other regularisation than the early stopping criterion (similar to $L_2$). We emphasise that regularisation and increased model depth can come at the expense of invariance of a CNN to pattern shifts (such as RV shifts of the planet in the cross-correlation), which provides the motivation to keep the model simple. To train and optimise the CNN, we use the stochastic gradient descent, with Nesterov momentum and an optimised learning rate, we set the hyperparameter bounds to the following: 

\begin{enumerate}

\item \textbf{Batch Size:} We let the optimiser choose among three possible values from the set: $B_{size}=\{16, 32, 64\}$.

\item \textbf{Epochs:} We let the network learn over a range of possible values: $Epoch = [100, 200]$ with $Epoch~\in~\mathbb{N}$.

\item \textbf{Learning rate:} The bounds of the learning rate of the stochastic gradient descent are set to $\eta=[0.0001, 0.01]$.

\item \textbf{Momentum:} The bounds for the momentum are widely set such that $Mom=[0.1, 0.9]$.

\item \textbf{Kernel size:} This hyperparameter regulates the size of the convolutional filter. The Kernel size is set as one same parameter being valid for both convolutional layers. The set is configured to be integers in $K_{size}=\{3, 5, 7\}$.

\item \textbf{Max-pooling} The maximum pooling take the maximum value over ranges of values of the convolutional layer's output, and pools them together. The maxpool parameters are defined as integers over the set $P_{max}=\{2, 3\}$.
\end{enumerate}

\subsubsection{Testing Regularisation Schemes in Dense Layers}
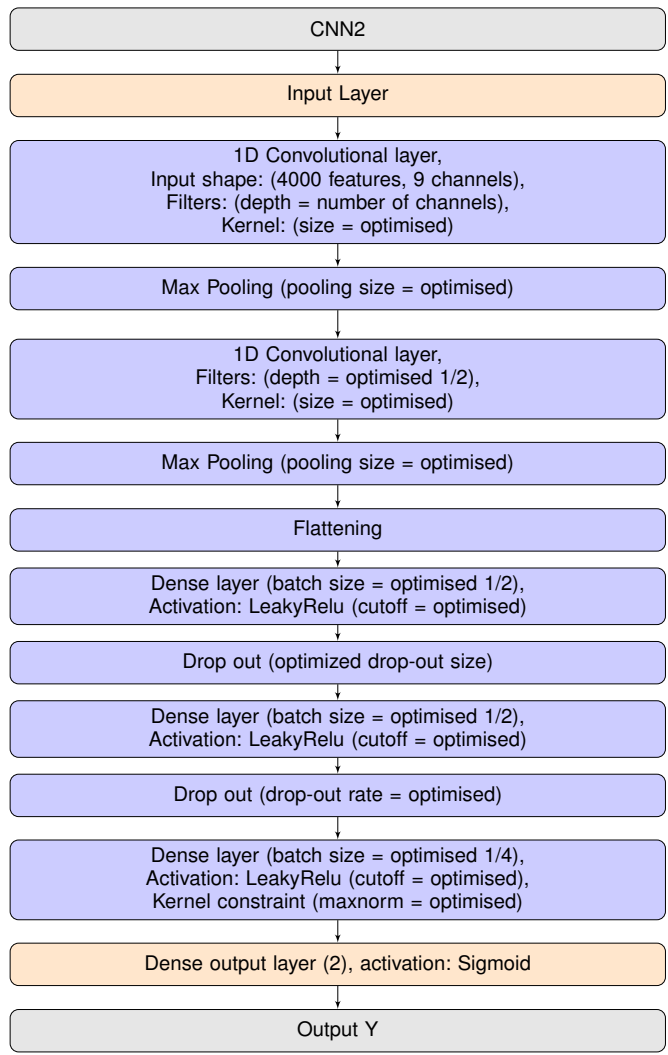
\begin{figure}[tbh!]
{\begin{center}
 \scalebox{0.8}{
 \begin{tikzpicture}[node distance=1.1cm, auto]
     \node [block2] (input) {CNN2};
     \node [block3, below of=input] (inputlay) {Input Layer};
     \node [block, below of=inputlay, node distance=1.6cm] (conv1) {1D Convolutional layer, \\Input shape: (4000 features, 9 channels), \\ Filters: (depth = number of channels), \\ Kernel: (size = optimised)};
     \node [block, below of=conv1, node distance=1.6cm] (pool1) {Max Pooling (pooling size = optimised)};
     \node [block, below of=pool1,node distance=1.5cm] (conv2) {1D Convolutional layer, \\ Filters: (depth = optimised 1/2), \\ Kernel: (size = optimised)};
     \node [block, below of=conv2,node distance=1.4cm] (pool2) {Max Pooling (pooling size = optimised)};
     \node [block, below of=pool2] (flat) {Flattening};
     \node [block, below of=flat] (hidden1) {Dense layer (batch size = optimised 1/2), \\ Activation: LeakyRelu (cutoff = optimised)};
     \node [block, below of=hidden1] (drop1) {Drop out (optimized drop-out size)};
     \node [block, below of=drop1] (hidden2) {Dense layer (batch size = optimised 1/2), \\ Activation: LeakyRelu (cutoff = optimised)};
     \node [block, below of=hidden2] (drop2) {Drop out (drop-out rate = optimised)};
     \node [block, below of=drop2,node distance=1.4cm] (hidden3) {Dense layer (batch size = optimised 1/4), \\ Activation: LeakyRelu (cutoff = optimised), \\ Kernel constraint (maxnorm  = optimised)};
     \node [block4, below of=hidden3,node distance=1.4cm] (dense) {Dense output layer (2), activation: Sigmoid};
     \node [block2, below of=dense] (output) {Output Y}; 
     \path [line] (input) -- (inputlay);
     \path [line] (inputlay) -- (conv1);
     \path [line] (conv1) -- (pool1);
     \path [line] (pool1) -- (conv2);
     \path [line] (conv2) -- (pool2);
     \path [line] (pool2) -- (flat);
     \path [line] (flat) -- (hidden1);
     \path [line] (hidden1) -- (drop1);
     \path [line] (drop1) -- (hidden2);
     \path [line] (hidden2) -- (drop2);
     \path [line] (drop2) -- (hidden3);
     \path [line] (hidden3) -- (dense);
     \path [line] (dense) -- (output);
 \end{tikzpicture}
 }
 \end{center}}
 \caption{\small{Architecture of CNN2. This figure illustrates the architecture of our second CNN. It includes regularised layers, to evaluate the benefit of regularisation for generalisation to different target noise regimes.}}
\label{fig:CNN2}
\end{figure}

Ultimately, we test a CNN with a regularisation structure (CNN2), to verify if the general CNN framework needs regularisation, or if overfitting is sufficiently controlled by our training scheme. Regularisation allows to reduce unwanted complexity of a model and thus overfitting, for example by dropping some features (drop-out), or by rescaling the weights of the features to a maximal L\textsubscript{2} norm. CNN2 includes two convolutional and max-pooling layers, as well as three dense layers which use Leaky-ReLu activation functions. We can combine several L\textsubscript{1} and L\textsubscript{2} regularisation types on the dense layers. We add two drop-out layers with individually tuned drop-out rates, combined with one kernel constraint on the activation of the last hidden layer. The final neurons are mapped into the probabilistic prediction space by a sigmoid function. The detailed architecture is presented in Fig.~\ref{fig:CNN2}. The model is optimised with stochastic gradient descent; we set the following bounds for the meta-heuristic optimiser:  
\begin{enumerate}

\item \textbf{Batch Size:} We let the optimiser search over the set $B_{size}=\{16, 32, 64\}$.

\item \textbf{Epochs:} We let the network learn over a range of possible epoch values to define the equivalent of an early stopping rule such that $Epoch = [100, 200]$ with $Epoch \in \mathbb{N}$ 

\item \textbf{Learning rate:} The bounds of the learning rate of the stochastic gradient descent are set as $\eta=[0.0001, 0.01]$

\item \textbf{Momentum:} The momentum's bounds are $Mom=[0.1, 0.9]$

\item \textbf{Kernel size:} The set of kernel sizes is defined once for both convolutional layers as $K_{size}=\{3, 5, 7\}$. 

\item \textbf{Max-pooling} The parameter of both maximum pooling layers are defined once as integers over the set $P_{max}=\{2, 3\}$

\item \textbf{Leaky relu:} In the case of the convolutional neural network, due to the high amount of tuning hyperparameters, we set the same bound for any Leaky ReLu activation as $\alpha=[0.1, 0.9]$

\item \textbf{Drop-out:} In this convolutional neural network, we have defined two layers of L\textsubscript{1} shrinkage, for which two hyperparameters were set so that the bounds are the same for $D1=[0.1, 0.8]$ and $D2=[0.1, 0.8]$.

\item \textbf{Kernel Maxnorm:} The Kernel maximum norm, equivalent to the L\textsubscript{2} regularisation, were set to be $K_{max}=[0.1, 5]$.

\end{enumerate}

Finally, each ML method is trained and tested on two datasets which are fed to the algorithms as a stack of cross-correlated series. The construction of both datasets is described in Sect.~\ref{sec:data}. For each of the MLCCS models, the hyperparameter tuning is automated using an evolutionary algorithm which performs a heuristic search over the hyperparameter space. The benefit of using this training scheme is that it provides high stability on the results. All models are trained and tested in a cross-validated fashion, with folds delimited by the temporal cubes. The algorithms predict a probability score for a molecular feature to be present. Hence, in order to separate the groups between signals and noise, we have to define a meaningful threshold. To preserve clarity and readability, we leave this last step to Sect.~\ref{sec:results}).

\section{Data}\label{sec:data}

In order to show that the MLCCS approach is flexible across planet types and instruments, we validate this proof of concept on two datasets described in this section. The first dataset is called the \textit{extracted spectra of companions} and represents a stack of individual spectra. Those spectra contain random insertions of synthetic gas giants and brown dwarfs among real instrumental noise. The purpose is to evaluate the capacity of the MLCCS methods to learn effective detection and classification on isolated spectra, for various planet types (e.g. for stacks of spectra without a relevant spatial structure). The second dataset is the \textit{directly imaged companions} dataset and evaluates the model's capacities to operate on imbalanced datasets where signals are scarce. It also investigates capacities to detect faint sub-stellar companions in structured data such as imaging spectroscopy. 

Both datasets are built out of a common basis of simulated planets embedded into real non-Gaussian i.i.d. noise. While the synthetic planets spectra are simulated using \texttt{petit\textsc{radtrans}} \citet{molliere2019petitradtrans}, we gather the instrumental noise using spaxels from noisy areas of real observations. Those are from GQ\,Lup\,B and PZ\,Tel\,B, observed in K-band at medium resolution R$\sim5000$ using SINFONI, the near-infrared spectrograph mounted on the Very Large Telescope (VLT). The data preparation steps are thoroughly described, as they are the determinant for quality and reproducibility of the results. We first outline the three steps which build the common basis for both datasets, namely, the IFS noise extraction, and the simulation of the planetary signals and molecular templates. Then, we explain how the synthetic planets were inserted into the \textit{extracted spectra of companions} and the \textit{directly imaged companions} datasets, and how these are prepared for the tests.

\subsection{Preparation of Instrumental Noise Cubes}\label{subsec:noise}

The results of this proof of concept rely on the calibration and quality of the datasets and noise. In fact, ground based spectral imaging noise can be non-identically distributed and non-independent (non i.i.d.) up to non-Gaussian, due to instrumental, telluric, stellar effects, and therefore difficult to simulate faithfully. The use of simplistic simulated i.i.d. Gaussian noise will lead to an inaccurate prediction of the performance of the ML methods, thus preventing reliable generalisation to real non-Gaussian noise. Therefore, it is preferable to use real preprocessed VLT/SINFONI noise, to guarantee accurate evaluation of the MLCCS methods. In this regard, we extract noise from a total of 19 uncombined integral field unit (IFU) cubes from GQ\,Lup\,B and PZ\,Tel\,B observations in K-band medium resolution spectroscopy, presented in Table~\ref{tab:obs}.

\begin{table}[tbh]
\setlength{\tabcolsep}{1.5pt}
\renewcommand{\arraystretch}{1.2}
\captionsetup{position=above}
\small
\caption{\small{IFU cubes used for noise extraction.}}
\centering 
\begin{tabular}{c|ccccccc}
\multicolumn{8}{c}{Observations}\\
\hline
Target&N\textsubscript{IFU}&Prog. ID & Date&DIT&NDIT&Airmass&Seeing\\ 
\hline
\hline
GQ\,Lup\,B&8&275.C-5033(A)&16.09.05&300&1&1.473&0.95\\
PZ\,Tel\,B&4&093.C-0829(B)&04.05.15&60&4&1.11&0.72\\
PZ\,Tel\,B&7&093.C-0829(B)&28.07.15&10&20&1.123&1.48\\
\hline
Total IFU&19&--&--&--&--&--&--\\
\hline
\end{tabular}
\\
\vspace{2mm}
\justifying 
{\small {\bf Notes.} The table summarises the 19 individual IFU cubes used for extraction of the real SINFONI noise, to train and test the ML models. We provide the targets, the number of sub-integrated cubes used from each target (N\textsubscript{IFU}), the program ID, observation date, integration time (DIT), number of DITs, average seeing and airmass across IFUs.}
\label{tab:obs}
\end{table}

Following \cite{cugno2021molecular}, the raw GQ\,Lup\,B and PZ\,Tel\,B datasets were first reduced with the \texttt{EsoReflex} pipeline for the SINFONI instrument \citep{abuter2006sinfoni}, which includes steps such as dark subtraction, bad pixel removal, detector linearity correction and wavelength calibration. It outputs 3D data cubes for each science observation. Hence, each science cube consists of two spatial dimensions and a wavelength dimension covering 1.929-2.472~$\mu$m. As NaN values were located at the waveband edges, the latter were removed by trimming the cubes to a wavelength dimension spanning from 1.97-2.45~$\mu$m for GQ\,Lup\,B and 2.00-2.44~$\mu m$ for both PZ\,Tel\,B datasets. Finally, a customised version of the \texttt{PynPoint} pipeline\citep{amara2012pynpoint, stolker2019pynpoint} is used to remove the stellar contribution (and the companion pseudo-continuum) from the frames. This step is performed applying high-resolution spectral differential imaging (HRSDI, \citealt{hoeijmakers2018medium, haffert2019two}); we modelled and subtracted the low-frequency spectral component in the data in order to leave only the high frequency components from the molecules in the planet's atmosphere. A detailed description of data pre-processing can be found in \cite{cugno2021molecular}. The resulting wavelength cubes are not mean or median combined in time for two reasons. First, we want to preserve a rough noise structure to ensure the robustness of the ML algorithm to variations in noise. Second, it also allows to use enough original data to train the ML algorithms, without having to use data augmentation techniques, which could increase risks of overfitting the data.

Before flattening of the $19$ pre-processed IFU cubes into a stack of spectra, the spaxels containing the true companion's signal were identified in each residual wavelength solution, using target centring coordinates. The signals were then confirmed using cross-correlation with a template. A wide aperture (radius of $5.5$ pixels) was drawn around the target's centre, such that the spaxels containing the true planets were removed from every cube with an additional margin of minimum 2 pixels. This is done to ensure an accurate and unbiased labelling of the training, validation and test sets, and avoid leakage of real molecular signal from the targets companions into the noise. After removal of the companions, the cubes are flattened and stacked, such that the noise spaxels from each and every cube were stacked along a unique spatial dimension to form an instrumental noise basis which we can sample from.

\subsection{Simulated Planets and Molecular Templates}\label{sec:simplan}

In this section, we describe the simulation of the templates and the planets which are inserted in the real noise spaxels. The synthetic companion spectra were simulated with \texttt{petit\textsc{radtrans}}~\citep{molliere2019petitradtrans}, as simplified gas giant and brown dwarf atmospheres. Their composition is made of molecules of interest such as water, carbon monoxide, methane and/or ammonia molecules (H\textsubscript{2}O, CO, CH\textsubscript{4}, NH\textsubscript{3}), with a hydrogen and helium dominated atmosphere. Overall, 10 possible singleton or pairs of molecules of interest are included in the atmospheres in addition to the hydrogen and helium rich environment. Realistic molecular abundance profiles are defined along the vertical extent according to the chemical equilibrium model yield by \texttt{easy\textsc{chem}} \citep{molliere2017observing}.
 
As for the atmosphere's structure, the radiative transfer routine \texttt{petit\textsc{radtrans}} relies on a parallel-plane approximation, and the simulations were made under the assumption of 100 layers atmospheres, between $10^2$ and $10^{-6}$ bar (equally distant in log space). The Guillot model was used to parameterise the P-T profiles \citep{guillot2010radiative}. This model assumes a double grey atmosphere with a characteristic opacity in the optical ($\kappa_{VIS}$) and in the thermal ($\kappa_{IR}$), thereby decoupling the incoming stellar irradiation from the outgoing planetary flux. The following values of $\gamma=\kappa_{VIS}/ \kappa_{IR}=0.4$, $\kappa_{IR} = 0.01 ~\mathrm{cm^2g^{-1}}$, and interior temperature $T_\mathrm{int} = 200~\mathrm{K}$ were used. The surface gravity and equilibrium temperature (translated into effective temperature) are the two last parameters of the Guillot model. Those were varied to create the planets grid, as $\logg$ and $\Teff$ may affect the spectral lines via the thermal structure and vertical mixing. Following \citet{stolker2021characterizing} and \citet{hoeijmakers2018medium}, the planets are simulated on a grid of $\Teff$ ranging from 1200$~\mathrm{K}$ to 3500$~\mathrm{K}$ with steps of 10$~\mathrm{K}$, and $\logg$ ranging from 2.5 to 5.5$~\mathrm{dex}$, with steps of 0.2$~\mathrm{dex}$. In this setting, the metallicity and C/O ratio are kept fixed to solar values ($Fe/H = 0.0$ and $C/O=0.55$). Overall, for each combination, we have a grid of $231$ $\Teff$ and $16$ $\logg$ values, thus $3\,696$ synthetic planet spectra per molecular combination and $36\,960$ spectra over all combinations. Finally, the continuum was approximated and removed using a Gaussian filter; a window size of 60 wavelength bins ($15$ nm) allowed to remove the continuum effectively without leaving any measurable trends or bumps in the residuals.

As for the molecular templates, high resolution emission spectra of H\textsubscript{2}O were generated with \texttt{petit\textsc{radtrans}}, using thermal structure grids. While the P-T profile model is the same as defined above, the molecular abundance of water was set to a constant $-2.0~\mathrm{dex}$ (i.e. mass fraction = $10^{-2}$) along the vertical extent of the atmosphere to produce well defined absorption lines. Then, the template is downsampled to the spectral resolution of the data and the same Gaussian filter was applied on the spectra to subtract the continuum emission. 

A selection of the single molecular templates is cross-correlated with each mock dataset outlined in the upcoming sub-sections; this returns several template channels per set, which we use to feed the CNNs. Since the focus is set on detection of new candidates rather than precise characterisation, we have to assume that the characteristics of the synthetic planets are unknown. Thus, we need to make minimal assumptions to maximise the number of detections. This means that it is preferable to use several template channels to vary the atmospheric characteristics. We did not identify a hard rule on the number of template channels for the CNNs, and those can be chosen relatively arbitrarily. Nevertheless, through our tests, we observed general trends helping the CNNs to perform well. First, expanding or refining the template channel grid has a benefit to cost trade-off between gaining in flexibility to find more planets from adding templates, and computational complexity. In addition,  the marginal benefits in agnosticity from adding one template will start to culminate. The best benefit to cost we noticed was between 5 and 10 template channels. Second, we did not identify any consistent or clear change in performance by changing templates. Still, the template parameters should be roughly spread over the parameter space of interest. 
Overall, those rules of thumb will depend on the dataset's characteristics, for instance its size, the variability of the planets it contains, or the available computational resources. As for the selection of the molecular composition, the most agnostic approach is to use a parallel combination of single molecular templates, to detect planets which have at least one of the stated molecules. Alternatively, one can use one molecule for all templates to apply a weaker constraint on composition. 

For the proof of concept, we focus on the search for planets with water features by using H$_2$O templates. Primarily, because water is a detectable spectral feature at K-band using CCS, and focusing the detection on a single molecule at a time makes it easier to evaluate the CNNs against benchmarks. Secondly, gas giants and brown dwarfs can be very rich in H\textsubscript{2}O depending on the location of their formation with respect to the snowlines and separation from the host star \citep{oberg2011effects, morley2014water, nixon2021deep}. Thus, water rich planets are relatively abundant in this population and hence convenient for broad search of such exoplanets. Finally, beyond the gas giant framework, it is also of high scientific interest to improve sensitivity to weak water features on smaller sub-Neptune and terrestrial planets in the habitable zone \citep{madhusudhan2021habitability, pham2022follow}. However, we show in Appendix~\ref{appendix:B} that the method can as well be trained using templates of single molecules, or combinations of those to find even more planets.

\subsection{The Extracted Spectra of Companions Mock Data}\label{subsec:planetpopds}

\begin{figure*}[htb!]
    \centering
    \includegraphics[width=16cm]{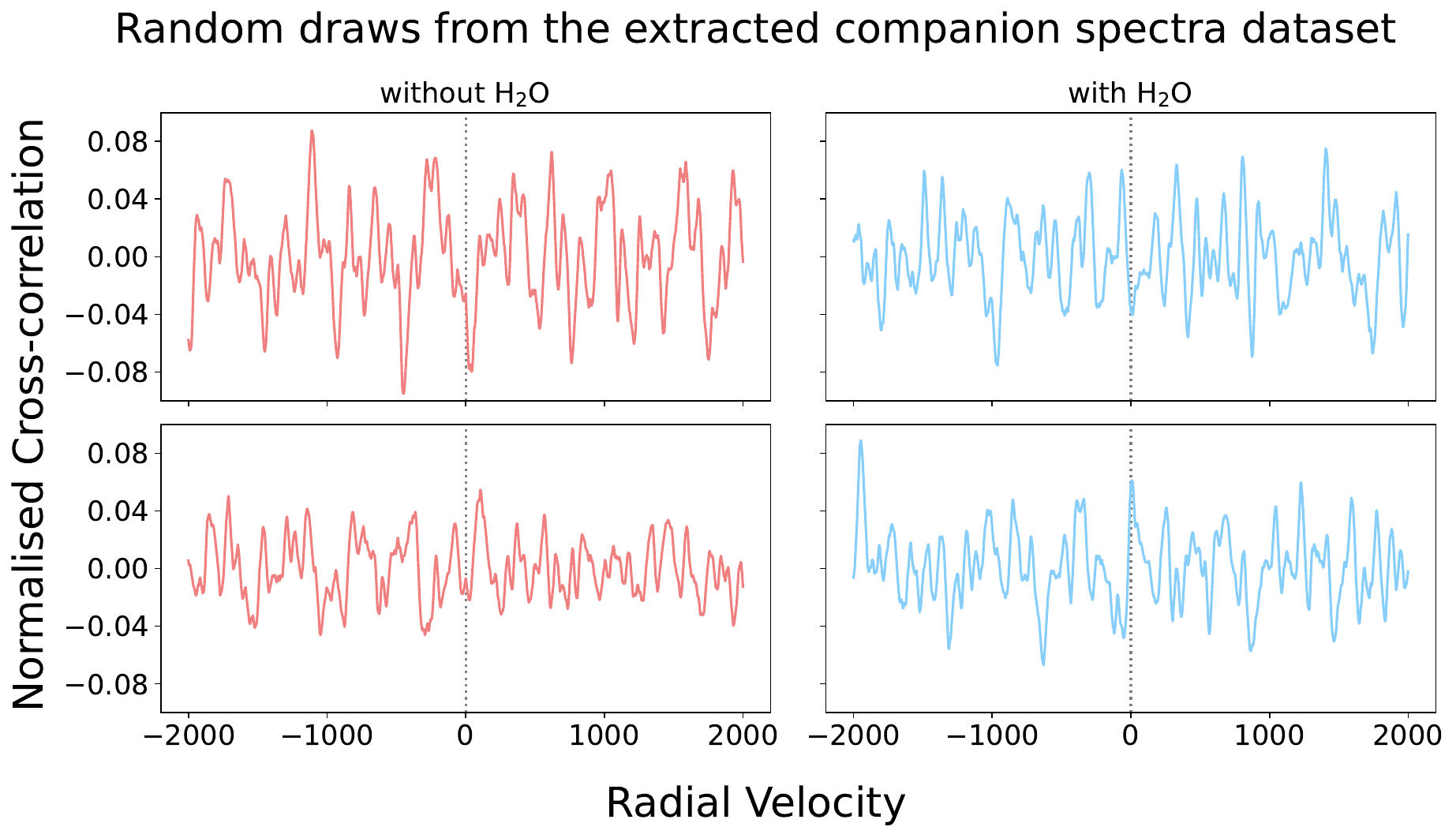}
    \caption{Four randomly picked cross-correlated spectra from the final \textit{extracted spectra of companions} dataset. For this particular case, the signals were inserted at $RV = 0$ (dashed vertical line) with a scale factor of 8 which corresponds to an average S/N of $\simeq0.633$. This plot aims to show how the transformed spectra with H$_2$O signals (blue) do not show a visible cross-correlation peak, nor any obvious patterns which differ from the negative group (red); separating the groups based on a S/N statistic is non-trivial.}
    \label{fig:distralpha8}
\end{figure*} 

Within this section, we present the first mock data, named the \textit{extracted spectra of companions} dataset, as it contains various types of planetary spectra embedded in instrumental noise. The goal of this dataset is to demonstrate the capacity of the MLCCS methods to improve planet detections via water signals, while using a minimal amount of prior information. It also proves the ability of the MLCCS approach to operate only in the RV dimension by treating the spaxels independently, i.e. without using any spatial information. Consequently, this one dimensional approach aims to prove the capacity of the MLCCS methods to operate on isolated spectra as well as in various spectroscopic modes, such as single and long slit spectroscopy (e.g. CRIRES+,  Keck/OSIRIS,  Keck/NIRSpec). Implementation of MLCCS methods could in principle be adapted to high resolution transmission spectroscopy by generalising the framework.

As explained in Sect.~\ref{sec:simplan}, we focus the search on water features, which makes H\textsubscript{2}O the planetary signal of interest; the rest is regarded as noise. The planets are randomly sampled without replacement, and an atmospheric variant can appear only once in the dataset. As a result, over a total of $24\,312$ spectral instances, $50\%$ of the dataset contains water-rich planets (the \textit{positive group}), and encompass simulated planets composed of combination of molecules including water. The remaining $50\%$ represents noise (the \textit{negative group}). Among the noise, $50\%$ of the spaxels are \textit{pure noise}, i.e. plain instrumental noise without any molecular spectrum. The other $50\%$ of the negative group is \textit{molecular noise}, i.e. any planets with combinations of molecules which do not include water, to represent water depleted planets. 

The instrumental noise was also randomly sampled without replacement from the eight GQ\,Lup\,B cubes (see Sect.~\ref{subsec:noise}). After preprocessing and removal of the real target as described in Sect.~\ref{subsec:noise}, each cube is left with $3\,039$ spaxels, which yields a total of $24\,312$ spaxels. To avoid creating spatial dependencies when training the MLCCS methods, the spaxels within each cube are shuffled to spatially decorrelate the noise. The shuffling is only applied within, but not across the cubes. This allows to split our data into training, validation and testing cube groups, without overfitting the noise. 

Inserting the planets in the noise without re-scaling would yield either too faint signals to be detected, or very strong signals leading to perfect classification. For feasibility of the study and comparability of the methods, all planets were inserted at rest frame with a radial velocity of $RV=0~\mathrm{km~s}^{-1}$. The noise is adjusted with respect to the signal, for a scaling factor $\alpha$: 
\begin{equation}\label{eq:alpha}
S = P + (1/\alpha) * \epsilon,
\end{equation}
where S represents the spectral series which is composed of the spectrum of a simulated companion atmosphere (P), and the instrumental noise series ($\epsilon$). We note that the scale factor corresponds to varying the average S/N and hence the average noise level, which encompass both variations in contrast and separation. As the planets are inserted randomly, it is not possible to match a specific scale factor to a contrast or separation for every case. The scale factor can be understood as a random variation of those two parameters, which is in line with our agnostic approach. However, specific tests on the influence of planet separation and contrast for a close category of CNN methods can be found in the companion paper \citep{nath2024}.

Finally, we sample random templates of water in a cluster design, over values of $\logg = \{2.9; 3.5; 4.1; 4.7; 5.3\}~\mathrm{dex}$ and $\Teff = \{1200; 1600; 2000; 2400; 2800\}~\mathrm{K}$. We cross-correlated the dataset into 9 cross-correlated template channels, which took about $30 s$ per channel when distributed over 32 CPUs. Fig.~\ref{fig:distralpha8} sets an example of four randomly picked cross-correlated spectra resulting from the final stacks. It is obvious that no cross-correlation peak is visible at the radial velocity of the planet for the positive group. Hence, separating and classifying each group is non-trivial for a S/N statistic.

\subsection{The \textit{Directly Imaged Companions} Mock Data}\label{sec:implands}

\begin{figure*}[tbh!]
    \centering
    \includegraphics[width=14cm]{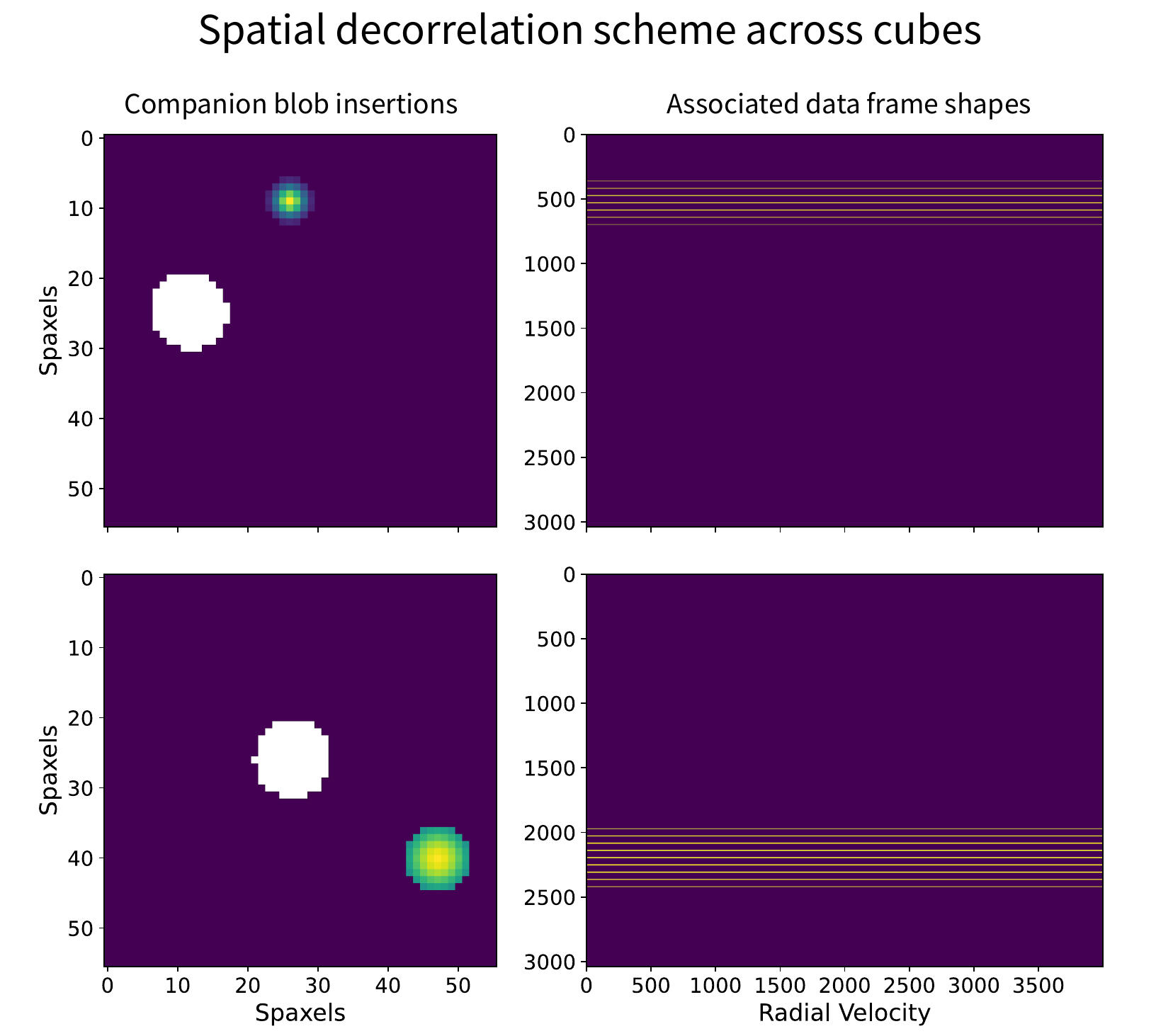}
    \caption{Spatial decorrelation in two IFU cubes of the structured mock dataset for directly imaged companions. \textit{Left panels:} brown dwarf signals are inserted in a structured manner, as varying Gaussian decaying ellipses within a delimited aperture, showing variations in position, size, and shape. The white areas indicate removal of real companions. \textit{Right panels:} The plots illustrate datasets after flattening and transformation, with the horizontal axis representing radial velocity and the vertical axis representing stacked spatial dimensions. Visible lines denote spaxels with inserted planetary signals, showing variations based on the inserted planets' properties. This method prevents the ML algorithm from learning redundant spatial artefacts; it emphasises learning from cross-correlation patterns in the spaxel dimension.}
   \label{fig:noiseoccurence}
\end{figure*}

This section describes the \textit{directly imaged companions} dataset, which is used to test and demonstrate that the spatially independent MLCCS methods can operate on structured data such as direct imaging, although it is highly imbalanced. The synthetic brown dwarf spectra were selected from the pool of simulated companions (Sect.~\ref{sec:simplan}) according to GQ\,Lup\,B and PZ\,Tel\,B's respective characteristics. They are inserted as faint Gaussian ellipses into the respective instrumental noise cubes. The spectra of each of the planets are selected with ranges of $\Teff=\{2450;2760\}~\mathrm{K}$ in steps of 10$~\mathrm{K}$ and $\logg=\{3.7;4.7\}~\mathrm{dex}$ in steps of 0.2$~\mathrm{dex}$ for the simulated GQ\,Lup\,B \citep[after characteristics reported by e.g.][]{seifahrt2007near, stolker2021characterizing} and ranges of $\Teff=\{2900;3100\}~\mathrm{K}$ in steps of 10K and $\logg=\{3.7;4.7\}~\mathrm{dex}$ in steps of 0.2$~\mathrm{dex}$ for the simulated PZ\,Tel\,B \citep[c.f.][]{jenkins2012benchmark}. All spectra are selected with a hydrogen and helium dominated atmosphere and composition variations of H\textsubscript{2}O and CO. 

The synthetic planets are inserted as aperture delimited Gaussian ellipses in the cubes, using a subset of selected spectra. Sampling subsets instead of single spectra allows the inclusion of signal variations within the aperture. This ensures robustness and avoids overfitting as the network encounters slight variations of spectra amid noise, it learns to handle encounters of new spectra in realistic test sets. Spectra subsets are sampled without replacement, ensuring that each cube presents a unique planet in terms of atmospheric structure and composition. The planets are inserted as Gaussian ellipses with random variations of elliptic ratio, size, luminosity decay, and locations, as shown in Fig.~\ref{fig:noiseoccurence}. This specifically prevents the neural networks from learning such deterministic dataset artefacts. This injection approach deviates from the point spread function (PSF) shape and size of SINFONI data, but allows objective testing of the ML methods' capability to recognise signals regardless of PSFs and data structure. This is crucial for versatility and generalisability to other spectrographs and instruments while maintaining spatial independence.

We insert faint simulated companions with H\textsubscript{2}O (mainly in the brown dwarf regime) at rest frame with an average S/N below the usual detection threshold of $5$. The goal is to show that the MLCCS methods can recover such embedded objects, even with a less conservative threshold. Finally, we cross-correlate our dataset according to Eq.~\ref{eq:ccf}, for a small grid of templates of water, spanning across several $\logg$ and $\Teff$ values, and roughly covering the parameter space of the inserted companions. We note that the cross-correlation of the whole dataset of 57782 spectra with 1436 wavelength bins into one template channel takes about 2 minutes when allowing distribution over 32 CPUs. This is performed for $5$ template channels for each dataset, respectively with $\Teff=\{2700;2800;2900;3000;3100\}~\mathrm{K}$ and $\logg=\{3.7;4.1;4.1;4.3;4.1\}~\mathrm{dex}$, and a total of $17$ cubes to train on. The two remaining cubes are kept for validation and testing.

\section{Results and Discussion}\label{sec:results}

In this section, we describe the metrics used to evaluate the models. Then, we present the results from those metrics on each dataset, and put the results in perspective with a sensitivity analysis. Finally, we discuss the implications of this work for the community, and propose further research.

After passing the cross-correlated stack through the ML algorithms, the trained models assign an output score to each instance, which represents the probability of a spectrum to belong to the positive group (i.e. spectra with water). We use the \textit{extracted spectra of companions} dataset to quantify how many of the inserted planets can be found despite the diversity of their spectra. Then, we verify the capability of those spatially independent methods to perform on structured data, such as the \textit{directly imaged companions} mock data. To do so, we evaluate the quality of the scoring against benchmarks, and quantify the resulting predictions. In order to divide the spectra into two predictive groups, namely the positive (exoplanets with H\textsubscript{2}O) and negative groups (no H\textsubscript{2}O detected), we need to separate the scores by using a threshold which yields predicted classes. We can then evaluate those predictions according to elements of the confusion matrix \citep[see e.g.][]{jensen2017new}, which include the amount of correct detections (TP), of false detections (FP), of missed detections (FN), and of correctly discarded spectra (TN). 
 
To formalise this, we use receiver operating characteristic (ROC) curves \citep{fawcett2006introduction} to show the performance on the balanced dataset (i.e. \textit{the extracted spectra of companions}). This evaluation metric allows us to explore the trade-off between correctly detecting exoplanets and incorrectly selecting false positives. 
As we vary either the classification threshold along the scores or some of the model tuning parameters, we can increase the true positive fraction, but will have to bear the cost of simultaneously increasing the false positives by a certain fraction (i.e. we are simply travelling along the curve). 
However, by using a better model, we take a stride to a higher curve. Hence, measuring the area under ROC curves (ROC AUC) allows to evaluate the overall gain over the trade-off in statistical sensitivity (or true positive rate, as TPR) relative to the false positive rate (FPR) which are defined in Eq.~\ref{eq:tpr} and ~\ref{eq:fpr} as follows:
{\begin{equation}\label{eq:tpr} TPR = \frac{TP}{(TP + FN)}  \end{equation} and}
{\begin{equation}\label{eq:fpr} FPR = \frac{FP}{(FP + TN)}. \end{equation}}

Although ROC curves are very useful to evaluate the scoring quality on balanced datasets, they tend to show over-optimistic results on imbalanced datasets. Therefore, when evaluating our methods on imbalanced data (e.g. full images), we consider another trade-off measure, namely the precision-recall (PR) curve. The recall corresponds to the true positive rate in Eq.~\ref{eq:tpr}, and is evaluated against precision, which is known as the positive predictive value or true discovery rate (TDR), as described in Eq.~\ref{eq:precision}: 
{\begin{equation}\label{eq:precision} TDR = \frac{TP}{(TP + FP)}. \end{equation}}
The precision-recall measure is more sensitive to improvements in the positive class and true detections \citep{davis2006relationship}, and is used to evaluate cases where the data is highly imbalanced \citep{saito2015precision}, such as our \textit{directly imaged companions} data. 

Finally, to evaluate the results in the light of a meaningful classification threshold, we introduce the false discovery rate \citep[FDR,][]{benjamini1995controlling}; another existing metric, employed for example in \citet{cantalloube2020exoplanet}:
{\begin{equation}\label{eq:fdr} FDR = \frac{FP}{(TP + FP)}. \end{equation}}
This metric is very meaningful for data analysis of surveys or archives, since it is able to control the impurity of the predicted positive sample, i.e. the proportion of false positive leakage into the detections, as shown in Eq.~\ref{eq:fdr}. It essentially answers the following question: \textit{"Among all claimed discoveries, what is the fraction of false detections which leaked in?"}. We emphasise here the difference with the false positive rate, which only controls for the amount of false positives leaking from the true negative population. Hence, the ROC curves and FDR metrics presented above will be used to evaluate the detection performance of the MLCCS methods against the S/N on the \textit{extracted spectra of companions}.

\subsection{Results on the \textit{Extracted Spectra of Companions} Dataset}\label{subsec:respp}

\begin{figure}[]
    \centering
    \includegraphics[width=8.9cm]{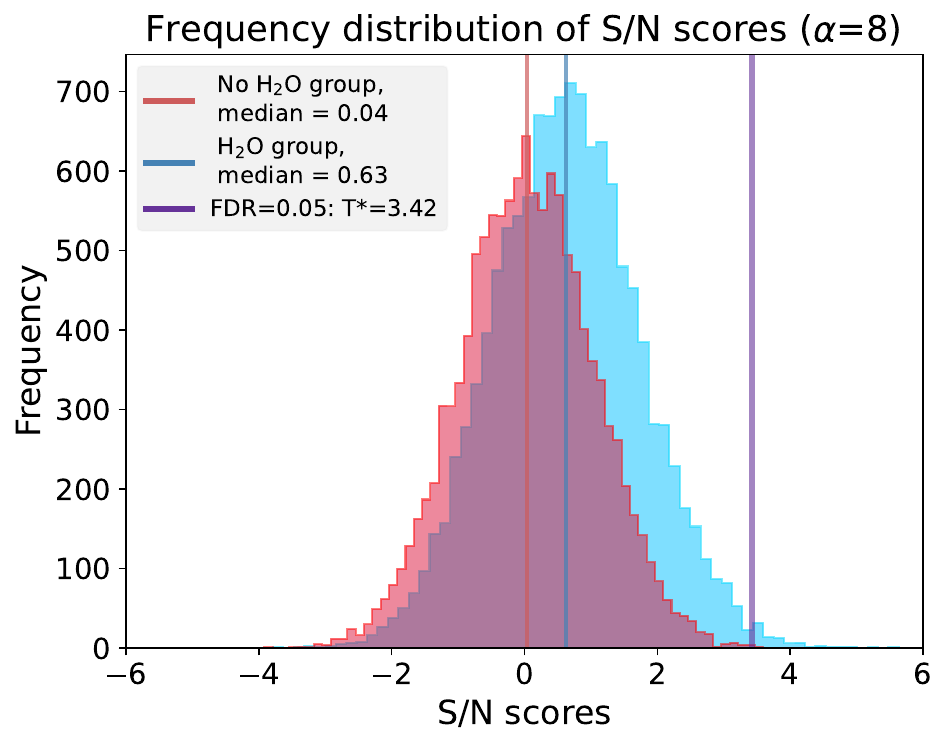}    
    \includegraphics[width=8.9cm]{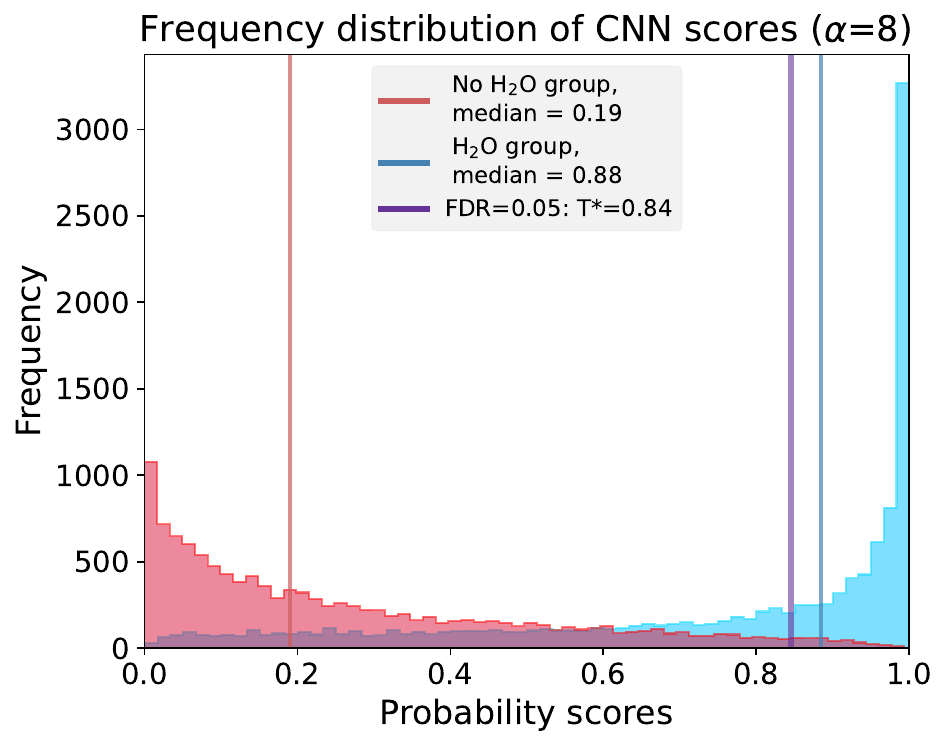}
    \includegraphics[width=8.9cm]{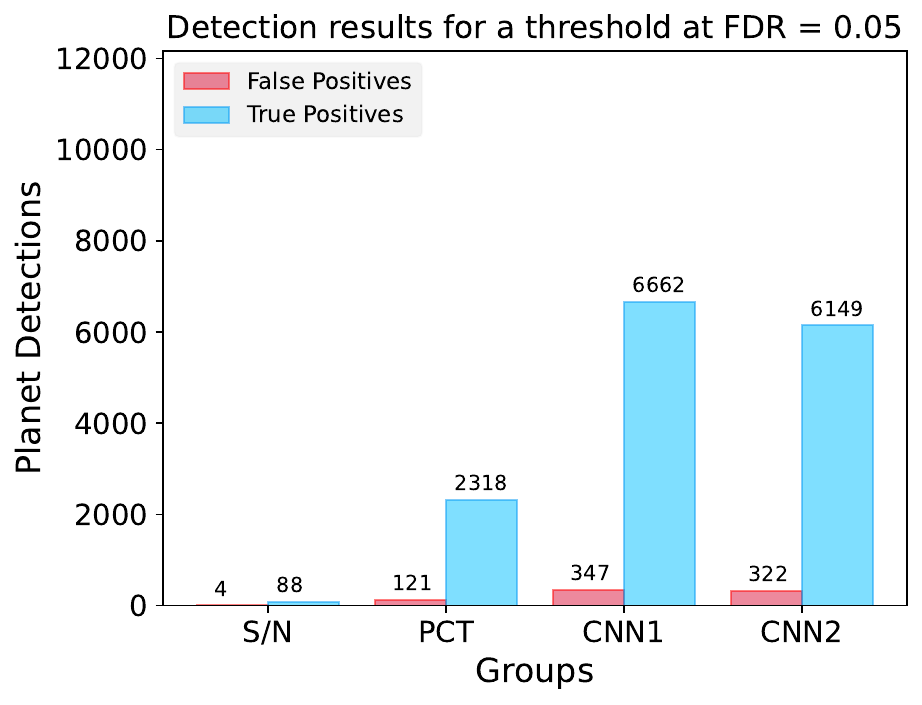}
    \caption{Scoring and Classification of the S/N and CNN. \textit{Top and middle}: Frequency distributions of aggregated scores assigned to the negative group (red) against the positive group (blue) for both methods. The scores represent the predicted likelihood of a given spectrum to belong to the positive group. The probabilistic scores assigned by the CNN provide a better separation of the groups (i.e. "conspicuity") than S/N scores. Classification predictions are set by a threshold for a FDR at $5\%$. \textit{Lower:} Maximal amount of planets recovered in the mock data by the S/N, perceptron (PCT) and both CNNs, within a maximal FDR of $5\%$.}
    \label{fig:distr8}
\end{figure}

Through those results, we evaluate the effectiveness of the MLCCS methods in finding water rich planets in the \textit{extracted spectra of companions} dataset. This test aims to demonstrate the MLCCS methods' ability to detect a variety of planets embedded in genuine instrumental noise. Thus, the main results are presented according to planet insertions which return extremely faint H\textsubscript{2}O signals in the positive group with an average of $S/N\simeq0.633$, against  $S/N\simeq0.044$ for the negative group (c.f. Fig.~\ref{fig:distr8}, top panel). Moreover, the CNNs take 9 channels as input, which relate to various templates of water spanning over combinations of $\Teff=\{1200;1600;2000;2400;2800\}~\mathrm{K}$ and $\logg= \{2.9,3.5,4.7,5.3\}~\mathrm{dex}$ (see Sect.~\ref{sec:simplan}). Note, that the S/N and perceptron baselines can only take a unique data template channel. The channel we use is issued from a template of $\Teff=1200~\mathrm{K}$ and $\logg=4.1~\mathrm{dex}$. We also ran a sensitivity analysis, by assigning different template channels to the baselines; the results did not show any significant difference in the performance evaluations, and are therefore left out of the study. 

\begin{figure}[tbh!]
    \centering
    \includegraphics[width=\columnwidth]{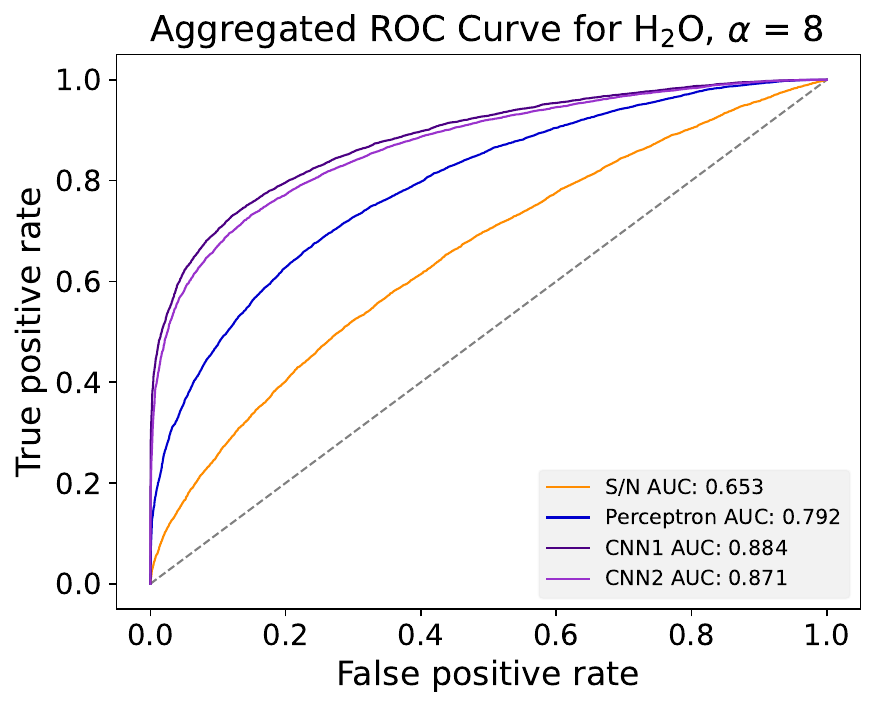}
    \caption{Quantification of scoring performance with receiver operating characteristic curves (ROC). The plot shows the improvements in the ROC trade-off between TPR and FPR. The improvement is measured in terms of area under the ROC curves (AUC). The CNNs outperform the baselines in finding true positives while limiting the increase in FPR.}
    \label{fig:accroc8}
\end{figure} 

The models were trained along the temporal dimension, specifically on six exposure cubes, then validated on one cube and tested on a remaining one. The training and testing routines were performed in a parallelised cross-validated fashion, with each fold running on its own GPU. The hyperparameter search process is automatised and will take about 4 hours end-to-end for one CNN (on a NVIDIA GeForce RTX 2080 Ti GPU). This provides stable results across folds and allows to get around lengthy hyperparameter fine-tuning for users with limited experience with ML algorithms. However, runtime will vary according to the dataset size, number of channels, hyper-parameter search and available computational resources. We show the aggregated ROC curves over all tested cubes in Fig.~\ref{fig:accroc8}. It shows the drastic improvements of the MLCCS methods in terms of scoring quality, with AUCs of $0.884$ and $0.871$ for the CNNs and $0.792$ for the perceptron, against $0.653$ for the S/N metrics, for a given scale factor of $\alpha=8$. This means that, for the same FPR, the CNNs can raise more true candidates. The scoring quality improvements can be explained by observing the frequency distributions on the aggregated scores over all tested cubes, as shown in both upper panels of Fig.~\ref{fig:distr8}. The scores assigned to the data by a classifier can be understood as the likelihood of a given spectrum to actually belong to the group of planets with water. The CNN exhibits high scoring confidence in distinguishing between the two groups, which translates into a strong contrast in scores' distributions. To avoid confusion with the usual meaning of contrast in astronomy (i.e. brightness of a planet relative to its host star), we favour the word "conspicuity". This effect makes the CNNs effective in detecting planets with water in the dataset and simultaneously reducing the occurrence of false positives. On the contrary, the S/N statistic scores show a poor conspicuity between both groups, forcing the use of high confidence thresholds (e.g. $T=5$), involving conservative results with higher FNs. 

To quantify classifications, we set thresholds to be comparable across scoring measures, aligning with a maximal FDR of 0.05 (Fig.\ref{fig:distr8}). We chose this value according to conservative  standards in the field of statistics \citep{benjamini1995controlling}, but it is of course possible to choose a more conservative value. The threshold should be adjusted for a desired confidence level before it can be used to evaluate detection performance of the models. Thus, the lower panel in Fig.\ref{fig:distr8} illustrates the maximum achievable detections when bounding false positives up to $5\%$ of the predicted detections. While S/N detects $88$ true planets, the perceptron finds 26 times more with $2\,318$ true detections. This renders a sensitivity (TPR) of $19.3\%$, against $0.7\%$ for the S/N. We note that the detection purity (TDR) always remains comparable, with $95.0\%$ for the perceptron against $95.6\%$ for the S/N, due to the upper bound on the FDR.
However, a key aspect of our methods is our presumption that precise knowledge of a planet's characteristics is not necessary for detection. This aspect is strongly connected to the significant improvements observed in those results. Indeed, even the use of a single molecular template with the perceptron enables the detection of hundreds of planets. This happens as we focus on finding weak signals rather than searching for strongest peaks which only occur for highly matching templates. Yet, the extra leap in detection sensitivity is offered by the CNNs, with the incorporation of multiple template channels as filters, which provide flexibility regarding composition uncertainties. In fact, the CNNs are capable of casting a wide net to discover more planets despite variations in atmospheric characteristics, thanks to the agnostic approach. Thus, each CNN achieves over $6\,000$ real detections out of $12\,000$ inserted planets (Fig.~\ref{fig:distr8}), representing a statistical sensitivity of $51.2\%$ and $55.5\%$ TPR respectively, with a purity of $95.0\%$ TDR. In other words, this test has proven that MLCCS can diversify planet searches in a single attempt, irrespective of the data structure. 

The results of this section were presented according to a scale factor of $\alpha=8$, which yields very strong noise with regards to signals. Yet, we note that the efficacy of MLCCS methods is negatively related to noise levels at the extremes, as illustrated in Fig.~\ref{fig:rocalphas}. If $\alpha$ is excessively small or large, the improvements over S/N metrics become marginal. For completeness of the study we discuss the interpretability of the results in the light of a sensitivity analysis over a range of scaling factors $\alpha$ in Appendix~\ref{disc:scale}. In addition, we also provide extended tests and discussion on the flexibility of MLCCS towards exoplanet compositions by incorporating  variations of molecules in the template channels (c.f. Appendix~\ref{appendix:B_supp}). We also provide an explainable framework by verifying that MLCCS is able to learn patterns in non-Gaussian noise (c.f. Appendix~\ref{app:nongaussian}) and molecular harmonics (c.f. Appendix~\ref{appendix_B_harmonics}). Finally, we prove in Appendix~\ref{app:rvshift} that MLCCS methods are generally robust and consistent in detecting planets despite small changes in the cross-correlation length and extent, and that CNN1 is able to achieve invariance to RV shifts of the planet in the series.

\subsection{Results on the \textit{Directly Imaged Companions} Dataset}\label{subsec:resultsimgplanet}

In real-world scenarios, datasets are often highly imbalanced due to various reasons, such as the limited number of planets found in surveys, or the small fraction of spaxels containing a planet in a full image. Consequently, the \textit{directly imaged companions} dataset (c.f. Sect.~\ref{sec:implands}) serves a dual purpose. Not only it demonstrates the ability of MLCCS to recover a planet in imaging data without relying on spatial information, but also shows its broader effectiveness in handling highly imbalanced data. This paper aims to show the general capacity of  MLCCS on structured and reconstructed cubes, and hence presents global performance results for given noise levels (may it be stellar, instrumental or telluric). For a high contrast imaging oriented analysis measuring achievable ML performances regarding different contrast and separations, we refer to our paper companion \citet{nath2024}. They apply multi-dimensional CNNs specifically for high contrast imaging spectroscopy and show results for a set of contrast and separations.

\begin{figure}[tbh!]
    \centering
    \includegraphics[width=\columnwidth]{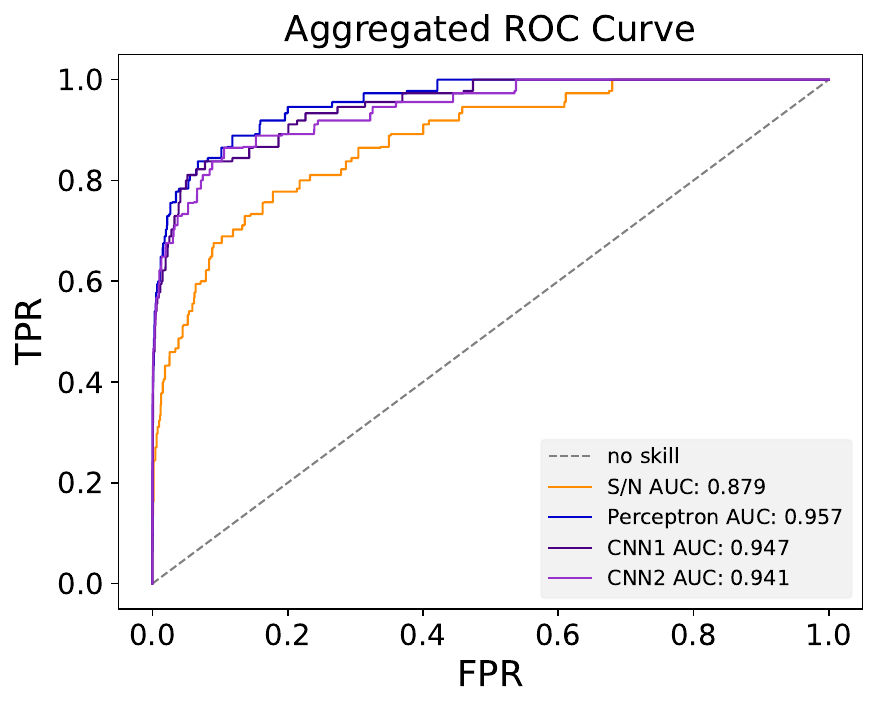}
    \includegraphics[width=\columnwidth]{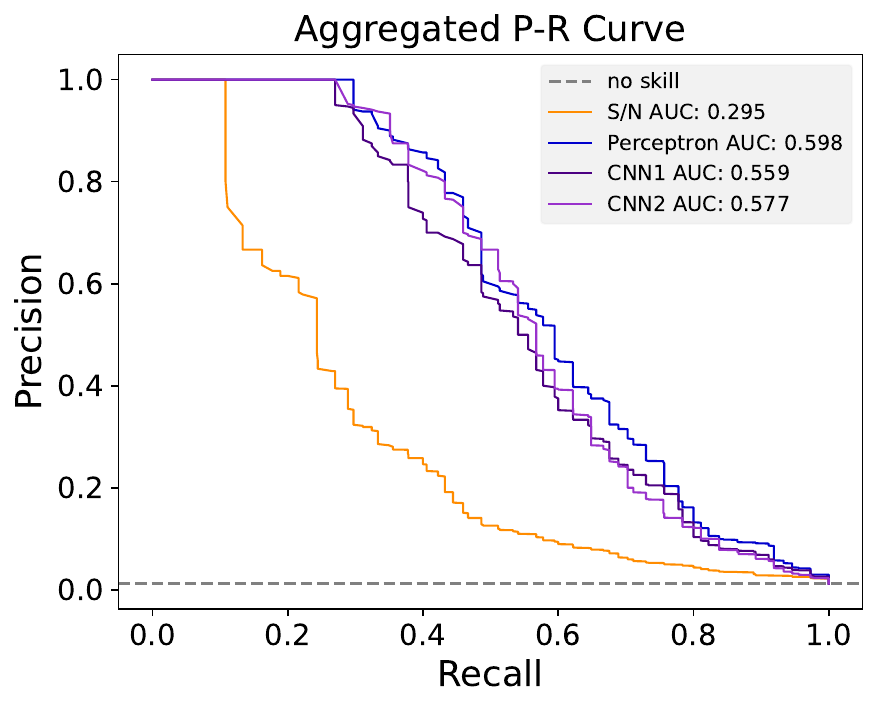}
        \caption{Aggregated ROC and PR Curves for the \textit{directly imaged companions} dataset. We use ROC and PR AUCs to quantify of the scoring quality of the ML methods relative to S/N. The ROC curves measure the trade-off between TPR and FPR, however, they tend to be over-optimistic in highly balanced frameworks such as in imaging data. The P-R curves measure the trade-off between precision and recall.}
    \label{fig:rocpr}
\end{figure} 

The models undergo training and validation on individual cross-correlated spaxels from $18$ cubes and are tested on the last cube in a cross-validated fashion of three tests. The end-to-end automated hyperparameter search process takes about 7 hours for a CNN. The baseline methods employ a single template channel ($\Teff=2900~\mathrm{K}$, $\logg=4.1~\mathrm{dex}$), while the CNN uses four channels simultaneously ($\Teff=\{2300,2500,2700,2900\}~\mathrm{K}$, $\logg=4.1~\mathrm{dex}$). Fig.~\ref{fig:rocpr} shows the scoring improvements performed by the MLCCS methods in comparison to the S/N metric, in terms of aggregated ROC and PR AUCs. For one image, the amount of positives (i.e. spaxels containing a planet in the image) are of the order of $\sim 1\%$, making it a highly imbalanced regime. The ROC curves are shown to enable comparison with plots related to the \textit{extracted spectra of companions}. Nevertheless, for quantification of the results, one should rely on the P-R Curves. While the PR AUC of the S/N equals $29.5\%$, the AUC of CNN1 achieves up to $57.7\%$, which represents a very big improvement for such imbalanced data. Results per test cubes are visible in Fig.~\ref{app:grid2} in Appendix~\ref{app:BDresults}; we can observe that the CNNs and the perceptron show rather equivalent performance results. This can be explained by the fact that the library of \textit{directly imaged companions} present less variations in atmospheric compositions and characteristics in comparison to the \textit{extracted spectra of companions}, making the use of template channels less relevant. While testing, we also noted that there was no significant performance improvement in P-R Curves by adding more than four template channels. Thus, a simple holistic approach such as the perceptron can be good enough, and should be favoured for computational efficiency when possible.

Fig.~\ref{app:grid1} shows visual results on three test cases from GQ\,Lup\,B noise cubes; S/N and probabilistic score results are presented together with detection grids. Once again, MLCCS methods offer a clearly enhanced conspicuity, as the score maps show a more confident separation between signal and noise, as previously discussed in regards to Fig.~\ref{fig:distr8}. Although S/N and probabilistic thresholds are difficult to compare as they obviously fold-in information differently, the improvement is still well visible in all panels of Fig.~\ref{app:grid1}. The perceptron and CNNs offer more TP for less FP leakage, already at lower thresholds (e.g. at $0.3$ Probability instead of $T=3$ or $T=5$ S/N). Overall, the CNNs exhibit a better detection sensitivity, by finding more pixels than the S/N. In addition, the higher confidence translates into a better conspicuity. This outcome highlights the dual capability of those ML methods, and corroborates the results on the \textit{extracted spectra of companions} mock data.

\subsection{Supplementary Results on a Simulation of PZ\,Tel\,B}\label{subsec:resultspztel}

\begin{figure*}[]
    \centering
    \includegraphics[width=\hsize]{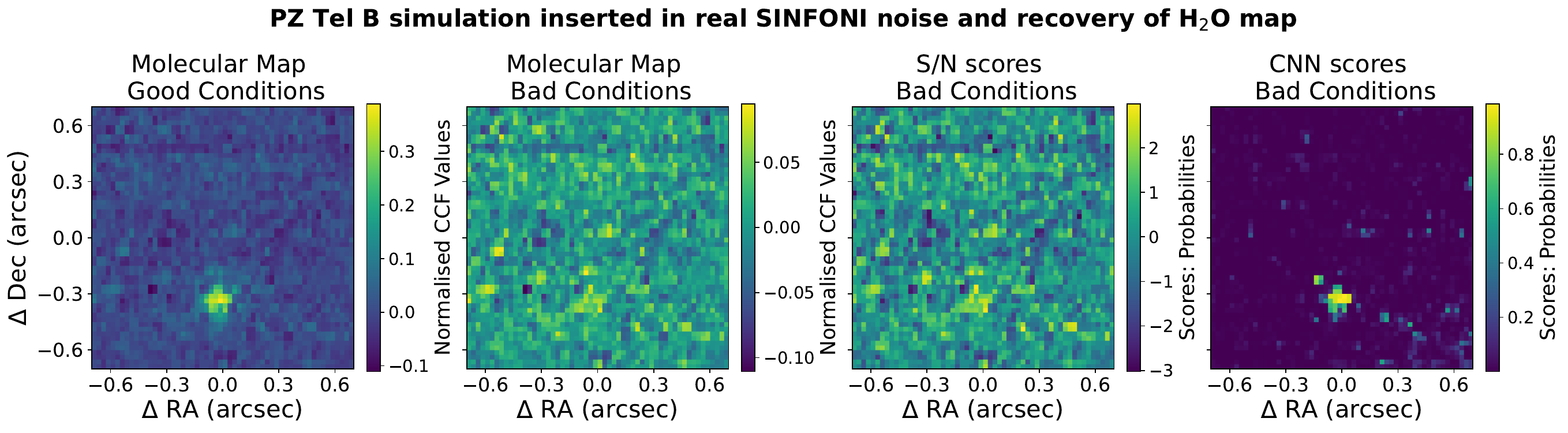}
    \includegraphics[width=\hsize]{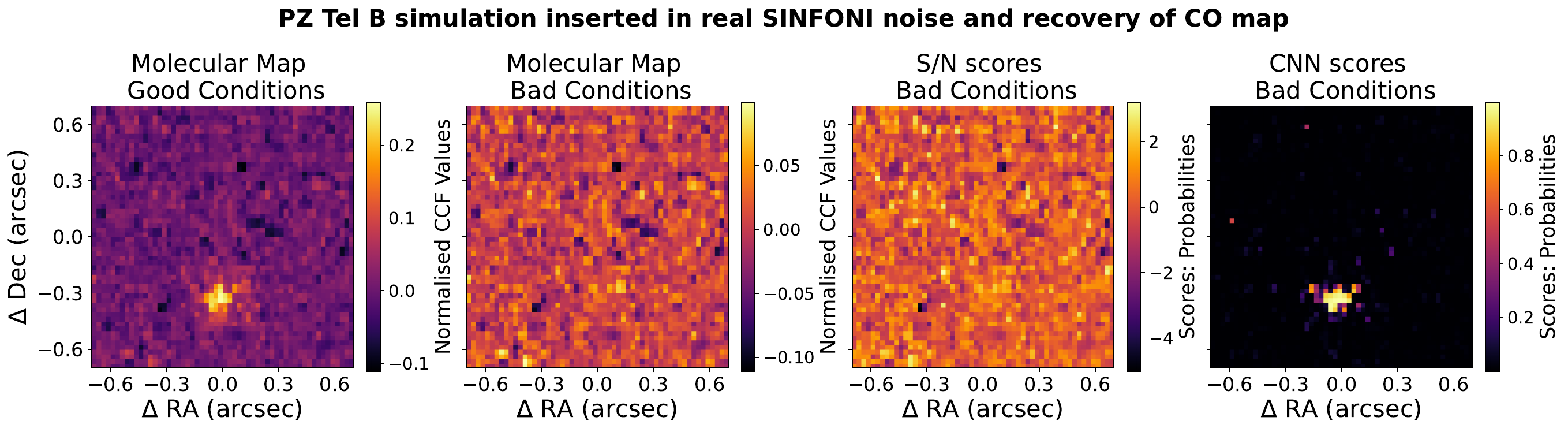}
        \caption{Simulated molecular maps and reconstruction of the predicted classification scores by the S/N and CNN. \textit{Left}: Insertion of a simulated PZ\,Tel\,B signal in the real PZ\,Tel\,B noise cubes, with same location, Gaussian decay and average $S/N=7.3$ as the real data shown as a molecular map in good seeing conditions. \textit{Middle-left}: Insertion of a signal in the PZ\,Tel\,B data with location and Gaussian decay of the original signal, but with an average $S/N=1.22$ to emulate a signal in bad seeing conditions. \textit{Middle-right}: S/N probability scores for bad seeing conditions. The planet is not detectable at $S/N = 5$. \textit{Right} CNN probability scores for bad seeing conditions. A very clear conspicuity improvement is observed in probabilistic map, clearly enhancing the planet's visibility.}
    \label{fig:pztelbreconstruct}
\end{figure*}

In this section, we conduct a final test using a simulated version of the molecular maps from Fig.~\ref{fig:molmap_PZTEL}. Thus far, all tests were applied to provide a quantitative measure on the performance of MLCCS methods, with the use of evaluation metrics (e.g. ROC curves, P-R curves and Confusion Matrix). Such metrics require clear labelling of the signals occurring in spaxels, to be able to distinguish true from false classifications. This requirement enforces the planet injections to be bounded inside a delimited aperture in the test set. However, in real observational cases, the signal simply decays spatially from the object into the rest of the frame, until it is too weak to be detected. Therefore, the motivation for this test is to find out if our MLCCS methods, while trained with spatially bounded signal injections, can still perform reliably on a realistic test case where the signal freely decays into the image. However, the lack of signal and noise labelling prevents the use of evaluation metrics to precisely quantify detections. Instead, those conclusive results are presented qualitatively, and require interpretation in conjunction with the quantitative outcomes from prior tests.

We perform such tests with synthetic variants of PZ\,Tel\,B spectra inserted at rest frame. We use three cubes of PZ\,Tel\,B in bad seeing conditions (from the third set of cubes in Table~\ref{tab:obs}) in which we insert planets with different signal strengths. The atmospheric characteristics of the inserted planets are assumed to be $\Teff=2800~\mathrm{K}$, $\logg=4.1~\mathrm{dex}$, and the composition includes H\textsubscript{2}O, CO in a hydrogen and helium dominated atmosphere (c.f. Sect.~\ref{sec:simplan}). As for the planet insertions, the centroid and Gaussian decay of the stellar PSF is calculated from the original data cubes in good and bad seeing conditions. The S/N are calculated within a $3.5$ pixel aperture. Then, in opposition to the training set, the realistic insertion of the simulations are made using a free signal decay with no aperture bound. The left panel of Fig~\ref{fig:pztelbreconstruct} represents a benchmark insertion as a bright companion with an average H$_2$O signal of $S/N=7.3$, matching the signal strength of the original data under good seeing conditions. Additionally, we insert dimmer planets in the noise at a lower signal strength corresponding to bad seeing conditions ($S/N=1.22$ for H$_2$O). This realistic scenario does not enable the use of PR curves or a confusion matrix due to the absence of aperture delimitation between signal and noise. Therefore, results are presented qualitatively, and require interpretation in conjunction with the quantitative outcomes from prior tests.

We perform the tests twice, once for the H$_2$O molecule, and a second time with the CO molecule. For each test, the CNNs are run using five template channels of the same molecule  (with $\Teff=\{2700;2800;2900;3000;3100\}~\mathrm{K}$ and $\logg= \{3.7;4.3;4.1;4.1;4.1\}~\mathrm{dex}$ respectively). The S/N is evaluated on an exact H\textsubscript{2}O template match ($\Teff=2800~\mathrm{K}$, $\logg= 4.1~\mathrm{dex}$). We trained and validated on the \textit{imaged companions} mock data, excluding the three cubes which we reserve for testing. For this setting, the full automated training process, including hyperparameter search, took about 8 hours. Note, nevertheless, that the final model could simultaneously evaluate all three flattened IFUs within seconds, generating scored data which we split and reshape back into three cubes. Fig~\ref{fig:pztelbreconstruct} shows scoring maps for one cube at various noise levels and for both molecules, along with S/N and CNN results under simulated poor seeing conditions. Notably, for an average S/N of 1.22 for the $H_2$0 map and 1.05 on the CO map (measured within a 3.5 pixels aperture), CNNs show a drastic enhancement on conspicuity in scoring maps. Additional results, presented in Fig.~\ref{app:scores_grid1} and Fig.~\ref{app:scores_grid2}, show successful detection of simulated PZ\,Tel\,B signals on bad seeing conditions, namely at scaled-down average H$_2$O S/N levels of 1/3\textsuperscript{rd} (top) and 1/6\textsuperscript{th} compared to the original good seeing conditions. Those results show that MLCCS is very be robust for planet detection in challenging noise and observing conditions, especially in non-Gaussian i.i.d. environments.

\subsection{Explainability of the Models by Ensuring Spatial Independence, and Generalisability to other Instruments}\label{disc:explain}

Through this work, we have put emphasis on providing a clear explainable and interpretable machine learning framework, to ensure that the model learns the right features. Thus, by preserving spatial independence, we force MLCCS to focus on the transformed spectral dimension. This strategy prevents the algorithms from learning structural artefacts from the spatial dimension, such as the PSF, spurious spatial dependencies, or local noise structures, which could be deleterious for the explainability and generalisability of some results. Thus, we train all supervised classification algorithms only on the RV extent of individual cross-correlated spaxels. Consequently, observing a clear tight aggregation of pixels in a spectroscopic image, instead of scattered pixels, increases confidence for the candidate detection, since we know that each pixel's score is independent from its neighbour; this is clearly exemplified in Fig.~\ref{fig:pztelbreconstruct}. We also made tests to verify what the models are learning. First, we show in Appendix~\ref{app:nongaussian} that in the presence of Gaussian noise structures, the CNNs would not be able to extract more information than the cross-correlation peak itself, and would yield very similar performances as the S/N \citep[e.g.,][]{gabbard2018matching}. Second, we ran tests on the CO molecule, which contains strong symmetries, to show that MLCCS methods do learn specifically the cross-correlation patterns such as harmonics and auto-correlation from molecular signals (c.f. Appendix~\ref{appendix_B_harmonics}). As \citet{hoeijmakers2018medium} and \citet{malin2023simulated} indicate, these patterns might be the result of overtones, related to the evenly spaced molecular lines in the template and spectrum of the exoplanet. These overtones are not used by the S/N statistic and may even decrease its score strength, but they do provide valuable information to the ML approaches. Third, although generally robust to small changes in the extent of the cross-correlation series, we report a light trend favouring more information in the cross-correlated series, which can be leveraged by MLCCS (c.f. Appendix~\ref{app:rvshift}). Finally, as CNNs can be implemented to be invariant to shift and stretches \citep{chaman2021truly}, we successfully validated preliminary tests on invariance to RV shifts of the planet in the cross-correlation series. Such tests have important implications for detection capabilities, which are further discussed in Appendix~\ref{app:rvshift}. 

In order to delve deeper into the explainability of the framework, we also trained and tested for a wider range of models, which included $L_1$ regularisation and variable selection oriented methods (e.g. Lasso, Random Forests etc) to understand how information is used in the RV dimension. We noticed that the variable selection methods would optimise towards using only the information at the cross-correlation peak and its immediate neighbourhood, but performed barely better than the S/N metrics. This indicates that the strong correlation between RV features prevents these models from performing a coherent variable selection \citep{tolocsi2011classification}. The methods which partially or fully include L\textsubscript{2} regularisation such as ElasticNet and Ridge are able to handle the correlations between RV features in a more consistent way: by setting a lower regularisation level and a higher $L_2$ to $L_1$ regularisation, we improved the results on ROC and PR AUC metrics. This means that if regularisation needs to be increased, for instance to improve the generalisability of the methods for noise from various targets, the $L_2$ type should be favoured.

With the spatially independent training scheme, the proof of concept is not only useful for isolated spectra or spectroscopic imaging modes; it also ensures the flexibility to train and test MLCCS with various spectroscopic modes (e.g. IFS and slit spectroscopy), as it does not depend on the structure nor the order of the given spaxels. For example, the tests on the \textit{extracted spectra of companions} dataset are valid for imaging spaxels as well as for long slit and single slit spectroscopy. In addition, CCS has proven to also work on transmission spectra \citep{de2013detection}. Thus, by replacing tests on medium-resolution planetary emission spectra to high-resolution transmission spectra, and by adapting the noise sources, our approach could undergo future investigations in this direction. Potential challenges related to stellar contamination and variability may need to be investigated and addressed in this case. 

Nevertheless, learning spatial information and the PSF could be a favourable by-product, according to the purpose and provided it is done and tested carefully. In this regard, the companion paper \citep{nath2024} applies a variant of the MLCCS approach specifically tailored for high contrast imaging spectroscopy. They demonstrate that they can achieve higher detection performance than non-ML methods by effectively leveraging features in space and time dimensions of a cross-correlated IFS. As a trade-off, lower emphasis is put on the RV series, which are included as a discrete set of filters gathered around the expected planet's RV, with the use of a unique full atmospheric template instead of a set of molecular template channels. Thus, the parallel and differentiated frameworks (i.e. learning molecular features vs. learning spatial features) emphasise different aspects which can be learned by CNNs. Together, both approaches consolidate each other in showing that ML can improve detection sensitivity to planetary signals in cross-correlated spectroscopic data. Future users should choose the framework according to the data and to the information they want to privilege and extract from it, or could use both to corroborate results.

\section{Conclusion}
\label{conclusion}

Through this paper, we have introduced a novel approach, namely MLCCS, to merge machine learning and cross-correlation spectroscopy for exoplanet detection, by leveraging molecular signatures in spectral data. As MLCCS techniques adopt a holistic approach in the cross-correlated spectral dimension, they are able to identify patterns from molecular signals, including harmonics and overtones. We show overall, that CNNs can operate effectively in non-Gaussian and non-i.i.d. environments to reveal sub-stellar companions embedded into complex noise structures. Thus, this approach addresses particular scenarios in which molecular mapping and CCS alone would fail to offer clear detections in the spectral dimension, while our companion paper by \citet{nath2024} address detections with emphasis in the spatial and temporal dimensions. Both papers strongly complement eachother in tackling detection problems, by leveraging dimensionalities differently in cross-correlated cubes, with the common purpose to detect planets embedded in spectroscopic noise.

Hence, through this work, we conducted two broad sets of experiments with dedicated test data sets to compare the performance of the MLCCS methods with a classical S/N statistic. In the first experiment we assess the performance of exoplanet detections under varying atmospheric characteristics in an unstructured stack of individual spectra. We found that MLCCS methods are able to discover $77$ times more planets than the S/N for a false discovery rate constrained with an upper bound at $5\%$. This achievement is attributed to the CNNs' use of multiple template filters, which enables an agnostic approach to exoplanet detection when atmospheric characteristics are unknown. Through additional tests, we also validated the capacity to combine different molecular templates to target a broader set of compositions. We also validated the invariance of our neural networks to RV shifts. Those two results have major implications for improving the flexibility in detecting unexpected planets in the data. In the second experiment, we showed that MLCCS enhances the detection of companions in structured data such as imaging spectroscopy. Indeed, we could strongly improve the conspicuity on scoring maps despite the spatially independent training scheme. 

By ignoring the spatial dimension, we focused on training MLCCS along the transformed spectral dimension. The goal is to ensure flexibility towards various spectroscopic instruments and observing modes for which cross-correlation methods have already proven useful (\citealp[e.g. with IFS data from VLT/SINFONI and JWST/MIRI][]{hoeijmakers2018medium, patapis2021direct}, \citealp[and slit spectroscopy with CRIRES, Keck/OSIRIS and CRIRES+][for both emission and transmission spectra]{de2014identifying, dit2018molecule, boldt2023optimising}). Indeed, the method could in principle be adapted and tested for transmission spectra after appropriate adjustments to the data and templates. On a wider scope, this technique would be crucial for spectroscopic observations which require relatively long observing times, as the signal strength may depend positively on integration time \citep[e.g.,][for IFS]{kiefer2021spectral} or even exposure and/or time resolution \citep[i.e.][on transmission spectra with CRIRES+]{boldt2023optimising}. Our MLCCS methods are able to detect planets even under challenging noise conditions. This ability could offer a significant reduction in telescope time. Moreover, we emphasise that all our tests were performed on individual or sub-combined exposure cubes; therefore we strongly encourage further investigations in this direction. For a better generalisability, we would also aim for full invariance towards noise cubes of different targets observed with the same instrument, to allow even for more robustness towards various seeing conditions.

Overall, our MLCCS methods improve the sensitivity to detect exoplanets and their molecules while offering robustness to systematic noise and sensitivity to molecular harmonics. This enhancement is crucial for exoplanet surveys in spectroscopic data \citep[e.g.,][]{agrawal2023detecting},  especially if they allow to reduce the required telescope time. We also expect this new approach to be beneficial to perform detection in cases where angular differential imaging \citep{marois2006angular} can not be employed, when strong systematic noise subsists after data reduction \citep[e.g.,][]{hoeijmakers2018medium}, and in poor observing conditions. Fortunately, existing archival data from VLT/SINFONI, CRIRES and new data inflow from VLT/ERIS, CRIRES+, JWST/NIRSpec will provide the chance to widely test, validate and calibrate MLCCS towards extensions and applications with various spectroscopic modes. Specifically, future work should investigate the benefits of using MLCCS methods with the newest and future instruments for which cross-correlation based methods have proven their potential, such as the detectability of trace species in cool companions with JWST/MIRI \citep{patapis2021direct, malin2023simulated}, the search for isotopologues in data from various instruments \citep{molliere2019detecting, zhang202113co, gandhi2023jwst}, or performance assessments of molecular mapping methods in challenging observation conditions with ELT/HARMONI \citep{houlle2021direct,bidot2023exoplanets, vaughan2024behind}.

\begin{acknowledgements}
E.O.G., J.H., and S.P.Q. gratefully acknowledge the financial support from the Swiss National Science Foundation (SNSF) under project grant number 200020\_200399. G.C. thanks the Swiss National Science 902 Foundation for financial support under grant number P500PT\_206785. R.N.-R.\ and O.A.\ are funded by the Fund for Scientific Research (F.R.S.-FNRS) of Belgium, and acknowledge funding from the European Research Council (ERC) under the European Union's Horizon 2020 research and innovation programme (grant agreement No 819155). The contributions of D.P. have been carried out within the framework of the NCCR PlanetS supported by the Swiss National Science Foundation under grants 51NF40\_182901 and 51NF40\_205606. D.P. acknowledges support of the Swiss National Science Foundation under grant number PCEFP2\_194576. We thank Rico Landman as well as Chloe Fisher for their helpful comments. We also thank Felix Dannert and Philipp Huber for many insightful discussions and their valuable feedback regarding the figures. We thank the editor for the useful suggestions which improved our manuscript. We thank the referee for the very careful review and propositinos which allowed us to greatly improve the quality of the manuscript.\\
\\
\textit{Author Contributions.} 
E.O.G. carried out all analyses, defined the optimisers and evaluation metrics and wrote the manuscript and codes. M.J.B. proposed the initial idea to use CNNs to detect exoplanets, and made propositions along the work to improve the structure of the manuscript. J.H. carried out the simulations of the planets and brown dwarfs and contributed to writing the related manuscript section. G.C. provided and reduced the real data cubes. J.S. tested the codes and the RV invariance under supervision of E.O.G. S.P.Q and N.F.M. provided guidance and administrative supervision along the project. All co-authors discussed the results and carefully commented on the manuscript.\\
\\
\textit{Code availability.} 
The codes are made publicly available on github in the form of a python library named MLCCS: \url{https://github.com/eogarvin/MLCCS}. Our codes make extensive use of \texttt{Keras} \citep{chollet2015}, \texttt{Matplotlib} \citep{Hunter:2007}, \texttt{Numpy} \citep{harris2020array}, \texttt{Pandas} \citep{mckinney2010data}, \texttt{Photutils} \citep{larry_bradley_2023_7946442}, \texttt{PyAstronomy} \citep{pya} and \texttt{Scikit-learn} \citep{scikit-learn}. The data preparation required the use of \texttt{petit\textsc{radtrans}}\citep{molliere2019petitradtrans}.\\

\end{acknowledgements}

%
\bibliographystyle{aa} 
\bibliography{references.bib} 
%
\clearpage

\onecolumn
\appendix

\section{Molecular Mapping: detections of CO and H$_2$O in PZ\,Tel\,B}\label{supp:molmap}

\begin{figure*}[htb!]
\centering
\includegraphics[width=16cm]{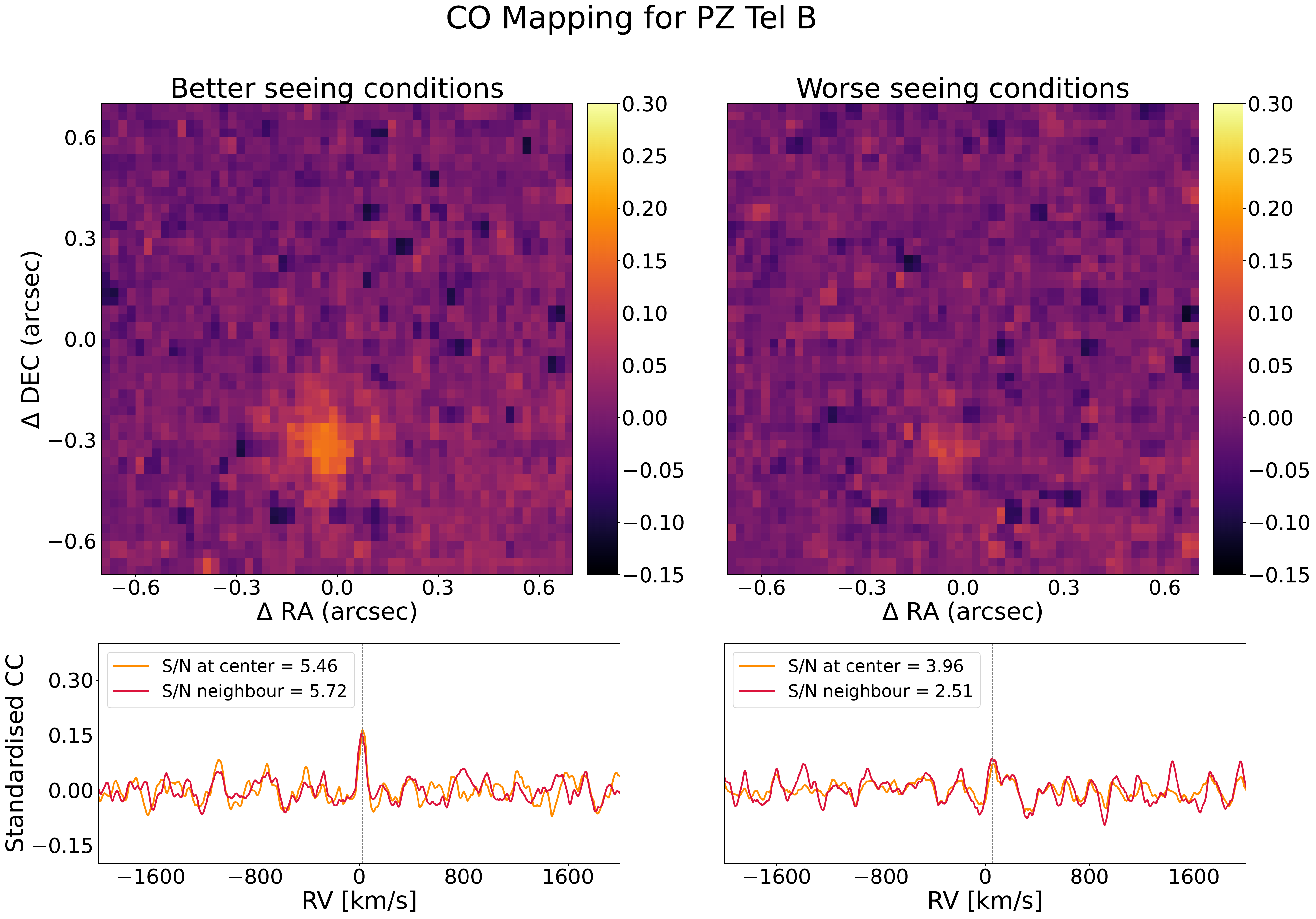}
\caption{Molecular mapping detection of CO for real PZ\,Tel\,B data using cross-correlation for spectroscopy. This figure is an analogue to Fig.~\ref{fig:molmap_PZTEL}, applied to the CO molecule instead of H$_2$O. It is even more obvious in the case of CO, that the cross-correlation and molecular maps fail to yield a clear peak under worse conditions.}
\label{fig:molmap_PZTEL_CO}
\end{figure*}

\begin{multicols}{2}
In this section, we report tentative "$5\sigma$" detections of H$_2$O (c.f. Fig.~\ref{fig:molmap_PZTEL}) and CO (c.f. Fig.~\ref{fig:molmap_PZTEL_CO}) for PZ\,Tel\,B data obtained with molecular mapping. The presence of H$_2$O and CO have been predicted by \citep{stolker2020miracles} according to calculated abundance profiles in relation to the measured temperature on the object. For both Fig.~\ref{fig:molmap_PZTEL} and~\ref{fig:molmap_PZTEL_CO}, the left IFU was observed on 04.05.2015 (Program ID: 093.C$-$0829 B), in 4 DITs of 60s with airmass 1.11, an average coherence time of 0.001931, and a start to end seeing from $0.77$ to $0.72$, which we define as the better observation conditions. The right IFU was observed on 28.07.2015 (Program ID: 093.C$-$0829 B) in 20 DITs of 10s with airmass 1.12, an average coherence time of 0.001931, and a start to end seeing from $1.73$ to $1.54$, which represent overall worse observation conditions relative to the right plot. The seeing seems to be a dominant factor in reducing the observation quality, but we also note that the \textcolor{orange}{exposure time} is lower.  
\end{multicols}

\section{Extended tests on the \textit{Extracted spectra of companions}}\label{appendix:B}

\subsection{Interpretability of the results: the scaling factor $\alpha$}\label{disc:scale}

\begin{figure*}[htbp!]
    \centering
    \includegraphics[width=\hsize]{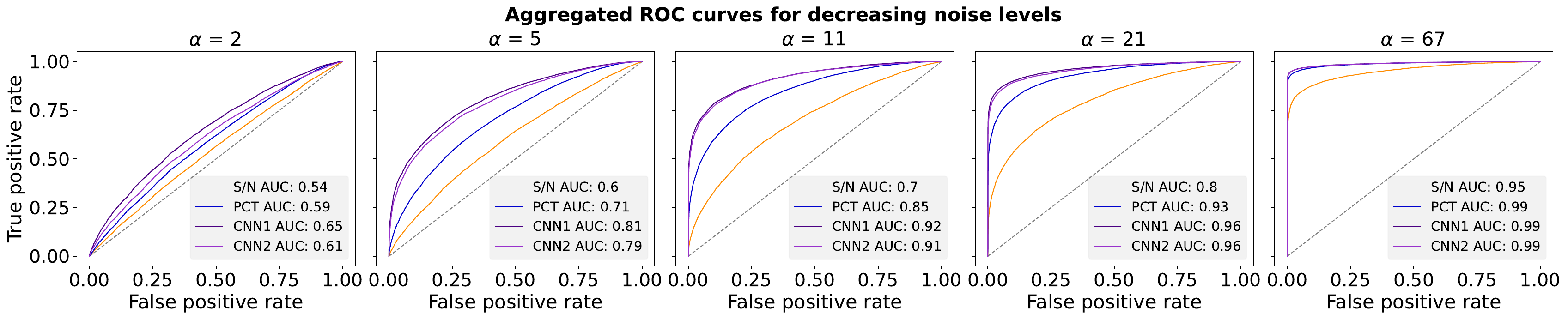}
    \caption{ROC AUC improvements and relative gains of MLCCS for a range of $\alpha$ values. The scale factors of $\alpha=[2,5,11,21,67]$ (from left to right), were selected to illustrate that MLCCS offers a smaller improvement relative to S/N for very small $\alpha$ (extreme noise) and very large $\alpha$ (extreme signal). The scaling factor $\alpha$ is inversely proportional to the strength of the noise.}
    \label{fig:rocalphas}
\end{figure*}

\begin{multicols}{2}
In order to test the sensitivity of the results to varying levels of noise, we implemented the methods over different scale factors on the \textit{extracted spectra of companions} dataset. The companions were inserted with different scale factors on the noise, as shown in equation~\eqref{eq:alpha}. We tested for the values of $\alpha = \{2;5;8;11;16;21;29;41;67\}$, which are respectively selected according to the following ROC AUC values of S/N: $[0.55;0.6;0.65;0.7;0.75;0.8;0.85;0.9;0.95]$. We depict the most illustrative cases in Fig.~\ref{fig:rocalphas}, to show the variations of the relative gains across methods, and that they depend on different levels of noise. 

The Fig.~\ref{fig:rocalphas} shows that the relative gain of using MLCCS methods do indeed depend on the noise level. For an excessively small or big scale factor, the improvement differential between the statistic metric (S/N) and the MLCCS methods will be almost null. If signals are drowned into the noise, all classifiers will lose their skills (c.f. $\alpha=2$ in Fig.~\ref{fig:rocalphas}). If signals are extremely strong, the noise has no influence and any statistic or algorithm tends towards a perfect classifier (c.f. $\alpha=67$ in Fig.~\ref{fig:rocalphas}). However, there is a range of noise levels, between the extremes, for which it is worth using MLCCS due to the gain in detection performance.
\end{multicols}

\subsection{Flexibility towards compositions of the planets: leveraging template multiplicity}\label{appendix:B_supp}

\begin{figure}[htb!]
    \centering
    \includegraphics[height=4.5cm]{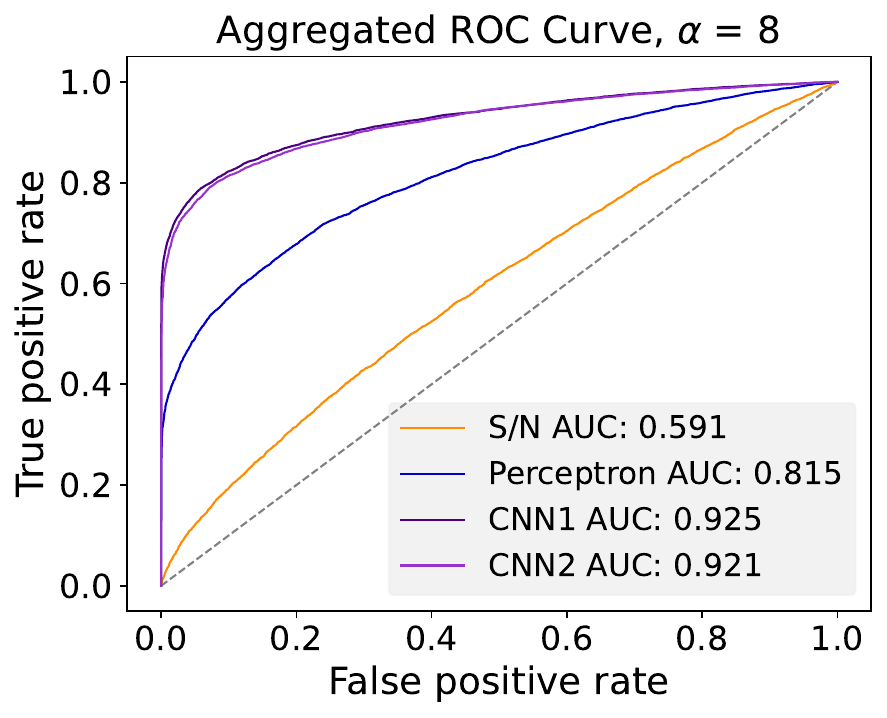}
    \includegraphics[height=4.5cm]{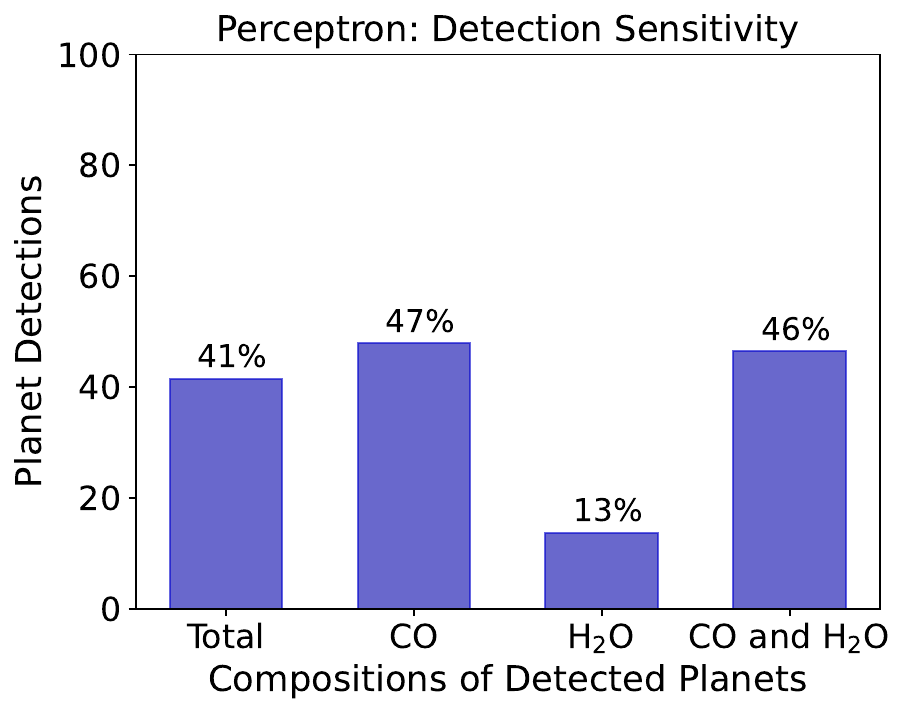}
    \includegraphics[height=4.5cm]{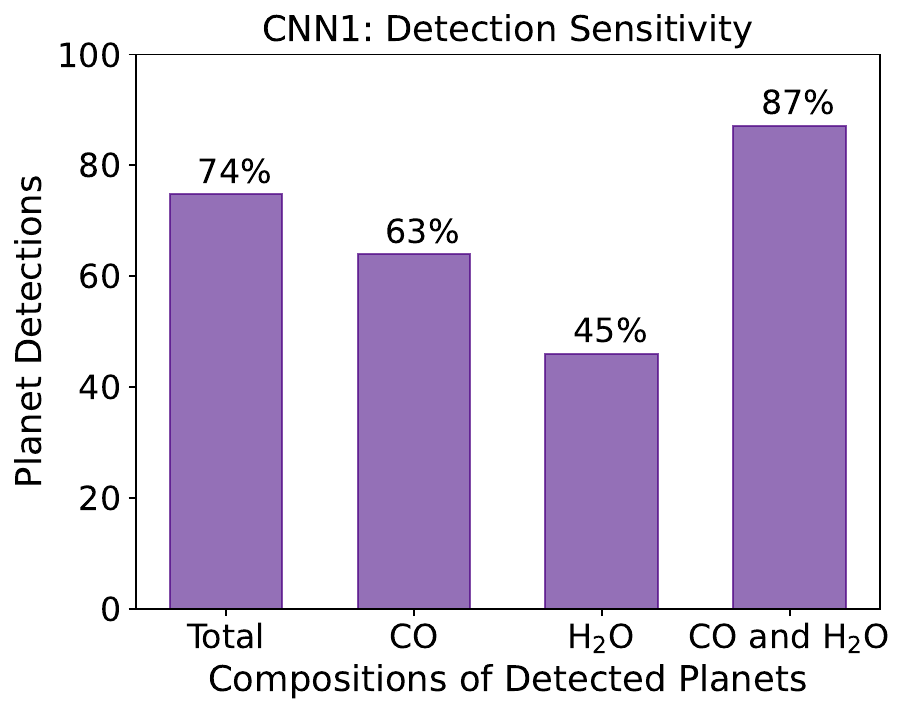}
    \caption{Scoring and Classification of the S/N and CNN, for planets cross-correlated with template channels of CO and H$_2$O. Overall, 12161 planets containing either CO, H$_2$O, or both, alongside with other molecules, were inserted with a scaling factor of 8. \textit{Left}: Aggregated ROC curves as already shown in Fig.~\ref{fig:accroc8} from section~\ref{subsec:respp}. \textit{Middle and Right:} Amount of planets which can be recovered by the perceptron and the CNN, for different characteristics (i.e. all planets, those containing either CO or H$_2$O, or both.}
    \label{fig:H2O_CO_test}
\end{figure}

\begin{multicols}{2}
In this section, we investigate how varying the composition of the template channels enables to search for a wider range of companions with CNNs, while increasing flexibility, agnosticity and detection sensitivity to planets. Thus, for this example, we use a total of 10 channels. We include 5 template channels of CO in parallel to 5 templates of H$_2$O for each CNN. Thus, for both molecules this yields $\Teff =[1200,1400,1600,2000,2500]$, while the CO templates have $\logg = [2.9,4.1,4.5,4.1,5.3]$ and the H$_2$O templates differ with $\logg = [4.1,3.5,4.1,4.1,5.3]$. As usual, the templates are chosen to cover the parameter space evenly (c.f. Sect.~\ref{sec:simplan}), while alternating the composition. No template contains both molecules together; we only use singletons. The first template (i.e. H$_2$O with $\Teff=1200$ and $\logg=4.1$) serves as "base template" for the S/N and perceptron which can only be fed with one channel and can detect only one molecule at a time. 

We run tests on the \textit{extracted spectra of companions} mock data. The test case uses a similar setting as in the main framework (as described in Sect.~\ref{subsec:planetpopds}), except for the fact that now, 50\% of the injected planets contain either H$_2$O, CO, or a mixture of both (amongst other molecules). The rest contains either pure noise or planets with other molecular mixtures, excluding H$_2$O and CO. For comparability with the main test setting, all planets are also injected with a scale factor $\alpha = 8$. The labels are provided to the algorithms to indicate that a detection should contain at least one of the two molecules. 

The ROC curves are shown in Fig.~\ref{fig:H2O_CO_test}. First, the ROC AUC of the S/N is low, with a value of 0.591, although we kept the same base H$_2$O template as for the main test in Fig.~\ref{fig:accroc8}. This is not surprising, since the S/N can only work with one template at a time and can only detect planets with water. On the other hand, the ROC AUCs show how the CNNs gain performance when allowed to search not only across atmospheric parameters, but also across molecular compositions. Such results are very promising for the implementation of agnostic frameworks in spectroscopic datasets. We also note that, at equal $\alpha$ value, the perceptron seems to have gained performance. It increased from a ROC AUC of 0.792 in Fig~\ref{fig:accroc8} to 0.815 in Fig.~\ref{fig:H2O_CO_test}. For this reason, we also investigated what types of planets the methods are able to find. 

The middle plot in Fig.~\ref{fig:H2O_CO_test} shows the percentage of total planets of interest that the perceptron is able to find. The perceptron finds $41\%$ of the planets in total. First, we observe that, although the perceptron is fed with the base molecular template of H$_2$O, it learns to detect CO more easily than H$_2$O. We emphasise that in main test (i.e. where we search only for planets with H$_2$O), the perceptron does not detect any planets which do contain CO without H$_2$O. Thus, those results are not accidental and are intrinsic to the fact that the perceptron learns some systematic and deterministic patterns of CO, even when the planet is cross-correlated with a different molecule. Then, we observe that by using multiple template channels with composition variations, that CNN1 is able to find up to $74\%$ of the total planets. It most consistently finds companions which contain both molecules. This result is very interesting, since all template channels consist of only one molecule at a time; this means the CNNs can combine the information provided separately to strengthen the detection of objects which satisfy more criteria. In addition, it also detects planets which contain either of each molecule, showing high flexibility for composition.    
\end{multicols}

\subsection{Explainability of the Framework: Idealistic Gaussian Noise}\label{app:nongaussian}

\begin{figure*}[htb!]
    \centering
    \includegraphics[height=3.5cm]{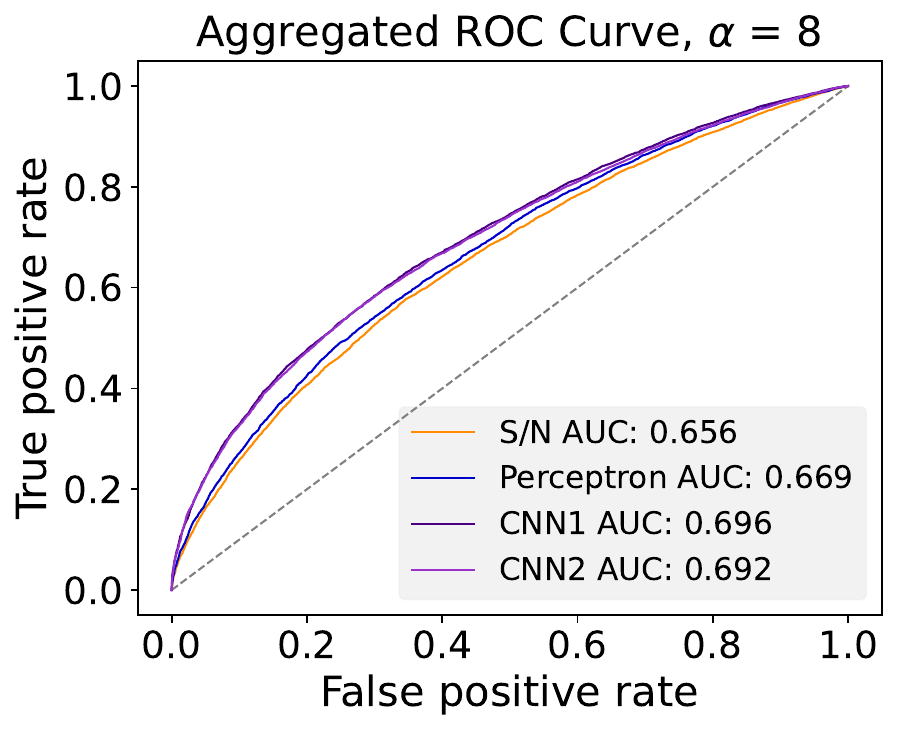}  
    \includegraphics[height=3.5cm]{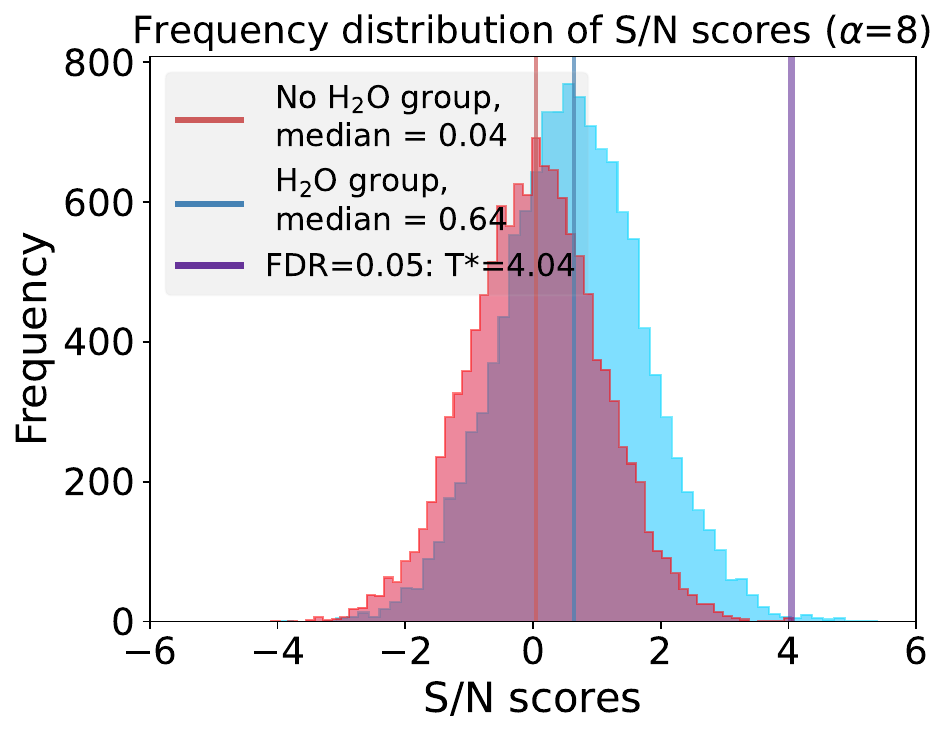}  
    \includegraphics[height=3.5cm]{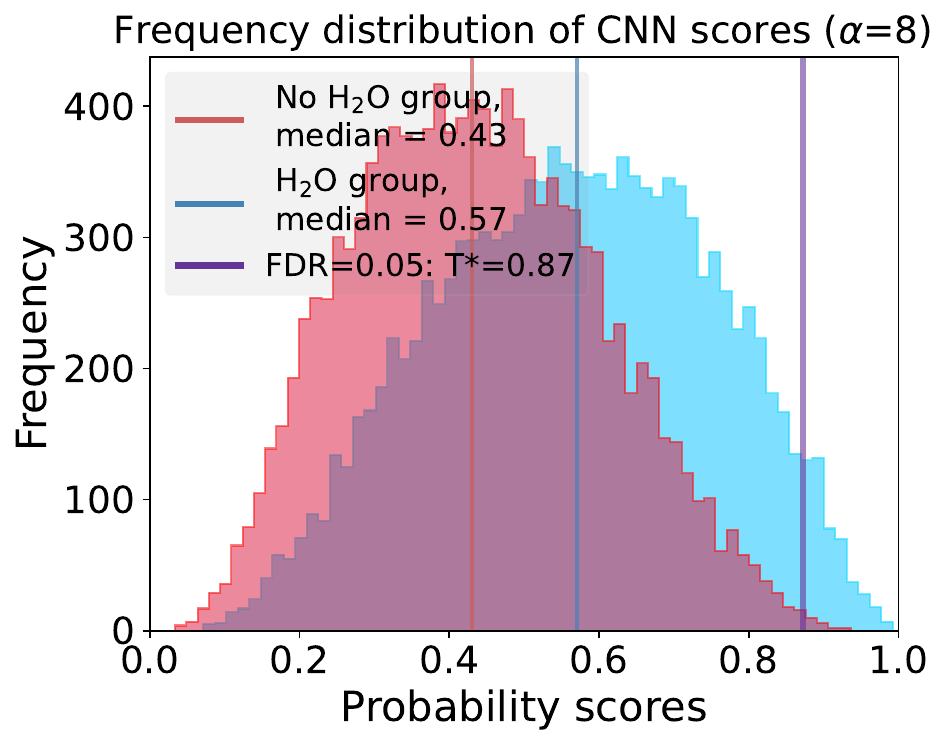}  
    \includegraphics[height=3.5cm]{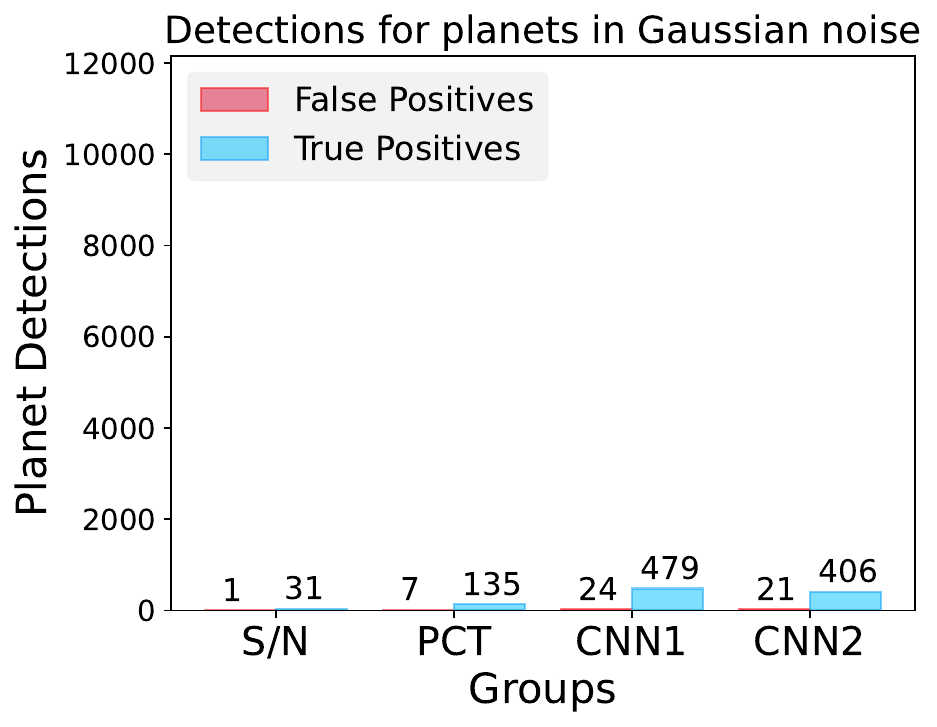}  
    \caption{Tests conducted on planetary signals (H$_2$O) embedded in idealistic simulated Gaussian noise. \textit{Left:} The ROC curves and AUC show that Neural Networks are barely able to perform better than the S/N. \textit{Middle-Left and Middle-Right:} Frequency distributions of aggregated scores assigned to the negative group (red) against the positive group (blue) for the S/N (left) and the CNN (right) in the presence of Gaussian noise. The CNNs are not able to separate the scores and improve conspicuity as well as in the case of non-Gaussian noise. \textit{Right:} Classification predictions under a Gaussian noise regime, given a FDR$\leq 5\%$. The use of MLCCS to extract signals becomes ineffective under Gaussian noise, as they are not able to extract more information than the S/N.}
    \label{fig:nongaussiantests}
\end{figure*}

\begin{multicols}{2}

In this section, we test and briefly discuss the results of the MLCCS methods for signal detection in an idealistic framework of identically and independently distributed Gaussian noise environment, as opposed to the main tests on real data from SINFONI. In order to do so, we build Gaussian replicates of the SINFONI noise. This means that for every SINFONI noise spectrum (c.f. Sect.~\ref{subsec:noise}, we estimate the mean and variance. Then, we simulate a Gaussian analogue of the noise series using the estimated mean and variance parameters. 
After this, all procedures we use to insert the planets in the noise are the same as in Sect.~\ref{subsec:planetpopds}. As we can observe from the left panel in Fig.~\ref{fig:nongaussiantests}, in the context of Gaussian i.i.d. noise environments, the MLCCS methods perform exactly as the S/N in terms of ROC AUC. This means that the algorithms are only able to learn as much information as the S/N provides. We can also observe this in the middle-right panel from Fig.~\ref{fig:nongaussiantests}. Indeed, the CNNs are not able to transform the data to separate the scores as efficiently as in the tests on non-Gaussian data (c.f. middle panel in Fig.~\ref{fig:distr8}).

In fact, such results are expected, since the cross-correlation provides the maximum likelihood solution for an optimal template in the case where the noise is Gaussian \citep{brogi2019retrieving}. This also means that the cross-correlation (as a generalisation of the Pearson's correlation) only provides a complete description of the similarity between a spectrum and a template when the random noise is Gaussian. Therefore, under pure Gaussian i.i.d. noise, methods such as matched filtering or S/N methods on cross-correlation would perform optimally in extracting signals. In such cases, CNNs would not be expected to perform better than the baseline, since there is no additional information to extract \citep{gabbard2018matching}. This is what we can observe with the results of our tests on Gaussian i.i.d. noise. We attribute the very small performance improvement of the CNNs here to remaining and persistent molecular harmonics which can overcome the Gaussian noise. Nevertheless, it is clear that if the signal is embedded in pure Gaussian noise, the MLCCS methods to not offer great improvements. This also acts as a proof of the non-Gaussianity of the realistic noise; the cross-correlation carries over the non-Gaussian i.i.d. noise pattern structures, which enable MLCCS to learn information from. We also note in the fourth panel of Fig.~\ref{fig:nongaussiantests}, that the S/N is detecting less planets than the non-Gaussian framework. This also shows that S/N detection significance and uncertainties are misestimated when we assume Gaussian noise on non-Gaussian data, because the detection capabilities change according to the nature of the noise. 

\end{multicols}

\subsection{Explainability of the results: Molecular Harmonics}\label{appendix_B_harmonics}

\begin{figure}[]
    \centering
    \includegraphics[height=5cm]{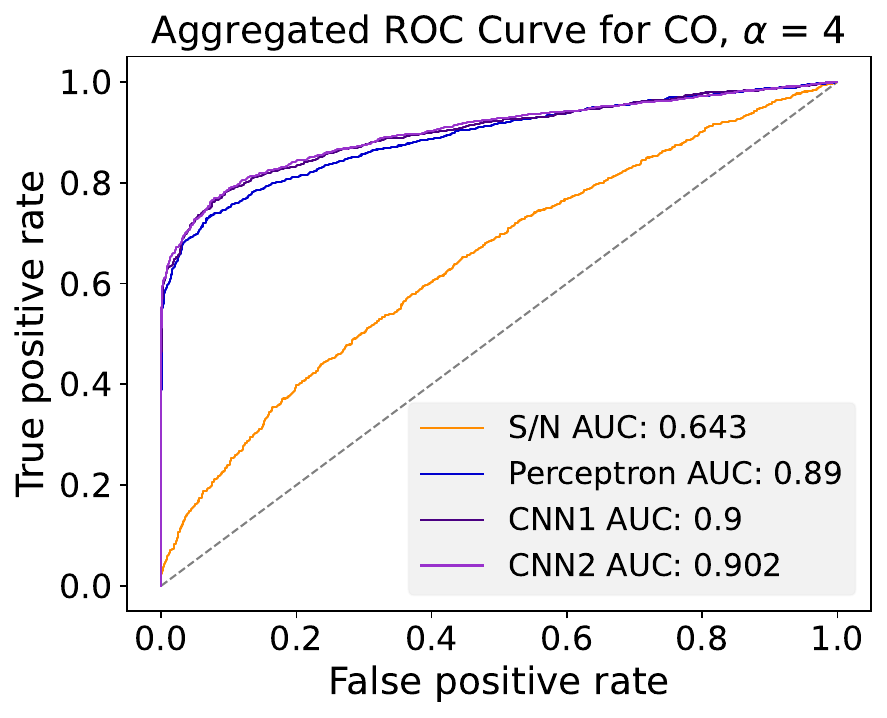}   
    \includegraphics[height=5cm]{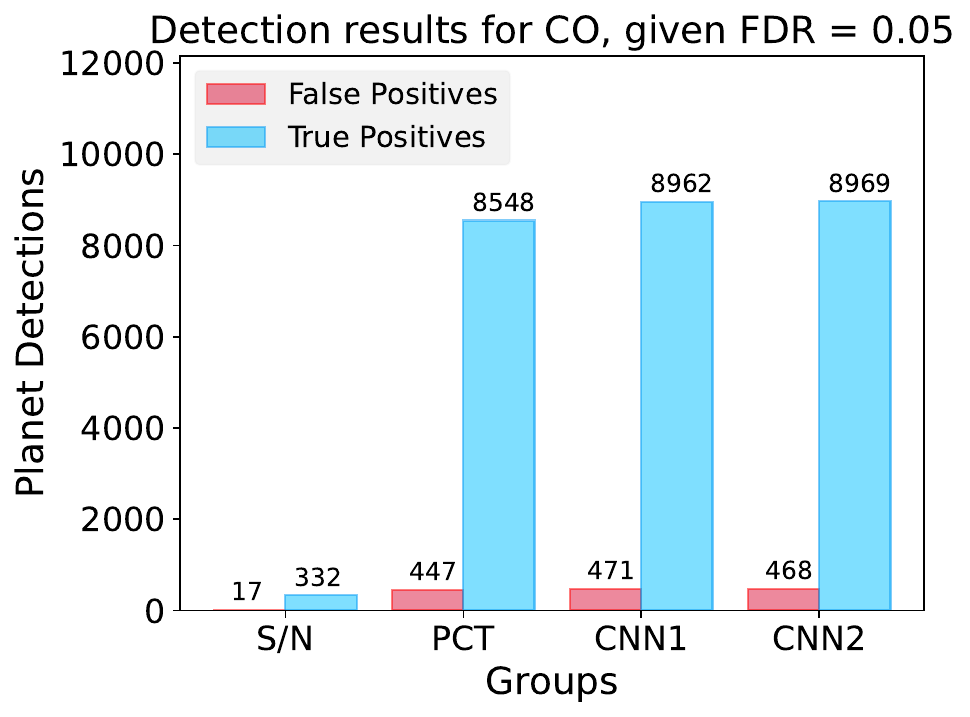}
    \caption{Scoring and Classification of the S/N and CNN for cross-correlated spectra with the CO molecule. 
    \textit{Left}: Quantification of scoring performance with receiver operating characteristic curves (ROC). The plot shows the improvements in the ROC trade-off between TPR and FPR. The improvement is measured in terms of area under the ROC curves (AUC). The CNNs provide even better performance in finding CO molecules while limiting increments in FPR, as compared to the main tests. \textit{Right:} Maximal amount of planets with CO which can be recovered in the mock data by the S/N, perceptron (PCT) and both CNNs, within a FDR $\leq 5\%$ bound.}
    \label{fig:distr_co}
\end{figure}

\begin{multicols}{2}

\begin{figure*}[]
    \centering
    \includegraphics[width=\hsize]{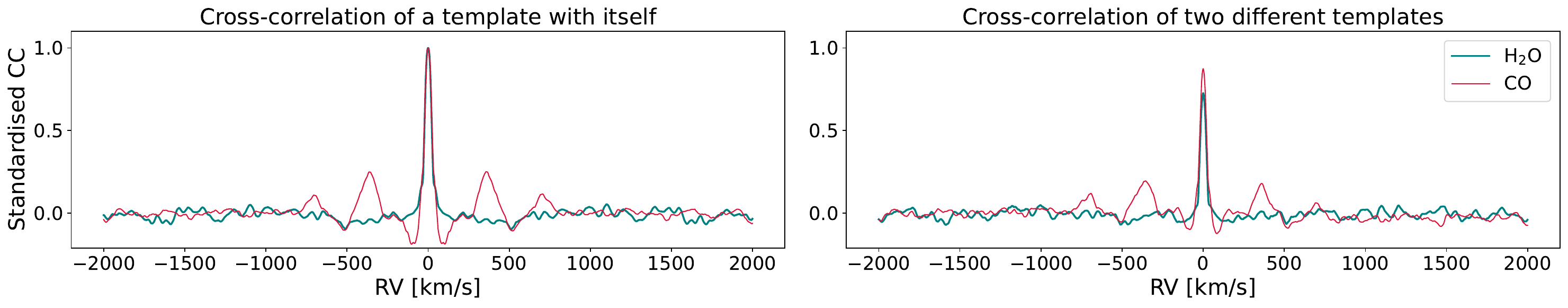}  
    \caption{Cross-correlated templates showing molecular harmonics and symmetric patterns for H$_2$O and CO. \textit{Left:} To verify how the molecular harmonics behave for H$_2$O and CO, we cross-correlated some templates with themselves, and representative instance in this panel, with a template of $\Teff=1200$ and $\logg=4.1$. The cross-correlation shows clear symmetric patterns for both molecules and strong harmonic patterns for CO. \textit{Right:} We also cross-correlated different templates with each other, for example one template of template of $\Teff=1200$ and $\logg=4.1$ and one template of $\Teff=2200$ and $\logg=4.1$ as shown in the right panel.}
    \label{fig:ccf_templates}
\end{figure*}

In order to improve the explainability of the framework, we investigated what the CNNs and the perceptron are able to learn. Such explainable frameworks in ML are fundamental to orient future research towards real data applications. As we suspected that MLCCS should be able to learn symmetries in the cross-correlated series as well as molecular harmonics, we extended the tests to the CO molecule only, which is known to host strong harmonics in its cross-correlated sets (Fig.~\ref{fig:ccf_templates}). We build a new dataset in the same way as in Sect.~\ref{subsec:planetpopds}. This time, it is constituted of $50\%$ of planets with CO. We observe in Fig.~\ref{fig:distr_co}, that for a similar S/N performance in terms of ROC AUC, it is easier to detect CO. Indeed, the perceptron achieves an AUC of 0.902 and the CNNs reach an AUC of 0.913. Such results proving the efficacy in detecting CO are corroborated by the molecular maps in Appendix~\ref{app:extended_molmapsim}.

The reason for this is that CO has clear and very strong harmonics, as shown in Fig.~\ref{fig:ccf_templates}. Those are easily learned using holistic approaches such as MLCCS on the whole cross-correlated series. Such harmonics are still visible and can be detected up to RVs of $\pm 750$, even in the case where the templates are not matching exactly with the planet's characteristics (c.f. example with the right panel of Fig.~\ref{fig:ccf_templates}). This relates to the fact that the ML algorithms are leveraging correlated patterns in the features of the cross-correlated series, as previously discussed in Sect.~\ref{disc:explain}. On the other hand, we observe that water has a clear cross-correlation peak which dies off at about $RV = \pm 200$. Its symmetric patterns are rather discrete but visible on the left panel. Thus, we see a clear difference in the strength of the auto-correlation patterns caused by harmonics and symmetries in the cross-correlated series for CO compared to H$_2$O molecules, explaining the performance improvement with CO templates.

\end{multicols}

\subsection{Robustness and Invariance: The Cross-correlation Series and Radial Velocity Shifts}\label{app:rvshift}

\begin{multicols}{2}
We ran tests to evaluate the robustness and invariance properties of our CNNs regarding variations in the cross-correlation series and shifts in RV. We first tested the effect of widening and shortening the extent of the cross-correlated series and widening the RV step. The methods' applicability is generally robust to such variations, as they involve only minor changes in the ROC AUC which consistently remains above 0.8 for the CNNs. Nevertheless, we generally observe that having more RV features offers better performance. For instance, lengthening the series up to $\pm 3000\kms$ slightly increases the ROC AUC (between 0.11 to 0.13 points for all ML methods compared to the main test). However, this comes with a cost in increased computational time. For tests where we shortened the series (e.g. $\pm 1200\kms$) or enlarge the RV steps (e.g. RV steps of 5 or 10), we observed an accelerated convergence of the optimisation process, but a slight decrease of the performance, although the AUC of the CNNs stay above 0.8. Finally, we tested more drastic changes, with a cross-correlation of $\pm 500\kms$ and RV steps of 10, overall reducing the RV features to only 100 points. In this case, the performance drops, with ROC AUCs reaching 0.724 and 0.728 for CNN1 and CNN2, against 0.685 for the perceptron and 0.652 for the S/N. Despite the relative robustness of the CNNs to changes in the amount of RV features, it  needs a reasonable amount of information to be able to learn structures in the cross-correlation.

 \begin{figure*}[]
    \centering
    \includegraphics[height=4.5cm]{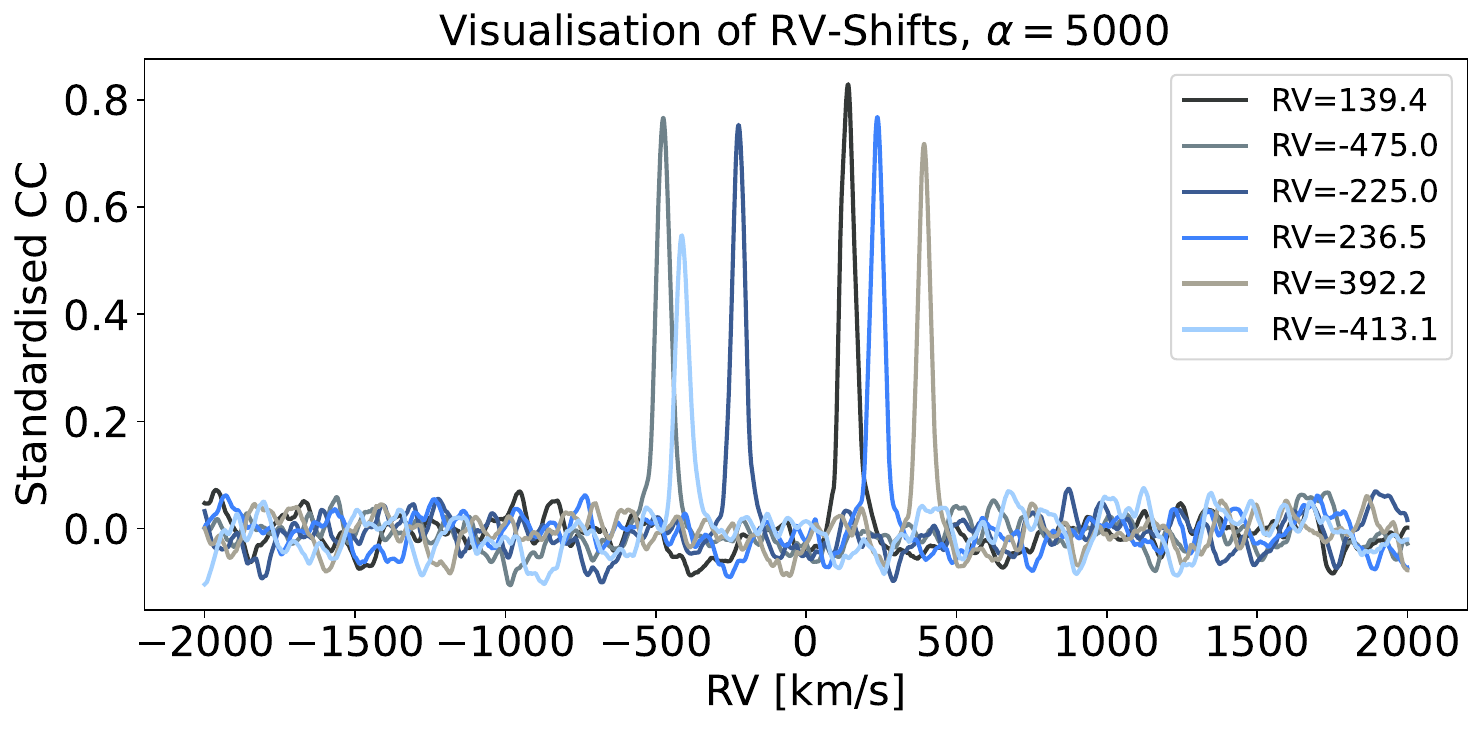}  
    \includegraphics[height=4.5cm]{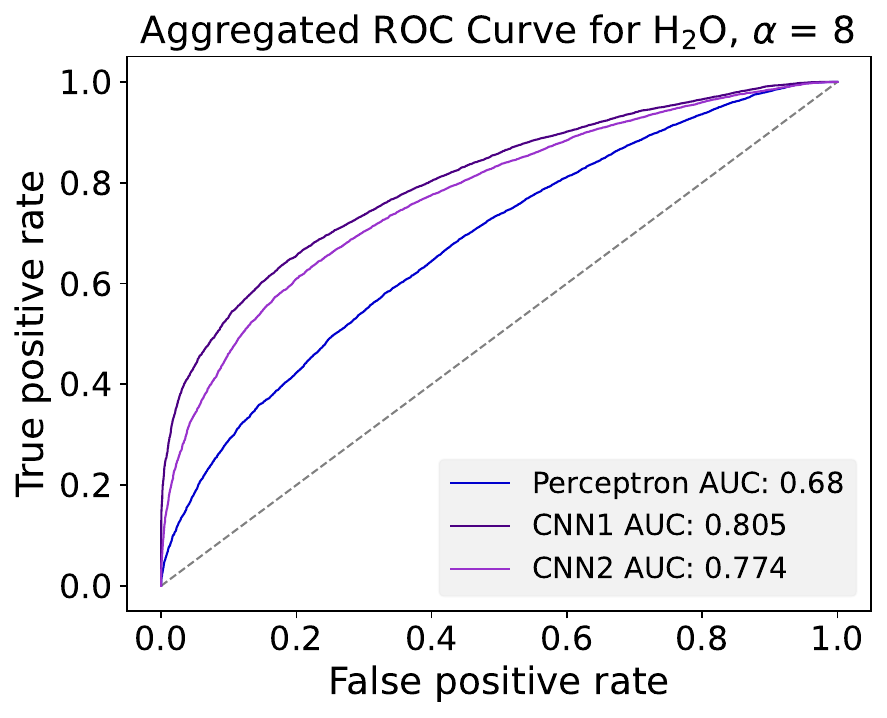}  
    \caption{Invariance tests on RV shifts. \textit{Left:} This panel shows an example of shifts in RV implemented between $\pm 500 \kms$. The H$_2$O signals were voluntarily pushed up to a scaling factor of $\alpha=5000$ only for visualisation purposes. \textit{Right:} The associated ROC AUC performance on RV shifted signals inserted with a scale factor of $\alpha=8$ yielding faint signals, as in the baseline framework.}
    \label{fig:rvshifts}
\end{figure*}

The second test was performed to evaluate invariance properties towards shifts in RVs. RV invariance is useful in the perspective that planets in real data have varying radial velocities which create shifts and stretches in fine pattern structures of the cross-correlated spectrum. So far, for comparability of the results with the S/N metric (which requires prior knowledge or visual inspection of the peak's RV), all tests were performed on planets present at rest frame. However, CNNs which were trained on a fixed RV could not generalise to varying RV shifts. Thus, we implemented variations of our dataset which include RV shifts occurring uniformly randomly around the rest frame. We tested realistic RV shifts with bounds at $\pm 100\kms$, as well as a more conservative scenario to challenge the CNNs with large RV shifts up to $\pm 500\kms$. The conservative results from the second test are shown in Fig.~\ref{fig:rvshifts}. Our results on both scenarios converged towards the conclusion that our CNNs are robust in detecting planets and their molecules in cross-correlations from Doppler-shifted spectra in the test set. Finally, true invariance \citep{Mouton_2020} to RV shifts is only proven when the test set is evaluated on a shift which was previously unseen in the training and validation sets. Thus, such tests were also evaluated; the most conservative scenario included training of MLCCS on RVs of [$-225.0$; $-276.8$; $236.5$; $176.7$; $392.2$; $-413.1$] $\kms$, validated on an RV shift of $139.4\,\kms$ and tested on an RV of $-475.0\,\kms$ and returned a ROC AUC of $0.805$ for CNN1 against $0.68$ for the perceptron. The lower performance of CNN2 is due to its more complex architecture and regularisation schemes mentioned in Sect~\ref{subsec:cnn}. Thus, CNN1 displays a high degree of invariance even in a conservative scenario. Such preliminary proofs on RV shifts have important implications for generalisation of the methods to real data, as shift invariant CNNs would be necessary to detect planets for which we do not have prior orbital constraints. On the other hand, training on fixed RVs can be preferred when prior orbital information is available and we focus the use of MLCCS for detection of molecular species.
\end{multicols}

\section{Extended test results on the \textit{Direct Imaging of Companions} dataset}\label{app:BDresults}

\begin{figure*}[tbh!]
    \centering
    \includegraphics[width=16.8cm]{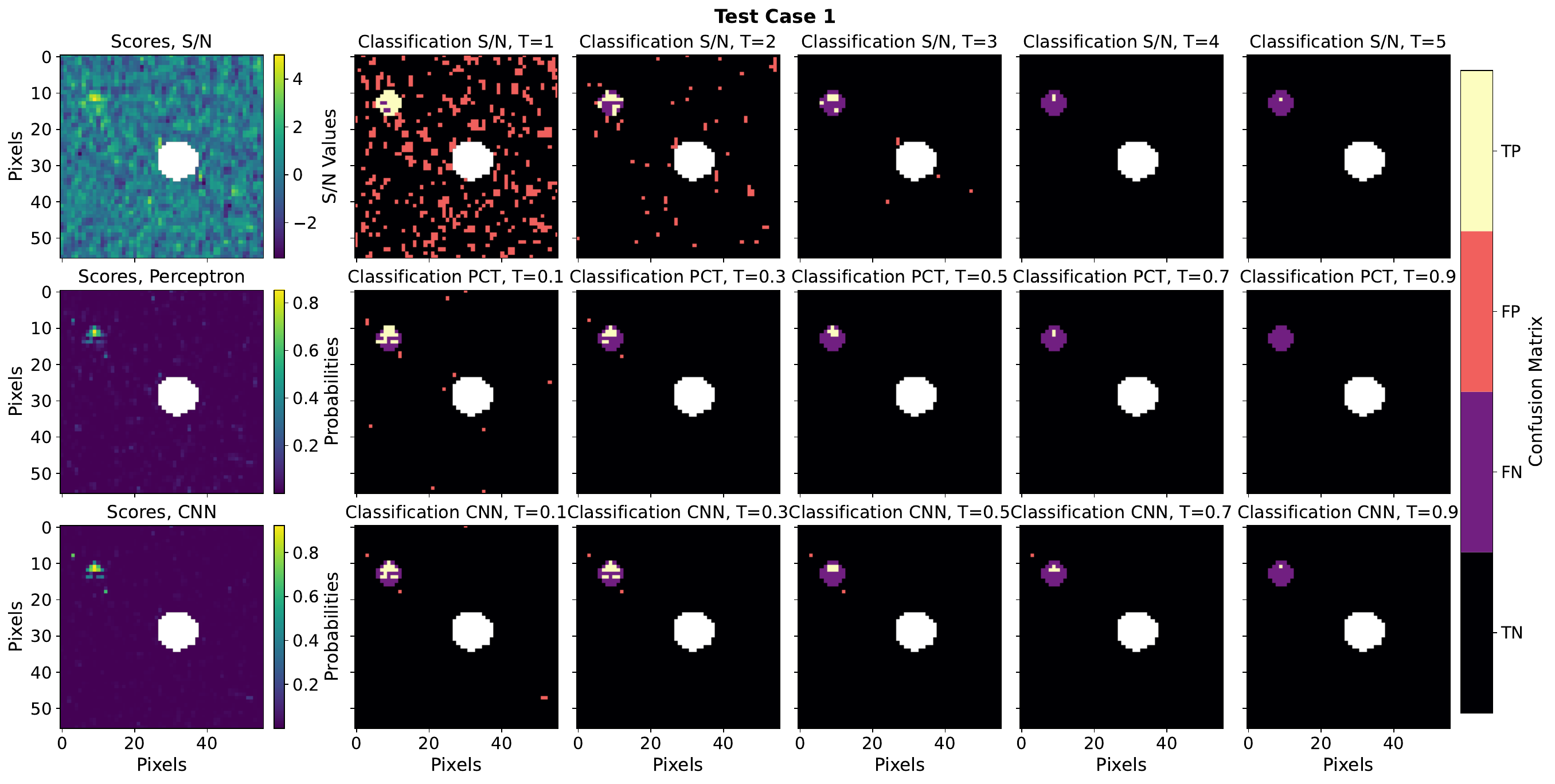}
    \caption{Reconstruction of classification scores by S/N, PCT, and CNN2 for various test cases (results of CNN1 are redundant). Equivalent benchmarking is challenging, but objectively applied thresholds highlight the distinct behaviour of each method's scores. \textit{First column:} Scores are displayed for the three methods in all test cases. \textit{Second to sixth column:} Classification maps shown according to increasing thresholds, set to $T=\{1; 2; 3; 4; 5\}$ for S/N and to $T=\{0.1;0.3;0.5;0.7;0.9\}$ for MLCCS. Four colours represent the confusion matrix elements: true positives (TP), false positives (FP), false negatives (FN) and true negatives (TN).}
    \label{app:grid1}
\end{figure*}

\begin{figure*}[tbh!]
    \centering
    \ContinuedFloat
    \includegraphics[width=16.8cm]{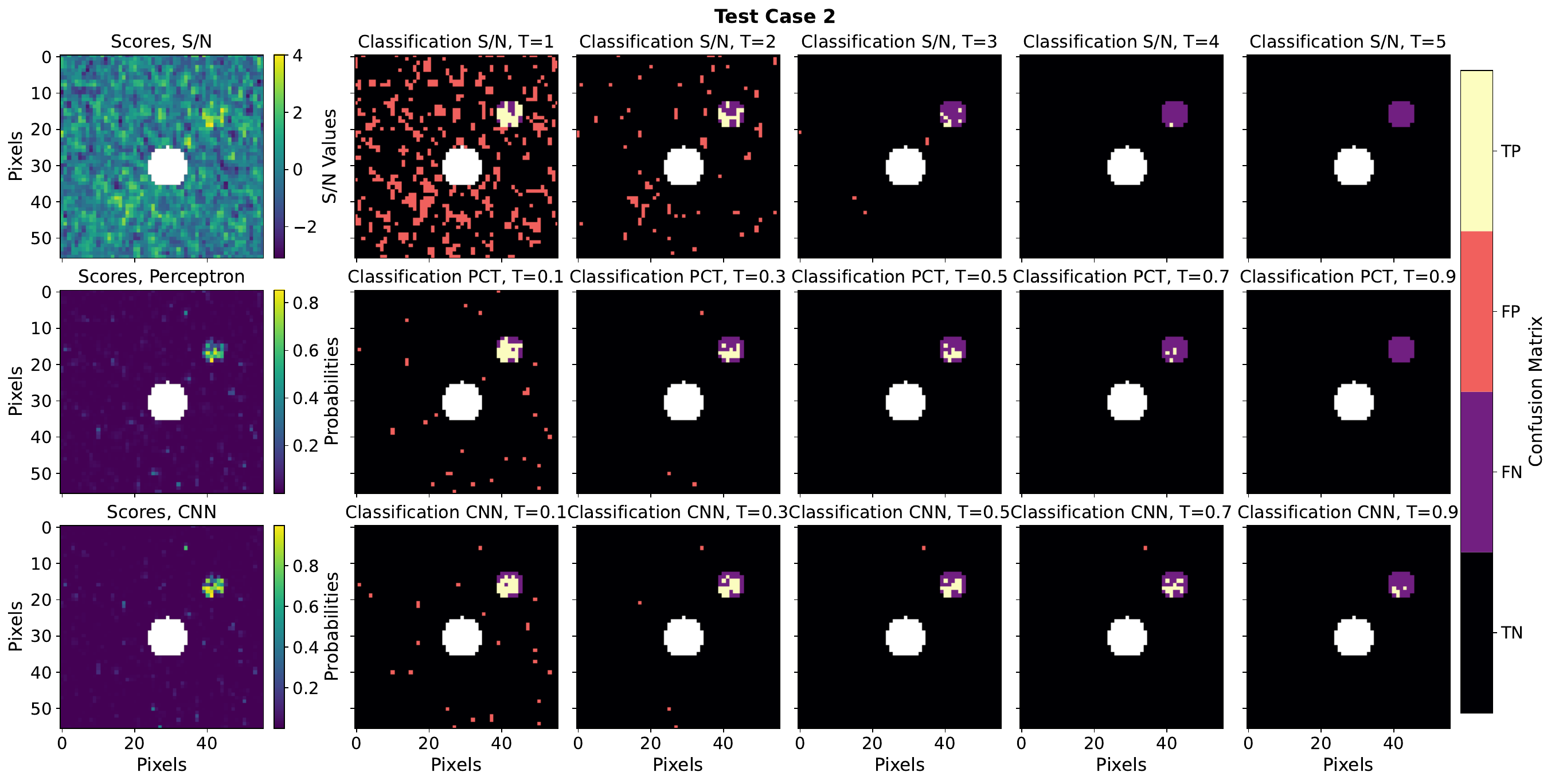}
    \includegraphics[width=16.8cm]{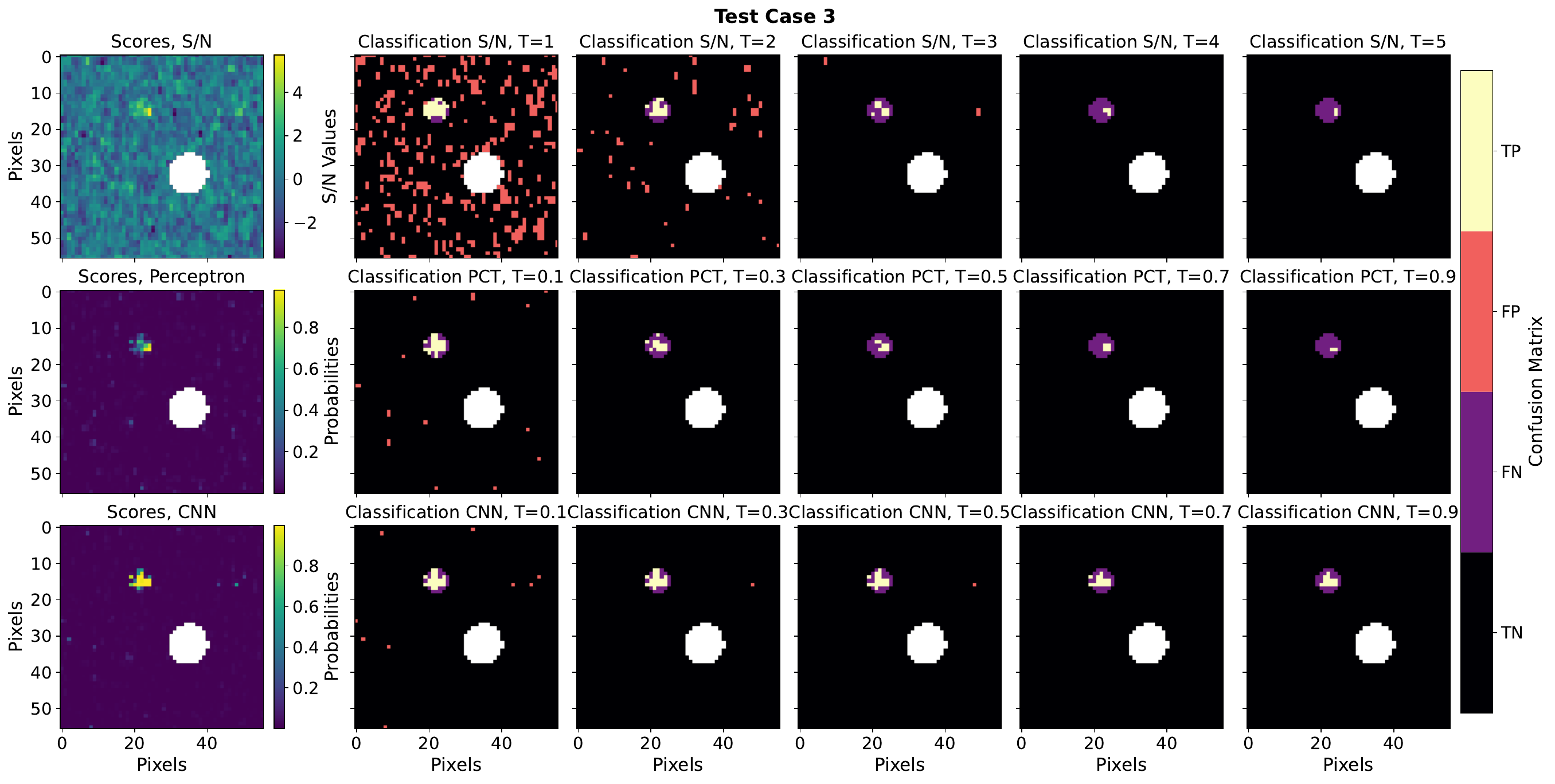} 
        \caption{Continued}
    \label{app:grid1}
\end{figure*}

\begin{multicols}{2}
In this section, we show the results of the three test cases of the \textit{Imaged brown dwarf} datasets. These test datasets have been constructed in the same way as their training and validation sets, namely by inserting aperture delimited planets. The synthetic planets are inserted with a random Gaussian ellipse size, shape and location, hence the aperture size and centre is adaptive to the insertion properties. The aperture delimitation allows to use the confusion matrix to evaluate the results, after the predictive classification is performed for a given threshold.

\begin{figure*}[tbh!]
    \centering
    \includegraphics[width=\hsize]{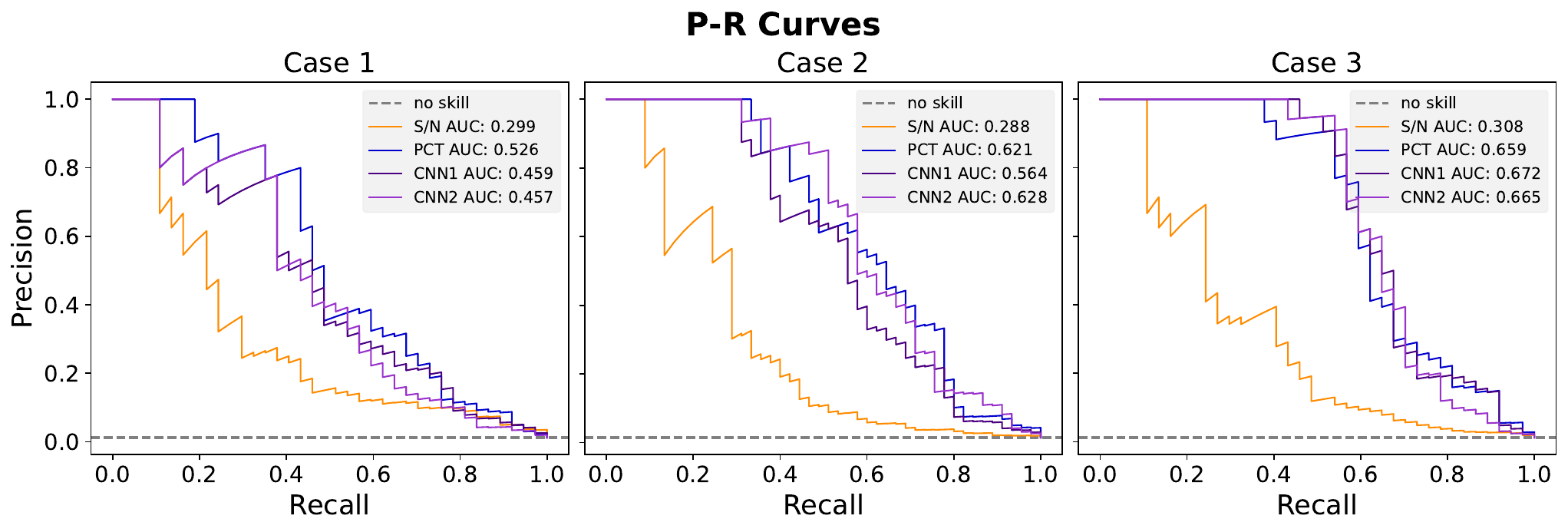}
    \caption{Precision-Recall curves related to the three test cases. Each curve is computed based on the scores attributed to each spaxel retrieved from an image cube. The curves represent the trade-off between precision and recall achieved by the scores. A classifier with no skill (gray dashed line) is proportional to the number of positive values in the data; it would perform random, uninformed classification of 0.1\% of the of the data to the positive class.}
    \label{app:grid2}
\end{figure*}

The results are visible in the three test cases in Fig.~\ref{app:grid1}, as scoring maps on the very left, and classification results according to increasing thresholds in the five columns on the right. Due to space constraints and to avoid redundancy, we only show the outcomes of CNN2; CNN1 results were very similar and equivalent. Overall, the plots show that the MLCCS classifiers generally enhance the prediction certainty. In terms of scoring results, there is a clear conspicuity improvement offered by the MLCCS methods. While the planet is barely visible in the S/N and really blends into the noise, it clearly appears in the probability scores of MLCCS methods. In fact, they are visible even if the detection score is low, because the noise is very clearly classified with high "negative" confidence (i.e. low probability to belong to the positive group). Fig.~\ref{app:grid2} aims to quantify the quality of the scores using P-R curves, described in Sect.~\ref{sec:results}. In case 1, the perceptron outperforms the CNNs, with an AUC of 0.526 against 0.459. In case 2, only CNN2 outperforms the perceptron, showing an AUC of 0.628 against 0.621; CNN1 however shows a weaker AUC. In case 3, both CNNs outperform the perceptron,  with AUCs of 0.672 and 0.665 for CNNs against 0.659 for S/N. Overall, all ML methods perform relatively equivalently, but show significant improvement compared to the S/N baseline which has an AUC between 0.288 and 0.308 over the plots.

The enhanced prediction certainty is also visible with the classifications at low thresholds (c.f. columns 2 to 5 in Fig.~\ref{app:grid1}). This fact stands out particularly at the two first thresholds. For example, $T=0.1$ on MLCCS probability scores shows significantly less FPs than $T=1$ and $T=2$ of the S/N for all cases. While increasing the thresholds neural networks are better able to reduce false positives while better preserving true positives. By comparing a threshold at $T=0.3$ for MLCCS methods against $T=3$ for the S/N metric, we can observe more TPs for less FPs; specifically, test case 1 presents 8 TPs for 5 FPS for S/N. Test cases 2 and 3 present about double amount of TPs for CNNs against S/N for similar to half the amount of FPs.  

To control that the low variability in the probability scores attributed to noise are not a result of overfitting, we conducted two categories of tests. Obviously, all tests cases were run on GQ\,lup\,B noise cubes, and trained on $17$ out of $19$ cubes as we exclude the validation and test sets, meaning that no noise spaxel is seen twice at any time. Yet, test cases 2 and 3 were output by a ML model which was validated on another GQ\,lup\,B noise cube, which implies that it was optimised according to the correct target noise. Now, for test case 1, we use a target noise cube of PZ\,Tel\,B to validate the model during training, to investigate if the results remain stable (i.e. if the model is not overfitting). This means that the model was optimised regarding a different noise distribution, and can still detect the simulated brown dwarf in a GQ\,lup\,B noise test cube. The probability scores attributed to the noise look less smooth, but the conspicuity is still very clear. This test is a first step towards investigating noise invariance across targets, for a given instrument. Nevertheless, it is crucial to emphasise that despite our precautions, the risk for moderate overfitting remains, even when the training set was not exposed to the spaxels of the validation or test set, due to the fact that other exposure cubes from the same target still remain in the training set. However, we highlight that the scenario of training and testing on similar noise cubes is not unlikely. In a real case, it is possible to train the models using noise from parts of exposure cubes that we expect to be empty, and test for regions of the leftover cubes where the planet is likely to be present. Yet, this procedure involves prior information about the orbit, inclination and expected location of the planet, from RV or astrometric data. If we rather focus on a blind search to discover new planets without prior information, we must indeed test the capacity of the networks to become invariant to noise across targets, which should be a focus for future work.
\end{multicols}

\section{Extended results on the PZ\,Tel\,B simulations}\label{app:extended_molmapsim}

\begin{figure*}[tbh!]
    \centering
    \includegraphics[width=17.5cm]{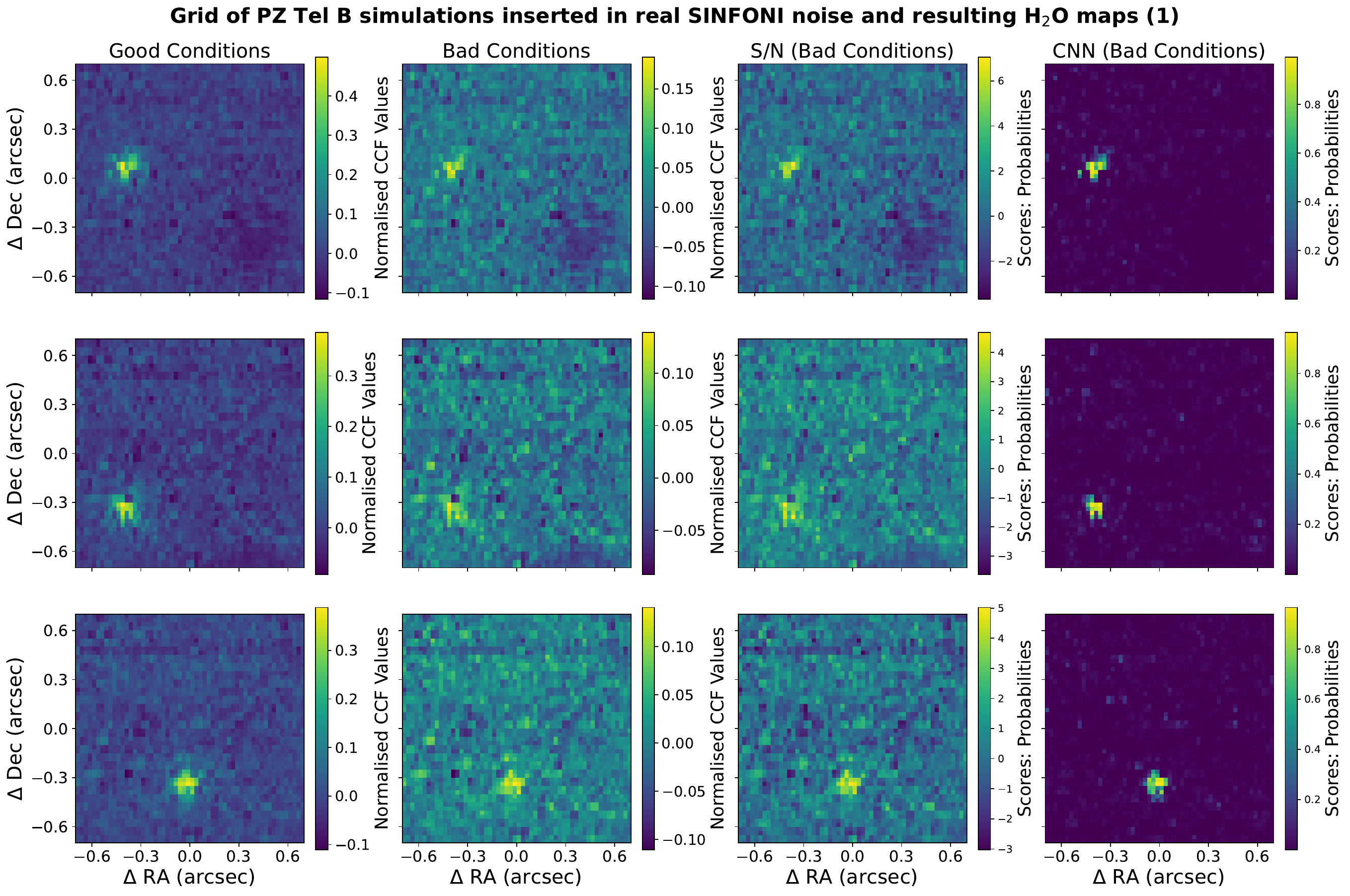}
    \includegraphics[width=17.5cm]{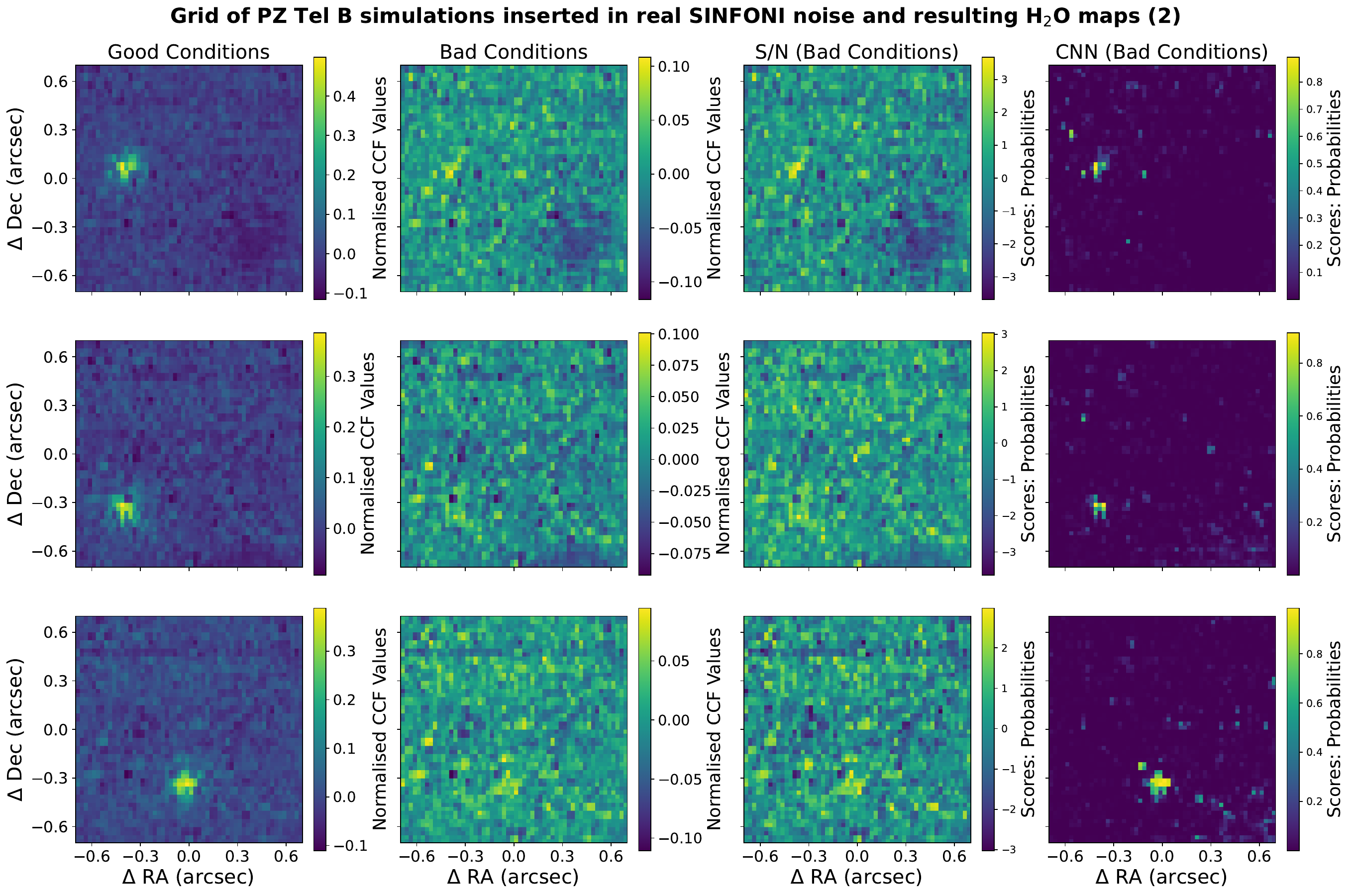}
    \caption{H$_2$O Scoring results for PZ\,Tel\,B simulations in its real noise. \textit{Subfigures 1 and 2, column 1:} The brown dwarf was inserted with its original average signal strength and decay to represent good seeing conditions. \textit{Subfigure 1 and 2, col. 2:} Then, it is inserted at $1/3$ of the original signal strength in good conditions (\textit{Subfigure 1}) and a second time at $1/6$ (\textit{Subfigure 2}). \textit{Subfigure 1,2; col. 3,4:} The S/N maps (\textit{col. 3}) and CNN maps (\textit{col. 4}) show the scoring results for the insertions in bad seeing conditions from col. 2.}
    \label{app:scores_grid1}
\end{figure*}

\begin{multicols}{2}
This section shows the full tests on the realistic PZ\,Tel\,B simulations inserted in its real SINFONI K-band noise at medium resolution. Details on the characteristics of the simulated planet are already described in Sect.\ref{subsec:resultspztel}. Fig.~\ref{app:scores_grid1} shows the results for two different noise levels (i.e. $\alpha$ scaling factor values) for (1) and (2). The first column of all plots shows a simulated reproduction of PZ\,Tel\,B under good conditions. For the three original real data cubes in good conditions, we measure the average S/N as $\beta = \{7.89; 5.893; 7.303\}$ in an aperture of 3.5 pixels, and calculate a decay rate of a Gaussian PSF as $\delta =\{3.595; 3.955; 3.701\}$. Then, we insert the simulated brown dwarf companions in respective noise cubes from bad observing conditions, with the corresponding S/N and Gaussian decay rate using a re-scaling factor of respectively $\alpha = \{121.23; 116.75; 113.85\}$. 

\begin{figure*}[tbh!]
    \centering
    \includegraphics[width=17.5cm]{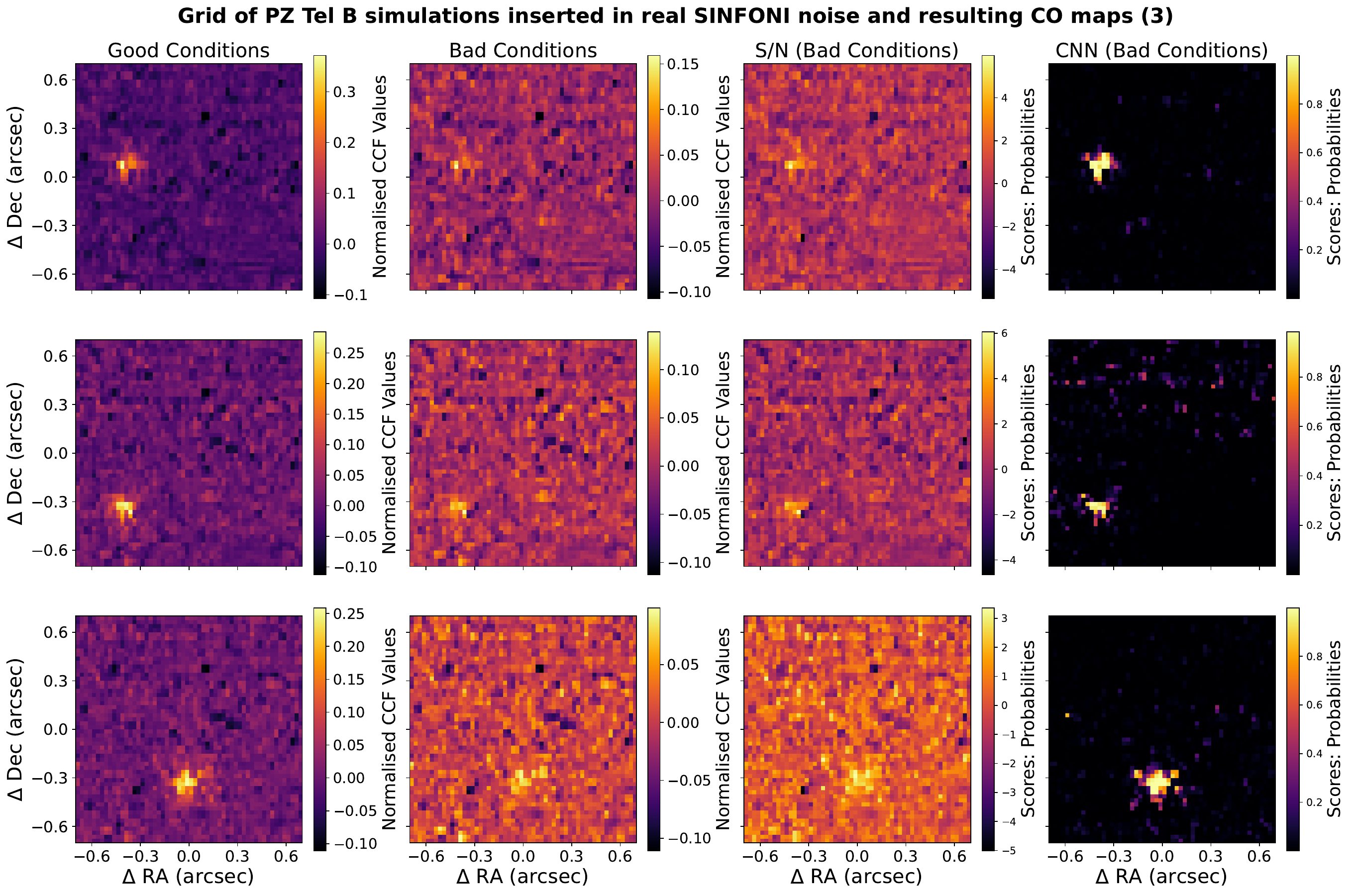}
    \includegraphics[width=17.5cm]{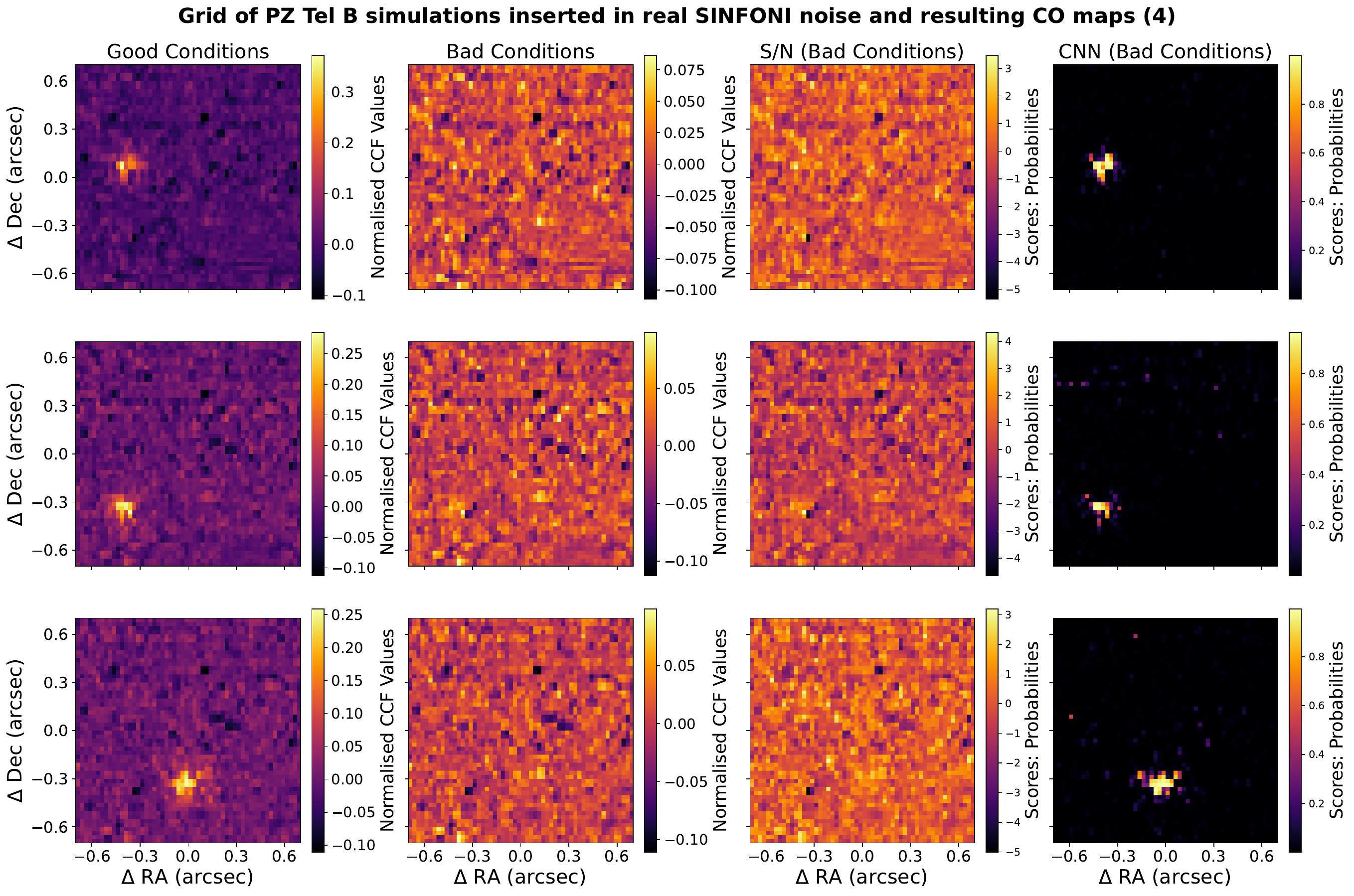}
    \caption{CO Scoring results for PZ\,Tel\,B simulations in its real noise. \textit{Subfigures 1 and 2, column 1:} The brown dwarf was inserted with its original average signal strength and decay to represent good seeing conditions. \textit{Subfigure 1 and 2, col. 2:} Then, it is inserted at $1/3$ of the original signal strength in good conditions (\textit{Subfigure 1}) and a second time at $1/6$ (\textit{Subfigure 2}). \textit{Subfigure 1,2; col. 3,4:} The S/N maps (\textit{col. 3}) and CNN maps (\textit{col. 4}) show the scoring results for planet insertions as bad seeing conditions from col. 2.}
    \label{app:scores_grid2}
\end{figure*}

The second columns of all plots show simulated insertions in bad conditions. For plot (1), the scaling factor $\alpha$ is cut down by 3 to align the parameters of $\beta$ and $\gamma$ with measurements on real data in bad conditions. This results in average S/N of $\beta = \{2.666;  2.07;  2.538\}$. For plot (2), we challenge the MLCCS models to a more extreme case; the simulated PZ\,Tel\,B was inserted at a scale factor $\alpha$ cut down by a factor of 6, yielding an extremely faint average S/N of $\beta = \{1.105; 0.997; 1.223\}$, hardly visible using the detection statistic. We apply CNN2 to obtain probability score maps on all cases (right column). Even when the S/N fails to find the brown dwarf in its scoring map, the CNN is able to reveal it. Analogue tests are run and confirmed for CO in Fig.~\ref{app:scores_grid2}.
\end{multicols}

\label{lastpage}
\end{document}